\documentclass[a4paper,11pt]{article} 
\usepackage{jheppub} 

\usepackage[table]{xcolor}                  
\usepackage{colortbl}
\usepackage{booktabs}                       
\usepackage{multirow,bigdelim}              
\usepackage{longtable}                      
\usepackage{arydshln}                       

\usepackage{units}                          
\usepackage{amsmath,amssymb,mathtools}      
\usepackage{slashed}                        
\usepackage{amsthm}
\usepackage{bbm,dsfont}                     

\usepackage{bm}                             
\usepackage{upgreek}                        
\usepackage[vcentermath]{youngtab}          
\usepackage{ytableau}                       
\usepackage{empheq}                         
\usepackage{suffix}	                        
\usepackage{lscape}	                        

\usepackage{graphicx}                       
\usepackage{subfig}	                        
\usepackage{tikz}                           
\usepackage{float}
\usepackage[export]{adjustbox}              

\usepackage{url}                            
\usepackage{makeidx}                       
\usepackage[numbers,sort&compress]{natbib}  
\usepackage[colorlinks=true,urlcolor=blue,anchorcolor=blue,citecolor=blue,filecolor=blue,linkcolor=blue,menucolor=blue,pagecolor=blue,linktocpage=true,pdfproducer=medialab,pdfa=true]{hyperref}	                    

\usepackage{ifthen}	                        

\usepackage{makeidx}
\usepackage{soul} 

\numberwithin{equation}{section}        
\numberwithin{table}{section}
\numberwithin{figure}{section}

\graphicspath{{figs/}}

\makeindex                              

\theoremstyle{plain}                    

\theoremstyle{definition}

\DeclareMathOperator{\Tr}{Tr}

\def\mc{\mathcal}
\def\md{\mathbf}

\def\ms{\mathsf}

\newcommand{\wh}[1]{\widehat{#1}}
\newcommand{\wt}[1]{\widetilde{#1}}

\def\IC{\mathbb{C}}

\def\IF{\mathbb{F}}

\def\IP{\mathbb{P}}
\def\IQ{\mathbb{Q}}
\def\IR{\mathbb{R}}
\def\IZ{\mathbb{Z}}

\def\cA{\mathcal{A}}

\def\cC{\mathcal{C}}

\def\cF{\mathcal{F}}

\def\cN{\mathcal{N}}
\def\cO{\mathcal{O}}

\def\dc{\mathbf{c}}

\def\dk{\mathbf{k}}

\def\dr{\mathbf{r}}

\def\dt{\mathbf{t}}

\def\dz{\mathbf{z}}

\def\sb{\mathsf{b}}

\def\vev#1{\left\langle #1 \right\rangle}

\DeclarePairedDelimiterX\opm[3]{\langle}{\rangle}{#1 \delimsize\vert #2 \delimsize\vert #3}
\DeclarePairedDelimiterX\ip[2]{\langle}{\rangle}{#1 \delimsize\vert #2}

\def\({\left(}
\def\){\right)}
\def\[{\left[}
\def\]{\right]}

\newcommand{\be}{\begin{equation}}
\newcommand{\ee}{\end{equation}}
\newcommand{\ba}{\begin{aligned}}
\newcommand{\ea}{\end{aligned}}
\newcommand{\ben}{\begin{eqnarray}\displaystyle}
\newcommand{\een}{\end{eqnarray}}

\newcommand{\pd}{\partial}

\newcommand{\re}{{\rm e}}
\newcommand{\ri}{{\mathsf{i}}}
\newcommand{\rd}{{\rm d}}

\newcommand{\nn}{\nonumber \\}

\makeindex

\def\IC{\mathbb{C}}

\def\IF{\mathbb{F}}

\def\IP{\mathbb{P}}
\def\IQ{\mathbb{Q}}
\def\IR{\mathbb{R}}
\def\IZ{\mathbb{Z}}

\newcommand{\CA}{\mc{A}}
\newcommand{\CC}{\mc{C}}
\newcommand{\CF}{\mc{F}}
\newcommand{\CO}{\mc{O}}
\newcommand{\CL}{\mc{L}}
\newcommand{\CM}{\mc{M}}
\newcommand{\CS}{\mc{S}}

\newcommand{\ep}{\epsilon}
\newcommand{\sC}{\ms{C}}
\newcommand{\sD}{\ms{D}}
\newcommand{\sS}{\ms{S}}
\newcommand{\sW}{\ms{W}}

\renewcommand{\dk}{\boldsymbol{k}}
\renewcommand{\dt}{\boldsymbol{t}}
\renewcommand{\dz}{\boldsymbol{z}}
\newcommand{\dS}{\boldsymbol{S}}

\allowdisplaybreaks[1]          

\title{\boldmath Resurgent Wilson loops in refined topological string}%
\author[a]{Jie Gu,}%
\author[a]{Gengbei Guo}
\affiliation[a]{School of Physics and Shing-Tung Yau Center\\
  Southeast University, Nanjing 210096, China}%

\emailAdd{jie-gu@seu.edu.cn}

\abstract{We study the resurgent structures of Wilson loops in refined
  topological string theory.  We argue that the Borel singularities
  should be integral periods, and that the associated Stokes constants
  are refined Donaldson-Thomas invariants, just like the free
  energies, except that the Borel singularities cannot be local flat
  coordinates.  We also solve the non-perturbative series in closed
  form from the holomorphic anomaly equations for the refined Wilson
  loops.  We illustrate these results with the examples of local
  $\IP^2$ and local $\IP^1\times\IP^1$.}


\begin{document}
\maketitle
\flushbottom


\section{Introduction}

It is now a general agreement that most asymptotic series in physics
are in want of a non-perturbative completion.  Topological string
theory as a subsector of type II superstring theory compactified on
Calabi-Yau threefolds has many asymptotic series, such as free
energies and Wilson loop vevs, which are both mathematically
well-defined and more amenable to calculations, and it is therefore a
perfect laboratory to explore non-perturbative completions of
asymptotic series.
In general, 
the non-perturbative completion would be ambiguous.
However, under
the assumption that the asymptotic series in topological string is
resurgent, the non-perturbative corrections are strongly constrained,
and the powerful method of resurgence theory \cite{Ecalle} can be used
to study them.

According to the resurgence theory, the non-perturbative corrections
to an asymptotic series of Gevrey-1 type
\begin{equation}
  \varphi(z) = \sum_n \varphi_n z^n,\quad \varphi_n\sim n!
\end{equation}
can be encoded in the form
\begin{equation}
  \varphi^{(*)} = \re^{-\CA_*/z}\sum_{n} \varphi^{(*)}_n z^{n+b_*}.
\end{equation}
Furthermore, these non-perturbative corrections are closedly related
to the perturbative series via Stokes transformations, so that much 
information of the non-perturbative corrections can be extracted from
the perturbative series $\varphi$.  For instance, the non-perturbative
\emph{action} $\CA_*$ that determines the magnitude $\re^{-\CA_*/z}$
of the non-perturbative correction is given as singularities of the
Borel transform of the asymptotic series, while the
\emph{coefficients} in the non-perturbative corrections, together with
the coefficients of the Stokes transformations known as the
\emph{Stokes constants}, can be read off from the large order
asymptotics of the perturbative coefficients.  These data are
sometimes collectively called the resurgent structure.

The resurgence method was first applied to study the perturbative free
energy $F(t;g_s)$ in topological string
\cite{Marino:2006hs,Marino:2007te,Marino:2008ya,Marino2009,Pasquetti2010} as an
asymptotic series in the string coupling constant $g_s$, 
and a rich structure of non-perturbative corrections was discovered. 

First of all, it was found that the non-perturbative actions or the
Borel singularities are integral periods of the mirror Calabi-Yau
threefolds
\cite{Drukker:2011zy,Aniceto2012,Couso-Santamaria2016,Couso-Santamaria2015,Gu:2023mgf}\footnote{See
  \cite{Marino:2006hs} for a possible conceptual understanding of this
  phenomenon, and also e.g.~\cite{Kazakov:2004du} for a similar
  understanding in minimal string.}, which are the central charges of
D-brane bound states in the type II superstring theory, corresponding
to D6-D4-D2-D0 bound states in type IIA superstring or equivalently to
D5-D3-D1-D(-1) bound states in type IIB superstring, giving the first
hint that the non-perturbative corrections in topological string are
related to D-branes.

Secondly, 
it was postulated and verified in
\cite{Couso-Santamaria2016,Couso-Santamaria2015,Couso-Santamaria2017a}\footnote{See
  \cite{Couso-Santamaria:2016vwq} also for important connection to
  another program of making non-perturbative completition to
  topological string free energy, the TS/ST correspondence
  \cite{Grassi:2014zfa}.}  that the non-perturbative trans-series are
constrained by the holomorphic anomaly equations
\cite{Bershadsky:1993ta,Bershadsky:1993cx}, the same set of partial
differential equations that constrains the perturbative free energies.
This idea was later further developed and exploited to full extent,
and the full non-perturbative trans-series in any non-perturbative
sector was solved in closed form \cite{Gu:2022sqc,Gu:2023mgf}.

Finally, important progress have been made regarding the Stokes
constants.  Although their exact calculation is still out of reach in
generic scenarios\footnote{Many Stokes constants can be obtained
  though in the special simplifying conifold limit
  \cite{Gu:2021ize}.}, there has been accumulating evidence
\cite{Bridgeland:2016nqw,Bridgeland:2017vbr,Alim:2021mhp,Gu:2021ize,Gu:2023wum,Iwaki:2023rst}
that the Stokes constant associated to each non-perturbative sector is
the Donaldson-Thomas invariant, the counting of \emph{stable} bound
states of D-branes.  It also elucidates further the nature of
non-perturbative actions: they are not only D-brane central charges,
but the central charges of stable D-brane configurations.  

Recently, these progress have been generalized
\cite{Alexandrov:2023wdj} (see also
\cite{Alim:2022oll,Grassi:2022zuk}) 
to refined free energies in topological string
$F(t;\ep_1,\ep_2)$\footnote{The Nekrasov-Shatashvili limit of the
  refined topological string is special, and the non-perturbative
  corrections to free energies were studied in
  \cite{Codesido:2017jwp,CodesidoSanchez:2018vor,Gu:2022fss}}.  The
refined free energy is treated as an asymptotic series in $g_s$
denoted as $F(t,\sb;g_s)$, where the parameters of Omega background
$\ep_1,\ep_2$ are related to $g_s$ via\footnote{Our convention differs
  from that in \cite{Alexandrov:2023wdj} by $g_s\rightarrow \ri g_s$.}
\begin{equation}
  \ep_1 = \ri\sb g_s,\quad \ep_2 = -\ri\sb^{-1} g_s,
\end{equation}
with the fixed parameter $\sb$, and it returns to the unrefined free
energy in the limit $\sb \rightarrow 1$.  For the refined free energy
$F(t,\sb;g_s)$, it was found that each non-perturbative sector with
action $\CA$ splits to two whose actions are $\CA/\sb$ and $\CA\,\sb$.
The trans-series in such a non-perturbative sector was also written
down in closed form \cite{Alexandrov:2023wdj}, as a trans-series
solution to the refined version of the holomorphic anomaly equations 
\cite{Huang:2010kf}.  Finally, it was argued that in this case, the
Stokes constants can be identified with the motivic or refined DT
invariants.

In this paper, we would like to study the non-perturbative corrections
of another important computable in refined topological string theory,
the Wilson loops. 
They play the role of eigenvalues in the quantum mirror curves
\cite{Aganagic:2011mi} in the Nekrasov-Shatashvili limit of the
refined topological string, and they are encoded in the $qq$-character
in the generic Omega background \cite{Nekrasov:2015wsu}.
Topological string on a local Calabi-Yau threefold engineers a 5d
$N=1$ SCFT, and when the latter has a gauge theory phase, the Wilson
loops are naturally defined.  The notion of Wilson loops was extended
to topological string on generic local Calabi-Yau threefolds
\cite{Kim:2021gyj} as insertion of additional non-compact 2-cycles,
and furthermore, holomorphic anomaly equations for the generalized
Wilson loops were written down in \cite{Huang:2022hdo,Wang:2023zcb}.

The resurgent structures of refined Wilson loops in the
Nekrasov-Shatashvili limit were discussed in \cite{Gu:2022fss}.  The
NS Wilson loops in different representations are proportional to each
other \cite{Huang:2022hdo}, and one only needs to study those in the
fundamental representation.  The generic refined Wilson loops in
different representations are very different, and their resurgent
structures could potentially be very rich.

We follow the idea of \cite{Grassi:2022zuk,Alexandrov:2023wdj}, and
treat the perturbative refined Wilson loops, which have similar
expansions as refined free energies in terms of the Omega deformation
parameters $\ep_1,\ep_2$, as aymptotic series in $g_s$ with fixed
deformation parameter $\sb$, which turn out also to be of Gevrey-1
type, so that the resurgence theory can be used to study the
non-perturbative corrections.  We find that rather than studying the
Wilson loops in different representations, it is more beneficial to
consider the \emph{generating series} of Wilson loops in all
representations, which can be regarded as the free energy of the
topological string on a new threefold with insertion of additional
non-compact two-cycles in the original Calabi-Yau threefold.  Although
a rigorous mathematical formulation is still lacking\footnote{With too
  much insertion, the threefold may cease to be a Calabi-Yau.  See
  \cite{Guo:2024wlg} for a mathematical theory of the Wilson loops.},
the similarity with the topological string free energy strongly
implies that the resurgent structures of the two are also similar.
Indeed, we find that for refined Wilson loops, the non-perturbative
actions are also integral periods, with the caveat that they cannot be
local flat coordinates (i.e.~A-periods); the non-perturbative series
can be solved in closed form from holomorphic anomaly equations for
Wilson loops; the Stokes constants must be the same as those of
refined free energy, and therefore they are also identified with
refined DT invariants.



The rest of the paper is structured as follows.  We review the
resurgent structure of free energy of refined topological string in
Section~\ref{sc:F}, the definition and the calculation of perturbative
Wilson loops in Section~\ref{sc:WL-pert}.  We present our results of
the resurgent structures of Wilson loops in Section~\ref{sc:WL-nonp}.
We show how to reduce to the unrefined limit and the NS limit in
Section~\ref{sc:limits}, and we are able to reproduce the results in
\cite{Gu:2022fss} and in particular prove some empirical observations,
especially the one that the Stokes constants of NS Wilson loops and
those of NS free energies were identical, when the former were not
vanishing.
Two examples of local $\IP^2$ and local $\IP^1\times \IP^1$
are given in Section~\ref{sc:examples}.



\section*{Acknowledgement}
We would like to thank Babak Haghighat, Albrecht Klemm, and Xin Wang
for useful discussions, and thank Marcos Mari\~no for a careful
reading of the manuscript. We also thank the hospitality of Fudan
University, the host of the workshop ``String Theory and Quantum Field
Theory 2024'', and the Tsinghua Sanya International Mathematics Forum,
the host of the workshop ``Modern Aspects of Quantum Field Theory'',
where the results of this work were presented.  J.G. is supported by
the startup funding No.~4007022316 of the Southeast University, and
the National Natural Science Foundation of China (General Program)
funding No.~12375062.

\section{Free energies and non-perturbative corrections}
\label{sc:F}

\subsection{Perturbative refined free energy}
\label{sc:F-pert}

The refined topologial string theory is defined over the complexified
Kahler moduli space of Calabi-Yau threefold $X$ in the A-model, or the
complex structure moduli space of the mirror Calabi-Yau threefold
$X^\vee$ in the B-model.
The perturbative free energy of refined topological string theory is a
formal power series in terms of two expansion parameter $\ep_1,\ep_2$
\cite{Huang:2010kf}
\begin{equation}
  \CF(\dt;\ep_1,\ep_2) = \sum_{n,g\geq 0}
  (\ep_1+\ep_2)^{2n}(\ep_1\ep_2)^{g-1} \CF^{(n,g)}(\dt).
\end{equation}
The coefficients $\CF^{(n,g)}(\dt)$ are sections of certain line
bundles of the moduli space, parametrized by flat coordinates $\dt$.

There are various ways to compute the coefficients of the refined free
energy, including instanton calculus \cite{Nekrasov:2002qd}, refined
topological vertex \cite{Iqbal:2007ii}, and blowup equations
\cite{Huang:2017mis} (see also \cite{Grassi:2016nnt} and
\cite[Sec.~8]{Gu:2017ccq}).  But for our purpose, it is more suitable
to use the refined holomorphic anomaly equations (refined HAE)
\cite{Huang:2010kf}, as it generates directly the coefficients of the
perturbative series, and in addition, it is applicable over the entire
moduli space.

In the framework of refined HAE, it is necessary to extend the free
energies $\CF^{(n,g)}(\dt)$ to non-holomorphic functions $F^{(n,g)}$,
and the failure of the holomorphicity is described by the refined HAE.
The non-holomorphic functions $F^{(n,g)}$ obtained from the refined
HAE reduce to $\CF^{(n,g)}$ in appropriate holomorphic limits, which
is equivalent to taking a local patch of the associated line bundle,
where $\dt$ are the flat coordinates on that local patch.  We follow
the convention in \cite{Gu:2022sqc,Gu:2023mgf} that we use Roman
capital letters for the non-holomorphic quantities obtained from HAE,
and curly captical letters for their holomorphic limit.  Note the
prepotential $F^{(0,0)}= \CF^{(0,0)} = \CF_0$ is always holomorphic.

To explain the refined HAE, we need to introduce a few properties of
the moduli space $\CM$.
The moduli space is Kahler, so that the metric is expressed in terms
of a Kahler potential, i.e.
\begin{equation}
  G_{a\bar{b}} = \pd_a\pd_{\bar{b}} K,
\end{equation}
where $\pd_a = \pd_{z^a}$ and the $z^a$ are a set of global complex
coordinates over the moduli space.  A set of covariant derivatives
$D_a$ can then be introduced with the Levi-Civita connection,
\begin{equation}
  \Gamma^a_{bc} = G^{a\bar{d}}\pd_bG_{c\bar{d}}.
\end{equation}
In addition, the moduli space is special Kahler (see
e.g.~\cite{klemm2018b} for more details of the special geometry.), in
the sense that in any local path of the moduli space, one can find a
symplectic basis -- known as a choice of frame -- of local flat
coordinates as well as their conjugates,
\begin{equation}
  \left\{t^a,\quad \frac{\pd \CF_0}{\pd t^a}\right\},
  \quad a=1,\ldots,\frac{1}{2}\dim\CM,
\end{equation}
both of which are classical periods of the mirror Calabi-Yau threefold
$X^\vee$, and they are related to each other via the preptential
$\CF_0 = \CF^{(0,0)}$.
We can then introduce the Yukawa coupling
\begin{equation}
  C_{abc}(\dz) = \frac{\pd t^\ell}{\pd z^a}\frac{\pd t^m}{\pd z^b}
  \frac{\pd t^n}{\pd z^c} \cdot
  \frac{\pd}{\pd t^{\ell}}\frac{\pd}{\pd t^m}\frac{\pd}{\pd t^n}\CF_0
\end{equation}
which, despite its definition, is independent of the choice of frame,
and is usually a rational function of $\dz$.
We also introduce the propagators $S^{ab}$ defined by
\begin{equation}
  \pd_{\bar{c}}S^{ab} =
  \re^{2K} G^{a\bar{d}} G^{b\bar{e}} \overline{C}_{\bar{c}\bar{d}\bar{e}}.
\end{equation}
which encode the non-holomorphic dependence.  The non-holomorphic free
energies $F^{(n,g)}$ are functions of both $z^a, S^{ab}$.

The refined topological string free energies then satisfy an
infinitely set of partial differential equations known as the refined
HAE \cite{Huang:2010kf},
\begin{equation}
  \label{eq:hae-ref}
  \frac{\pd F^{(n,g)}}{\pd S^{ab}} = \frac{1}{2}\left(D_aD_b F^{(n,g-1)} +
    \sideset{}{'}{\sum}_{n',g'\geq 0}D_a F^{(n',g')}D_b
    F^{(n-n',g-g')}\right),\quad n+g >2,
\end{equation}
where $\sum{}'$ means excluding $(n',g') = (0,0)$ or $(n,g)$.  These
infinitely many equations can be encoded in a single master equation
\begin{equation}
  \label{eq:hae-Z}
  \frac{\pd}{\pd S^{ab}} \wt{Z} = \frac{\ep_1\ep_2}{2}D_a D_b \wt{Z},
\end{equation}
with the partition function
\begin{equation}
  \wt{Z} = \exp \wt{F} = \exp \sum_{n+g>0}
  (\ep_1+\ep_2)^{2n}(\ep_1\ep_2)^{g-1} F^{(n,g)}.
\end{equation}
These equations are supplemented with the initial conditions that the
free energy $F^{(0,1)}$ is related to the propagator by
\begin{equation}
  \pd_a F^{(0,1)} = \frac{1}{2}C_{abc} S^{bc} + f_a(\dz)
\end{equation}
while $F^{(1,0)}$ is given by
\begin{equation}
  F^{(1,0)} = -\frac{1}{24}\log(f(\dz) \Delta(\dz)).
\end{equation}
Here $f_a(\dz)$ and $f(\dz)$ are model-dependent holomorphic functions
of $z^a$, and $\Delta(\dz)$ is the discriminant, the equation of
singular loci in the moduli space.

As the equations \eqref{eq:hae-ref} are recursive in genus $(n,g)$,
the free energies $F^{(n,g)}$ can be solved by direct integration up
to an integration constant, which is independent of the
non-holomorphic propagator $S^{ab}$, and which is purely holomorphic,
known as the holomorphic ambiguity.  The holomorphic ambiguity is
fixed by imposing the boundary conditions that at a conifold
singularity of the moduli space, the free energies $\CF^{(n,g)}(t_c)$
in the holomorphic limit with the local flat coordinate $t_c$
(appropriately normalised) that vanishes at the singularity satisfy
the so-called \emph{gap condition} \cite{Huang:2010kf,Krefl:2010fm}
\begin{align}
  \CF(t_c;\ep_1,\ep_2) =
  &\left(-\frac{1}{12} +
    \frac{1}{24}(\ep_1+\ep_2)^2(\ep_1\ep_2)^{-1}\right)\log(t_c)\nn +
  &\frac{1}{\ep_1\ep_2}\sum_{n_1,n_2\geq
    0}\frac{(2n_1+2n_2-3)!}{t_c^{2n_1+2n_2-2}}\hat{B}_{2n_1}\hat{B}_{2n_2}\ep_1^{2n_1}\ep_2^{2n_2}
    + \CO(t_c^0),
    \label{eq:gap-ref}
\end{align}
where
\begin{equation}
  \hat{B}_n = (1-2^{1-n})\frac{B_n}{n!}
\end{equation}
and $B_n$ denoting the Bernoulli numbers, and that the free energies
are regular everywhere else\footnote{In particular, the free energies
  should be regular at a pure orbifold point.  But in some models such
  as massless local $\IF_0$, a conifold singularity may be hidden
  inside the orbifold point so that gap also appears there.}
\footnote{We only consider local Calabi-Yau.  For topological string
  compact Calabi-Yau, there can be other types of singularities such
  as K-point singularities.}.

Another important universal feature of the refined free energy is that
near the large radius point \cite{Hollowood:2003cv,Huang:2010kf}, it
has the integrality structure
\begin{equation}
  \label{eq:GV-ref}
  \CF(\dt;\ep_{1,2}) = \sum_{w\geq 1}\sum_{\gamma\in
    H_2(X,\IZ)}\sum_{j_{L,R}} (-1)^{2j_L+2j_R}
  N^\gamma_{j_L,j_R}
  \frac{\chi_{j_L}(q_L^w)\chi_{j_R}(q_R^w)}
  {w(2\sinh \frac{w \ep_1}{2})(2\sinh \frac{w \ep_2}{2})}\re^{-w\gamma\cdot \dt}.
\end{equation}
generalising the Gopakumar-Vafa formula of unrefined free energy
\cite{Gopakumar:1998jq}.  Here $q_{L,R} = \re^{\ep_{L,R}}$ and
$\ep_{L,R} = \frac{1}{2}(\ep_1\mp \ep_2)$, and $\chi_j(q)$ is the
character of $su(2)$ of spin $j\in\frac{1}{2}\IZ_{\geq 0}$,
\begin{equation}
  \chi_j(q) = \frac{q^{2j+1} - q^{-2j-1}}{q - q^{-1}}.
\end{equation}
$N^\gamma_{j_L,j_R}$ are non-negative integer numbers, and they count
the numbers of stable D2-D0 brane bound states wrapping curve class
$\gamma \in H_2(X,\IZ)$ with spins $(j_L,j_R)$ in the little group
$SU(2)_L\times SU(2)_R$ in five dimensions, and they are known as the
BPS invariants.

\subsection{Non-perturbative corrections}
\label{sc:F-nonp}

It is more convenient to study non-perturbative corrections to
asymptotic series with a single expansion parameter.  It is therefore
suggested in \cite{Grassi:2022zuk,Alexandrov:2023wdj} to use the
parametrisation\footnote{The conventions in
  \cite{Grassi:2022zuk,Alexandrov:2023wdj} are slightly different.  We
  take in the majority of the paper the convention in
  \cite{Alexandrov:2023wdj} which is more symmetric, and revert in
  Section~\ref{sc:Fm-nonp-NS} to the convention in
  \cite{Grassi:2022zuk} which is more suitable for taking the NS
  limit.}
\begin{equation}
  \label{eq:ep12gs}
  \ep_1 = \ri \sb g_s, \quad \ep_2 = -\ri \sb^{-1} g_s
\end{equation}
to convert the refined free energy $F(\ep_1,\ep_2)$ to a univariate
power series in terms of the string coupling $g_s$,
\begin{equation}
  \label{eq:Fags}
  F(\sb;g_s) = F(\ri \sb g_s, -\ri \sb^{-1} g_s) =
  \sum_{g\geq 0} g_s^{2g-2} F_g(\sb),
\end{equation}
where the coefficients $F_g(\sb)$ are not only functions of the moduli
$\dt$ but also of the deformation parameter $\sb$, given by
\begin{equation}
  F_g(\sb) = \sum_{n=0}^g (-1)^n
  (\sb-\sb^{-1})^{2n} F^{(n,g-n)}.
\end{equation}


It is noted then in \cite{Alexandrov:2023wdj} that the HAE
\eqref{eq:hae-ref} can be re-cast as equations of the deformed free
energies $F_g(\sb) := F_g(\dz,\dS,\sb)$
\begin{equation}
  \label{eq:hae-alpha}
  \frac{\pd F_g(\sb)}{\pd S^{ab}} = \frac{1}{2}\left(
    D_a D_b F_{g-1}(\sb) + \sum_{h=1}^{g-1} D_a F_h(\sb) D_b F_{g-h}(\sb)
  \right),\quad g\geq 2.
\end{equation}
The initial condition is given by
\begin{equation}
  F_1(\sb) = F^{(0,1)} - (\sb-\sb^{-1})^2 F^{(1,0)}
\end{equation}
while at conifold points, the boundary condition \eqref{eq:gap-ref}
becomes \cite{Alexandrov:2023wdj}
\begin{equation}
  \label{eq:gap-alpha}
  \CF_g(t_c,\sb) = \frac{c_g(\sb)}{t_c^{2g-2}} + \CO(1),\quad
  g\geq 2,
\end{equation}
where the coefficients $c_g(\sb)$ are
\begin{equation}
  c_g(\sb) = -(2g-3)!\sum_{m=0}^{g} \wh{B}_{2m}\wh{B}_{2g-2m} \sb^{2(2m-g)}.
\end{equation}
This gap condition reduces to the universal conifold behavior of the
unrefined topological string \cite{Ghoshal:1995wm} in the limit
$\sb \rightarrow 1$.

The non-perturbative corrections to the refined free energies was
studied in \cite{Alexandrov:2023wdj}, based on previous works
\cite{Gu:2022sqc,Gu:2023mgf,Iwaki:2023rst}, and right now a fairly
good understanding has been obtained for all the ingredients,
including the non-perturbative action, the non-perturbative series, as
well as the Stokes constants, which we will quickly review here.

First of all, the non-perturbative actions were already systematically
studied
\cite{Pasquetti2010,Drukker:2011zy,Aniceto2012,Couso-Santamaria2016,Couso-Santamaria2015,Gu:2023mgf},
in the unrefined case.  Note that actions of non-perturbative sectors
appear as Borel singularities of the perturbative free energies.  It
was found that the Borel singularities always appear in pairs
$\pm \CA$, as the perturbative series is resonant in the sense that
$\CF(\dt;-g_s) = \CF(\dt;g_s)$.  Furthermore, they seem to appear not
alone but always in sequences $\CA,2\CA,3\CA,\ldots$.  And most
importantly, it was argued that the action $\CA$ is holomorphic, and
it is in fact an integer period of the mirror Calabi-Yau threefold
$X^\vee$; equivalently, it coincides with the central charge of a
D-brane bound state in either type IIA superstring compactified on $X$
or type IIB superstring compactified on $X^\vee$.  More concretely,
the action $\CA$ can be written as
\begin{equation}
  \label{eq:A-gamma}
  \CA_\gamma = -\ri c^a \frac{\pd \CF_0}{\pd t^a} + 2\pi d_a t^a + 4\pi^2 \ri d_0.
\end{equation}
Here $(\pd_{t^a}\CF_0, 2\pi t^a,4\pi^2 \ri)$ is a frame-dependent
basis of integer periods; $\gamma = (c^a,d_a,d_0)$ are integers, and
they are the D-brane charges\footnote{These are D4-D2-D0 charges in
  type IIA, or D3-D1-D(-1) charges in type IIB, up to an integer
  linear transformation.  As $X$ or $X^\vee$ is non-compact, there is
  no D6 or D5 brane charge.}, so that $\CA_\gamma$ equals the central
charge $Z(\gamma)$ of the corresponding D-brane bound state, up to
some overall factor.  In the case of refined free energy
$\CF(\dt,\sb;g_s)$, it was found that
\cite{Grassi:2022zuk,Alexandrov:2023wdj} each Borel singularity
$\CA_\gamma$ splits to two $\sb^{-1}\CA_\gamma$ and $\sb\CA_\gamma$.

The non-perturbative series are more complicated, but they can still
be written down in closed form.  One first notices that the
non-perturbative series is much simpler in the holomorphic limit of
the so-called $\CA$-frame, where the action $\CA$ is a local flat
coordinate on the moduli space, i.e.~the coefficients $c^a$ vanish
identically in \eqref{eq:A-gamma}.  By studying the genus expansion of
the gap condition \eqref{eq:gap-alpha}, as well as that of the refined
GV formula \eqref{eq:GV-ref}, it was argued in
\cite{Alexandrov:2023wdj} that the non-perturbative series associated
to the non-perturbative action $\ell\sb^{-1}\CA$ or $\ell\sb\CA$ has
only one-term, and it is given respectively by
\begin{subequations}
  \label{eq:FA}
\begin{align}
  \CF^{(\ell)}_{\CA,\sb} =
  &\frac{(-1)^\ell}{\ell}
    \frac{\pi}{\sin\left(\pi\ell/\sb^{2}\right)}\re^{-\ell \CA/(\sb g_s)},
    \label{eq:FA1}\\
  \CF^{{(\ell)}}_{\CA,1/\sb} =
  &\frac{(-1)^\ell}{\ell}
    \frac{\pi}{\sin\left(\pi\ell \sb^2\right)}\re^{-\ell \sb\CA/g_s}.
    \label{eq:FA2}
\end{align}
\end{subequations}

Next, following the idea in
\cite{Couso-Santamaria2016,Couso-Santamaria2015}, it is postulated
\cite{Alexandrov:2023wdj} that the non-perturbative series is also a
solution to the refined HAE \eqref{eq:hae-alpha}, and more
importantly, it can be solved exactly as in
\cite{Gu:2022sqc,Gu:2023mgf}, where the simple solutions \eqref{eq:FA}
are used analogously to the gap conditions as boundary conditions to
help fix holomorphic ambiguities.  To write down the solution, it is
more convenient to use the non-perturbative partition function that
encodes a sequence of non-perturbative free energies
\begin{equation}
  Z_r(\sb) = \exp \sum_{\ell=1}^\infty \CC^\ell \left(F^{(\ell)}_\sb
  + F^{(\ell)}_{1/\sb}\right)
\end{equation}
where $\cC$ is a bookkeeping parameter for the non-perturbative
sectors that can be set to one, and that
\begin{equation}
  \label{eq:Fl-ref}
  F^{(\ell)}_{\sb}=
  \re^{-\ell \CA/(\sb g_s)}\sum_{n\geq 0} g_s^{n}  F^{(\ell)}_{n,\sb},\quad
  F^{(\ell)}_{1/\sb}=
  \re^{-\ell \sb\CA/g_s}\sum_{n\geq 0} g_s^{n} F^{(\ell)}_{n,1/\sb}.
\end{equation}
In the holomorphic limit of the $\CA$-frame, \eqref{eq:FA} implies
that the non-perturbative partition function reads
\begin{equation}
  \label{eq:ZrA}
  Z_{r,\CA} = \exp\left[\sum_{\ell\geq 1} \cC^\ell
    \left(\CF_{\CA,\sb}^{(\ell)} + \CF_{\CA,1/\sb}^{(\ell)}
    \right)\right] = 1 + \sum_{n+m > 0}
  C_{n,m}\exp\left(-\frac{n\CA}{\sb g_s} -
    \frac{m\sb\CA}{g_s}\right)
\end{equation}
where $C_{n,m}$ can be read off from the expansion.  This serves as
the boundary conditions for solving the refined HAE
\eqref{eq:hae-alpha}.  Then in general, the non-perturbative partition
function reads
\begin{equation}
  \label{eq:Zr}
  Z_r = 1 + \sum_{n+m>0} C_{n,m} \exp \Sigma_{\frac{n}{\sb}+m \sb}
\end{equation}
where
\begin{equation}
  \Sigma_{\lambda} = \sum_{k\geq 1} \frac{(-\lambda)^k}{k!} \sD^{k-1} G.
\end{equation}
The derivative is
\begin{equation}
  \sD = g_s(\pd_b\CA)\left(S^{ab} - \CS^{ab}_{\CA}\right)\pd_a
\end{equation}
with $\CS^{ab}_{\CA}$ being the holomorphic limit of the propagator in
the $\CA$-frame, and the function $G$ is
\begin{equation}
  G = \sD F^{(0)} =  \frac{\CA}{g_s} + \sum_{g\geq 1} g_s^{2g-2}\sD F_g
\end{equation}
where we demand that $\sD F_0 = g_s\CA$ in any frame.  With this
convention, \eqref{eq:Zr} can also be written as
\begin{equation}
  Z_r = \frac{1}{Z^{(0)}}\left[\exp\sum_{\ell\geq 1} \frac{(-\cC)^{\ell}}{\ell}\left(
    \frac{\pi}{\sin(\pi\ell\sb^{-2})}\re^{-\ell\sb^{-1}\sD}
    +\frac{\pi}{\sin(\pi\ell \sb^2)}\re^{-\ell\sb\sD}
  \right) \right]Z^{(0)}
\end{equation}
In other words, it is lifted from $Z_{r,\CA}$ by substituting $\sD$
for $\cA/g_s$.
Here $Z^{(0)}$ is the perturbative partition function
\begin{equation}
  Z^{(0)} = \exp F^{(0)} = \exp \sum_{g\geq 0} g_s^{2g-2} F_g(\sb).
\end{equation}


The holomorphic limit of the general results can be easily obtained.
In an $\CA$-frame, the derivative $\sD$ vanishes in the holomorphic
limit, and we recover \eqref{eq:ZrA}.  If we are not in a $\CA$-frame,
so that the coefficients $c_a$ do not vanish identically, we can shift
the definition of the prepotential $\CF_0$ so that $d_a,d_0$ all
vanish.  Then in the holomorphic limit, the derivative $\sD$ reduces
to
\begin{equation}
  \label{eq:D-hol}
  \sD \rightarrow -\ri g_s c^a \frac{\pd}{\pd t^a}.
\end{equation}
and $\Sigma_\lambda$ becomes
\begin{equation}
  \label{eq:Sigma-hol}
  \Sigma_\lambda \rightarrow \CF(\dt + \ri \lambda g_s\dc,\sb;g_s)
  - \CF(\dt,\sb;g_s).
\end{equation}
For instance, the one-instanton amplitude is
\begin{subequations}
  \label{eq:F1}
  \begin{align}
    \CF^{(1)}_{\sb} =
    &\frac{\pi}{\sin(\pi/\sb^2)}
      \exp\left[\CF(\dt+\ri g_s\dc/\sb) - \CF(\dt)\right],
      \label{eq:F1a}\\
    \CF^{(1)}_{1/\sb} =
    &\frac{\pi}{\sin(\pi \sb^2)}
      \exp\left[\CF(\dt+\ri \sb g_s\dc) - \CF(\dt)\right].
      \label{eq:F1b}
  \end{align}
\end{subequations}
where $\CF(\dt) = \CF(\dt,\sb;g_s)$.

Finally, it was conjectured \cite{Alexandrov:2023wdj} (see also
\cite{Gu:2023wum,Iwaki:2023rst} as well as
\cite{Bridgeland:2016nqw,Bridgeland:2017vbr,Alim:2021mhp,Gu:2021ize}
for discussion in unrefined case) that the Stokes constant
$\sS_\gamma(\sb)$ (resp.~$\sS_\gamma(1/\sb)$) associated to
$\ell \CA_\ell/\sb$ (resp.~$\ell \sb\CA_\ell$) are all identical, and
it is identified with the refined (or motivic) Donaldson-Thomas
invariant.  More precisely, it is given by
\begin{equation}
  \label{eq:S-Omega}
  \sS_\gamma(\sb) = \Omega(\gamma,-\re^{-\pi\ri/\sb^2}).
\end{equation}
The refined DT invariants $\Omega(\gamma,y)$ is a $SU(2)$ character
given by 
\begin{equation}
  \Omega(\gamma,y) = \sum_{j}\chi_j(y) \Omega_{[j]}(\gamma),
\end{equation}
where the integers $\Omega_{[j]}(\gamma)$ count BPS multiplets due to
the stable D-brane bound state of charge $j$ with angular momentum
$j\in\IZ_{\geq 0}/2$.  For D2-D0 bound states, the refined DT
invariants are related to the BPS invariants $N^\gamma_{j_L,j_R}$
through
\begin{equation}
  \Omega(\gamma,y) =
  \sum_{j_L,j_R}\chi_{j_L}(y)\chi_{j_R}(y)N^\gamma_{j_L,j_R} =
  \sum_{j}\chi_j(y) \Omega_{[j]}(\gamma)
\end{equation}
with
\begin{equation}
  \Omega_{[j]}(\gamma) = \sum_{|j_L-j_R|\leq j \leq j_L+j_R} N^\gamma_{j_L,j_R},
\end{equation}
and they reduce to the genus zero Gopakumar-Vafa invariant
$n_{\gamma,0}$ wtih $\sb=1$ and $y=1$
\begin{equation}
  n_{\gamma,0} = \Omega(\gamma,1).
\end{equation}


\section{Perturbative Wilson loops in topological string}
\label{sc:WL-pert}


The refined topological string theory compactified on a local
Calabi-Yau threefold $X$ engineers \cite{Katz:1996fh,Katz:1997eq} a 5d
$\mc{N}=1$ SCFT $T[X]$ in the Coulomb branch on the Omega background
$S^1\times \IR^4_{\ep_1,\ep_2}$ \cite{Nekrasov:2002qd}, such that that
the partition functions of the two theories are the same, once we
identify appropriately the moduli spaces as well the parameters
$\ep_1,\ep_2$ of the two theories.

Some of these 5d SCFTs have a gauge theory phase.  The simplest
example is when the Calabi-Yau threefold is the canonical bundle over
$\IP^1\times \IP^1$, known as the local $\IP^1\times \IP^1$, and the
corresponding gauge theory is a 5d $G = SU(2)$ SYM.  In these cases,
one can define the vev $W_{\dr}$ of the half-BPS Wilson loop operators
$\sW_{\dr}$,
\begin{equation}
  W_{\dr} = \vev{\sW_{\dr}},
\end{equation}
where the operator is given by \cite{Young:2011aa,Assel:2012nf}
\begin{equation}
  \sW_{\dr} = \Tr_{\dr} \ms{T}
  \exp\left(\ri\oint_{S^1}\rd t (\ms{A}_0(t)-\ms{\phi}(t))\right).
\end{equation}
Here $\ms{T}$ is the time-ordering operator, $\dr$ is a
representation of the gauge group.
$\ms{A}_0(t) = \ms{A}_0(\vec{x} = 0,t)$ is the zero component of the
gauge field, and $\ms{\phi}(t) = \ms{\phi}(\vec{x}=0,t)$ is the scalar
field that accompanies the gauge field.  Both of them are fixed at the
origin of $\IR^4$ and integrated along $S^1$ to preserve half of the
supersymmetry.

However, most of the 5d SCFTs are non-Lagrangian and they do not have
a gauge theory phase.  Nevertheless, the definition of half-BPS Wilson
loops can be generalized through geometric engineering
\cite{Kim:2021gyj}.  From the topological string point of view, the
Wilson loops arise from the insertion of a collection of non-compact
2-cycles $J = \{\sC_1,\sC_2,\ldots\}$ with infinite volume
intersecting with compact 4-cycles in $X$, and the partition function
of the topological string on the new threefold $\hat{X}$ with
insertion now reads
\begin{equation}
  Z_J = Z_\emptyset\cdot\left(1+\sum_{\emptyset\neq I\subset J }W_I
    M_I\right),\quad
  M_I = \prod_{\sC_i\in I} M_{\sC_i}
\end{equation}
Here $Z_{\emptyset}$ is the partition function without insertion,
$M_{\sC_i}$ accounts for the infinite volume of the inserted
non-compact 2-cycle
\begin{equation}
  M_{\sC_i} = \frac{\re^{-t_{\sC_i}}}{2\sinh(\ep_1/2)\cdot 2\sinh(\ep_2/2)},
\end{equation}
where we have absorbed the momentum in $\IR^4$ into the denominator so
that the Wilson loop is localised at the origin, and $W_I$ is the
Wilson loop vev $W_{\md{r}}$, where the non-negative intersection
number of $I$ with compact 4-cycles give the highest weight of the
representation $\md{r}$ \cite{Kim:2021gyj}.  Obviously, this
definition of Wilson loops can be generalised to non-Lagrangian SCFTs
without a gauge theory phase, such as the $E_0$ theory engineered by
local $\IP^2$.

As proposed in \cite{Wang:2023zcb}, from the point of view of
topological string, it is more convenient to consider the so-called
\emph{Wilson loop BPS sectors} $F_I$ defined by
\begin{equation}
  \label{eq:ZJ}
  Z_J = \exp \sum_{I\subset J} F_I M_I,
\end{equation}
which are analogues of free energies of topological string without
insertion.  They are related to the Wilson loop vevs by
\begin{equation}
  W_I =  \sum_{I = \cup_j I_j \neq\emptyset} \prod_{j} F_{I_j}.
\end{equation}
The special case of $F_{\emptyset}$ is the refined topological string
free energy without insertion.  For $I\neq\emptyset$, the BPS sector
$F_I$ has a GV-like formula
\begin{equation}
  \label{eq:FI-GV}
  F_I = (2\sinh(\ep_1)\cdot 2\sinh(\ep_2))^{|I|-1}
  \sum_{\gamma\in H_2(\hat{X},I,\IZ)}\sum_{j_L,j_R} (-1)^{2j_L+2j_R}
  N^\gamma_{j_L,j_R}\chi_{j_L}(q_L)\chi_{j_R}(q_R)\re^{-\gamma\cdot\dt}.
\end{equation}
Here $|I|$ is the number of non-compact 2-cycles in $I$.  The integers
$N^\gamma_{j_L,j_R}$ count the numbers of stable D2-D0 brane bound
states wrapping compact 2-cycles $\gamma$ in $\hat{X}$ which intersect
with $I$.  Their more rigorous mathematical definition will be
discussed in \cite{Guo:2024wlg}.  The formula \eqref{eq:FI-GV} can be
derived by applying the GV formula \eqref{eq:GV-ref} on $\hat{X}$ but
keeping only the multi-wrapping number $w=1$ for curve classes that
include the non-compact 2-cycles $\sC_i$ as ingredients for they are
infinitely heavy.
The GV-like formula for the BPS sectors then also indicate the genus
expansion
\begin{equation}
  F_I = \sum_{n,g\geq 0} (\ep_1+\ep_2)^{2n}(\ep_1\ep_2)^{g+|I|-1} F^{(n,g)}_I.
\end{equation}

The Wilson loop BPS sectors can be computed from their own set of HAEs
\cite{Huang:2022hdo,Wang:2023zcb}
\begin{equation}
  \label{eq:hae-FI}
  \frac{\pd}{\pd S^{ab}}F_I^{(n,g)} = \frac{1}{2}\Big(
  D_aD_bF_I^{(n,g-1)} + \sideset{}{'}\sum_{I'\subset I,n',g'\geq 0} D_a F_{I'}^{(n',g')}
  D_b F_{I\backslash I'}^{(n-n',g-g')}
  \Big).
\end{equation}
The summation $\sum{}'$ means exclusions of
$(I',n',g') = (\emptyset,0,0)$ and $(I,n,g)$.  These equations can be
derived by assuming that the total free energy of topological string
on $\hat{X}$,
\begin{equation}
  \wt{F} = \wt{F}_\emptyset + \sum_{I\subset J} F_I M_I,
\end{equation}
where the tilde means the genus zero part of the free energy
$F_\emptyset$ is removed, also satisfies the refined HAE
\eqref{eq:hae-Z}.

We will be interested in the case that all non-compact 2-cycles
$\sC_i$ are birational, so that $F_I$ only depends on the cardinality
of $I$, and we can denote $F_I$ by $F[m]$ with $m=|I|$.  By including
infinitely many copies of $\sC_i$ in $J$, we can formally write
\eqref{eq:ZJ} as
\begin{equation}
  Z_\infty = Z_\emptyset\cdot\exp \sum_{m=1}^\infty \frac{1}{m!}F[m] M^m,
\end{equation}
by noticing that
\begin{equation}
  \lim_{N\rightarrow \infty} {N\choose m}F[m] M_{\sC_1}^m =
  \lim_{N\rightarrow \infty}\frac{1}{m!}F[m] (NM_{\sC_1})^m
  =:\lim_{N\rightarrow \infty}\frac{1}{m!}F[m] M^m.
\end{equation}
%
%
The Wilson BPS sector $F[m]$ has the genus expansion
\begin{equation}
  F[m] = \sum_{n,g \geq 0} (\ep_1+\ep_2)^{2n}(\ep_1\ep_2)^{g+m-1} F^{(n,g)}[m].
\end{equation}
By similarly considering the total free energy,
\begin{equation}
  F = \sum_{m=0}^\infty \frac{1}{m!}M^m F[m]
  = \sum_{n,g\geq 0}(\ep_1+\ep_2)^{2n}(\ep_1\ep_2)^{g-1}
  \sum_{m=0}^g\frac{1}{m!}M^m F^{(n,g-m)}[m],
\end{equation}
one can find that the components $F^{(n,g)}[m]$ are subject to the
refined HAEs \cite{Huang:2022hdo,Wang:2023zcb},
\begin{equation}
  \label{eq:hae-Fm}
  \frac{\pd}{\pd S^{ab}}F^{(n,g)}[m] = \frac{1}{2}\left(
    D_aD_bF^{(n,g-1)}[m] + \sideset{}{'}\sum_{m',n',g'} {m\choose m'}
    D_a F^{(n',g')}[m'] D_b F^{(n-n',g-g')}[m-m']
  \right).
\end{equation}
The summation $\sum{}'$ means exclusion of $(m',n',g') = (0,0,0)$ and
$(m,n,g)$.  Analogous to \eqref{eq:hae-alpha}, the Wilson BPS sectors
$F[m]$ can be solved recursively from \eqref{eq:hae-Fm}, with the
additional initial condition,
\begin{equation}
  F^{(0,0)}[1] = z^{-\sigma},\quad \sigma \in \IQ,
\end{equation}
which is the model dependent classical Wilson loop,
as well as the boundary condition that the holomorphic limit
$\CF^{(n,g)}[m]$ for $m\geq 1$ is regular everywhere in the moduli
space \cite{Wang:2023zcb}.

\section{Non-perturbative corrections to Wilson loops}
\label{sc:WL-nonp}

We would like to treat the Wilson loop BPS sectors
$F[m](\ep_1,\ep_2)$ ($m\geq 1$) as univariate power series in
terms of $g_s$ using the parameterisation \eqref{eq:ep12gs}, i.e.
\begin{equation}
  F[m](\sb;g_s) = F[m](\ri \sb g _s,-\ri\sb^{-1} g_s)
\end{equation}
and consider the correponding non-perturbative corrections.  In terms
of genus components, the asymptotic series $F(\alpha;g_s)[m]$ reads
\begin{equation}
  \label{eq:Fm-pert}
  F[m](\sb;g_s) = \sum_{g\geq 0} g_s^{2g+2m-2} F_g[m](\sb),
\end{equation}
where
\begin{equation}
  F_g[m](\sb) = \sum_{n=0}^g (-1)^n(\sb-\sb^{-1})^{2n} F^{(n,g-n)}[m].
\end{equation}

\subsection{Solutions to non-perturbative corrections}
\label{sc:Fm-nonp}

From the derivation of the HAEs for Wilson loops \eqref{eq:hae-FI} or
\eqref{eq:hae-Fm}, it is clear that instead of considering the
non-perturbative corrections of individual Wilson loop BPS sectors
$F[m]$, we should study the corrections of the total free energy of
$\hat{X}$, which is the generating functions of all $F[m]$.  With the
parametrisation \eqref{eq:ep12gs}, the total free energy reads
\begin{equation}
  \label{eq:FagsM}
  F(\sb;g_s,M)= \sum_{m\geq 0} \frac{1}{m!} M^m F[m](\sb;g_s)
  = \sum_{g\geq  0}g_s^{2g-2} F_g(\sb;M)
\end{equation}
where
\begin{equation}
  F_g(\sb;M) = \sum_{n=0}^g (-1)^n (\sb-\sb^{-1})^{2n}  
  \left( \sum_{m=0}^{g-n} \frac{1}{m!} M^m F^{(n,g-n-m)}[m]
  \right).
\end{equation}
Now, the results on non-perturbative corrections in
Section~\ref{sc:F-nonp} should also apply but with the free energies
$F(\sb;g_s)$ for $X$ without insertion replaced by the free energies
$F(\sb;g_s,M)$ for $\hat{X}$ with insertion, and for each BPS sector
$F[m]$, we only need to extract the corresponding coefficients in the
resulting generating series.  We will examine the non-perturbative
action $\CA$, the non-perturbative series $F^{(\ell)}$, and the Stokes
constants $\sS$ in turn.

As indicated by \eqref{eq:A-gamma}, without Wilson loop insertion, the
non-perturbative action $\CA$ is given by integral periods, which are
the complexified volumes of compact 2-cycles and 4-cycles in $X$, and
they remain the same in the new threefold $\hat{X}$ with insertion of
additional non-compact 2-cycles.  Alternatively, the integral periods
are closedly related to the prepotential
$\CF_0(\dt) = \CF^{(0,0)}(\dt)$ or equivalently the genus zero
component of the total free energy $\CF(\sb;g_s)$, i.e.
\begin{equation}
  \CF_0(\sb) =  \CF^{(0,0)}.
\end{equation}
For the threefold $\hat{X}$ with Wilson loop insertion, the
non-perturbative action should still be related to the genus zero
component of the total free energy $\CF(\sb;g_s,M)$.  The genus zero
free energy is not changed after the Wilson loop insertion, as
\begin{equation}
  \CF_0(\sb;M) = \CF^{(0,0)}.
\end{equation}
Therefore,  we conclude that the non-perturbative actions of Wilson
loop BPS sectors are still given by integral periods of the CY3 $X$
without insertion, i.e.~in the form of \eqref{eq:A-gamma}.

The non-perturbative series $F^{(\ell)}(\sb;g_s)$ with action
$\ell\CA$ ($\ell=1,2,\ldots$) for topological string on the CY3 $X$
without Wilson loop insertion is given by \eqref{eq:Zr}, which are
functions of the perturbative free energy $F(\sb;g_s)$ in
\eqref{eq:Fags}.  The non-perturbative series $F^{(\ell)}(\sb;g_s,M)$
for topological string on the threefold $\hat{X}$ with insertion
should be given by the same formulas but with $F(\sb;g_s)$ replaced by
the generating series of Wilson loop BPS sectors, i.e.~the
perturbative free energy $F(\sb;g_s,M)$ of $\hat{X}$ given in
\eqref{eq:FagsM}.  To find the non-perturbative corrections in
particular to the BPS sector $F[m]$, we merely need to extract the
coefficients of $M^m/m!$, in other words
\begin{equation}
  F^{(\ell)}[m](\sb;g_s) = \frac{\pd^m}{\pd M^m}
  F^{(\ell)}(\sb;g_s,M) \Big|_{M = 0}.
\end{equation}

The non-perturbative corrections reduce to the holomorphic limit in
any frame with the rules of substitution
\eqref{eq:D-hol},\eqref{eq:Sigma-hol}.  Note that in particular, in an
$\CA$-frame, where the non-perturbative action $\CA$ is a local flat
coordinate, the non-perturbative corrections should be
\eqref{eq:FA1},\eqref{eq:FA2}.  Yet again both formulas only involve
the genus zero free energy, which is not changed by the Wilson loop
insertion, and has no higher $M$ powers.  This implies that
\begin{equation}
  \CF^{(\ell)}_\CA[m](\sb;g_s) = 0,\quad m\geq 1,\ell\geq 1.
\end{equation}
In other words, Wilson loop BPS sectors have no non-perturbative
corrections in the holomorphic limit of any $\CA$-frame.

In any frame other than an $\CA$-frame, the non-perturbative
corrections in general do not vanish in the holomorphic limit.  To
give an example, we consider the non-perturbative correction
$\CF^{(1)}$ with actions respectively $\CA/\sb$ and $\sb \CA$ for the
generating function $F(\sb;g_s,M)$, and they read
\begin{subequations}
  \label{eq:Fm1inst-ref}
  \begin{align}
    \CF^{(1)}_{\sb}(\dt;M) =
    &\frac{1}{2\sin(\pi/\sb^2)}\exp\left(
      \CF(\dt +\ri g_s \dc/\sb;M) - \CF(\dt;M)\right),\\
    \CF^{(1)}_{1/\sb}(dt;M) =
    &\frac{1}{2\sin(\pi \sb^2)}\exp\left(
      \CF(\dt +\ri \sb g_s \dc;M) - \CF(\dt;M)\right),
  \end{align}
\end{subequations}
where $\CF(\dt;M)$ is the shorthand for $\CF(\sb;g_s,M)$.  The
non-perturbative corrections to individual BPS sectors $\CF[m]$
$(m\geq 1)$ can be read off using Fa\`{a} di Bruno's formula, and we
obtain
\begin{subequations}
  \begin{align}
    \CF^{(1)}_{\sb}[m](\dt) =
    &\frac{\re^{\left(\CF(\dt +\ri g_s \dc/\sb) - \CF(\dt)\right)}}
      {2\sin(\pi/\sb^2)}
      \sum_{d(\dk) = m}
      \frac{m!}{\Pi_j
      k_j!(j!)^{k_j}}\Pi_j(\CF[j](\dt+\ri g_s\dc/\sb) - \CF[j](\dt))^{k_j},\\
    \CF^{(1)}_{1/\sb}[m](\dt) =
    &\frac{\re^{\left(\CF(\dt +\ri \sb g_s \dc) - \CF(\dt)\right)}}
      {2\sin(\pi\sb^2)}
      \sum_{d(\dk) = m}
      \frac{m!}{\Pi_j
      k_j!(j!)^{k_j}}\Pi_j(\CF[j](\dt+\ri \sb g_s\dc) - \CF[j](\dt))^{k_j}.
  \end{align}
\end{subequations}
where $\dk = (k_1,k_2,\ldots)$ is an integer partition of $m$ with
$d(\dk) = \sum_j j k_j$.
They have the genus expansion
\begin{equation}
  \CF^{(1)}_{\sb}[m] = \re^{-\CA/(\sb g_s)}\sum_{n\geq 0}  g_s^{n+m}\CF^{(1)}_{\sb,n}[m],\quad
  \CF^{(1)}_{1/\sb}[m] = \re^{-\sb\CA/g_s}\sum_{n\geq 0}  g_s^{n+m}\CF^{(1)}_{1/\sb,n}[m].
\end{equation}

Finally, the Stokes constant $\sS_\gamma(\sb,M)$ for the generating
function associated to the non-perturbative action $\CA_\gamma$ should
in principle depend on both $\sb$ and $M$.  But as the insertion mass
$M$ is infinitely heavy while the Stokes constant should be a finite
number, the dependence of the Stokes constant on $M$ should drop out.
In fact, the Stokes constants of the topological string free energies
are generally conjectured to be constants independent of the K\"ahler
moduli of the Calabi-Yau threefold \cite{Gu:2021ize,Gu:2022sqc}, the
Stokes constants for the generating series, which is the topological
string free energy of the new threefold $\hat{X}$, should also not
depend on the K\"ahler moduli, $M$ included.  We thus conjecture that
\begin{equation}
  \label{eq:SM}
  \sS_\gamma(\sb,M) = \sS_\gamma(\sb),
\end{equation}
which is conjectured to be given by the refined DT invariant,
cf.~\eqref{eq:S-Omega}.  As the BPS sectors are linear coefficients of
the generating function as a power series of $M$, they all share the
\emph{same} Stokes constant, given by \eqref{eq:SM}.

Before we illustrate these results with examples, we discuss their
implication in the two special limits of the Omega background.

\subsection{Limiting scenarios}
\label{sc:limits}

\subsubsection{The unrefined limit}
\label{sc:Fm-nonp-unref}

We first consider the unrefined limit, where
\begin{equation}
  \ep_1 = \ri g_s,\quad \ep_2 = -\ri g_s.
\end{equation}
This limit is obtained by simply  taking
\begin{equation}
  \sb\rightarrow 1.
\end{equation}
In this limit, the refined free energy becomes the conventional free
energy of unrefined topological string
\begin{equation}
  \label{eq:F-unref}
  F(g_s) = F(\sb=1;g_s) = \sum_{g\geq 0} g_s^{2g-2} F^{(0,g)}.
\end{equation}
Likewise, the generation function of the refined Wilson loop BPS
sectors becomes that of the unrefined Wilson loop BPS sectors
\begin{equation}
  \label{eq:FM-unref}
  F(g_s,M) = F(\sb=1;g_s,M) = \sum_{m\geq 0}\frac{1}{m!}M^m F[m](\sb=1;g_s).
\end{equation}

In the non-perturbative sectors, the pair of Borel singularities
$\CA/\sb$ and $\sb\CA$ merge and become a single Borel singularity.
The non-perturbative series associated to $\CA$ is then the
$\sb\rightarrow 1$ limit of the sum of the refined non-perturbative
series $F^{(\ell)}_\sb$ and $F^{(\ell)}_{1/\sb}$.  For instance, in
the 1-instanton sector, the non-perturbative series for the unrefined
string free energy in the $\CA$-frame is, cf.~\eqref{eq:FA}
\begin{equation}
  \CF^{(1)}_\CA = \lim_{\sb\rightarrow 1}
  \left(\CF^{(1)}_{\CA,\sb} + \CF^{(1)}_{\CA,1/\sb}\right)
  = \lim_{x\rightarrow 0}
  \left(\frac{\re^{-\frac{\CA}{g_s}\re^{-x}}}{2\sin(\pi\re^{-2x})}
    + \frac{\re^{-\frac{\CA}{g_s}\re^{x}}}{2\sin(\pi\re^{2x})}\right)
  = \re^{-\CA/g_s}(1+\CA/g_s),
  \label{eq:F1A-unref}
\end{equation}
while in a non-$\CA$-frame, the non-perturbative series is,
cf.~\eqref{eq:F1}
\begin{align}
  \CF^{(1)} =
  &\lim_{\sb\rightarrow 1} \left(\CF^{(1)}_\sb +
    \CF^{(1)}_{1/\sb}\right)  = \lim_{x\rightarrow 0}
    \left(\frac{\pi}{\sin(\pi\re^{-2x})}
    \exp(\CF(\dt+\ri g_s \dc \re^{-x}) - \CF(\dt))
    + (x\leftrightarrow -x)\right) \nn=
  &(1-\ri c^a g_s \pd_{t^a}\CF(\dt +\ri \dc g_s)\exp(\CF(\dt+\ri
    \dc g_s)-\CF(\dt))),
    \label{eq:F1-unref}
\end{align}
where $\CF(\dt) = \CF(\dt;g_s)$ is the unrefined topological string
given in \eqref{eq:F-unref} (taking appropriate holomorphic limit).
Both \eqref{eq:F1A-unref} and \eqref{eq:F1-unref} agree with
\cite{Gu:2022sqc}.
Note that in the unrefined case, the leading exponent of the
non-perturbative series is $-1$, i.e.
\begin{equation}
  \CF^{(1)} = \re^{-\CA/g_s}\sum_{n\geq 0}g_s^{n-1} \CF^{(1)}_n,
\end{equation}
in constrast to the refined case \eqref{eq:Fl-ref}.

In the case of Wilson loop BPS sectors, there are only Borel
singularities $\CA$ which are not flat coordinates.  The generating
function of the non-perturbative series
\begin{equation}
  \CF^{(1)}(M) = \sum_{m\geq 0} \frac{1}{m!}M^m \CF^{(1)}[m]
\end{equation}
is given by \eqref{eq:F1-unref} with $\CF(\dt) = \cF(\dt;g_s)$
replaced by $\CF(\dt;g_s,M)$ given by \eqref{eq:FM-unref} (taking
appropriate holomorphic limit); in other words,
\begin{equation}
  \CF^{(1)}(M) = 
  \left(1-\ri c^a g_s\pd_{t^a}\CF(\dt+\ri \dc g_s;M)\right)
  \exp\left(
  \left(\CF(\dt + \ri \dc g_s;M) - \CF(\dt;M)\right)
  \right).
    \label{eq:Fm1inst-unref}
\end{equation}
The non-perturbative series for individual BPS sectors are the
coefficients $M^m/m!$, and they have the genus expansion
\begin{equation}
  \CF^{(1)}[m] = \re^{-\CA/g_s}\sum_{n\geq 0} g_s^{n+m-1}\CF^{(1)}_n[m].
\end{equation}

\subsubsection{The NS limit}
\label{sc:Fm-nonp-NS}

Next, we would like to consider the Nekrasov-Shatashvili limit and
recover the results in \cite{Gu:2022fss}.  For this purpose, we choose
a slightly different parametrisation, in accord with
\cite{Grassi:2022zuk}
\begin{equation}
  \ep_1 = \ri\hbar,\quad \ep_2 = -\ri\alpha\hbar.
\end{equation}
The advantage of this $\alpha$-parametrisation is that the NS limit is
obtained simply by taking $\alpha\rightarrow 0$.  The
$\alpha$-parametrisation is related to the $\sb$-parametrisation by
\begin{equation}
  g_s = \sqrt{\alpha}\hbar,\quad \sb = 1/\sqrt{\alpha}.
\end{equation}
The reparametrisation of the refined free energy is then
\begin{equation}
  F(\dt;\ep_1,\ep_2) = F(\dt,\alpha;\hbar) = \sum_{g\geq
    0}\hbar^{2g-2} F_g(\dt,\alpha)
\end{equation}
where
\begin{equation}
  F_g(\dt,\alpha) = \sum_{n=0}^g(-1)^n \alpha^{g-n-1}(1-\alpha)^{2n}F^{(n,g-n)}(\dt).
\end{equation}
The genus components are related to those in the $\sb$-parametrisation
by
\begin{equation}
  F_g(\sb) = \alpha^{1-g}F_g(\alpha).
\end{equation}
In particular, the genus zero component is proportional to the
prepotential
\begin{equation}
  F_0(\alpha) = \alpha^{-1}F_0(\sb) = \alpha^{-1}\CF^{(0,0)}.
\end{equation}
In the NS limit with $\alpha\rightarrow 0$, we find that
\begin{equation}
  \label{eq:F0NSlim}
  F(\dt,\alpha;\hbar) = \alpha^{-1}\hbar^{-2} F_{\text{NS}}(\dt;\hbar) + \CO(\alpha^0),
\end{equation}
where $F_{\text{NS}}(\dt;\hbar)$ is the NS free energy given by
\begin{equation}
  F_{\text{NS}}(\dt;\hbar) = \sum_{g\geq 0}\hbar^{2g} (-1)^gF^{(g,0)}(\dt).
\end{equation}

The perturbative series $F(\dt,\alpha;\hbar)$ should be promoted to
the full trans-series with non-perturbative corrections.  In the
$\sb$-parametrisation, the non-perturbative actions, or equivalently
Borel singularities, are located at $\CA/\sb$ and $\sb\CA$ where $\CA$
are integral periods, and the exponential suppressing factors in the
non-perturbative corrections are respectively
\begin{equation}
  \label{eq:exp-fac}
  \re^{-\CA/(\sb g_s)},\quad \re^{-\sb\CA/g_s}.
\end{equation}
In the $\alpha$-parametrisation, they become
\begin{equation}
  \re^{-\CA/\hbar},\quad \re^{-\CA/(\alpha\hbar)},
\end{equation}
and the Borel singularities are now instead located at $\CA$ and
$\CA/\alpha$ \cite{Grassi:2022zuk}.  In the NS limit with
$\alpha\rightarrow 0$, the second singularity runs away to infinity
and we are left with the first singularity.

To find the non-perturbative series associated to this Borel
singularity, we should take the refined non-perturbative series for
the Borel singularity $\CA/\sb$ in the $\sb$-parametrisation, convert
it to the $\alpha$-parametrisation, and finally take the
$\alpha\rightarrow 0$ limit.  As both the perturbative and
non-perturbative refined series should scale at the same rate in the
last step, for otherwise there will be no finite non-perturbative
corrections to the NS free energy, \eqref{eq:F0NSlim} indicates that
the non-perturbative refined series should have the asymptotics
$\sim \CO(\alpha^{-1})$, and the non-perturbative NS series is the
coefficient of $\alpha^{-1}$ in the leading term; more concretely,
\begin{equation}
  F^{(\ell)}(\dt,\alpha;\hbar) = \alpha^{-1}
  \hbar^{-2} F_{\text{NS}}^{(\ell)}(\dt;\hbar) + \cO(\alpha^0),\quad
  \alpha\rightarrow 0.
\end{equation}

For instance, if the action $\CA$ is a local flat coordinate, the
associated non-perturbative refined series in the $\ell$-instanton
sector in the $\sb$-parametrisation is given by \eqref{eq:FA1}, which
is converted to the $\alpha$-parametrisation as
\begin{equation}
  \CF^{(\ell)}_{\CA,\sb} =
  \frac{(-1)^{\ell}}{\ell}\frac{\pi}{\sin(\pi\ell \alpha)}
  \re^{-\ell\CA/\hbar}.
\end{equation}
In the small $\alpha$ limit, we find
\begin{equation}
  \CF^{(\ell)}_{\CA,\sb}
  \sim \alpha^{-1}
  \frac{(-1)^{\ell}}{\ell^2}\re^{-\ell\CA/\hbar}\quad\Rightarrow
  \quad
  \CF_{\text{NS},\CA}^{(\ell)}  = \hbar^2\frac{(-1)^\ell}{\ell^2}\re^{-\ell\CA/\hbar},
\end{equation}
which agrees with \cite{Gu:2022fss}.

More generically if the $\CA$ is not a local flat coordinate, the
non-perturbative refined series in the generic $\ell$-instanton sector
in the $\sb$-parametrisation are given by \eqref{eq:Zr} with the
boundary condition \eqref{eq:ZrA}.  To derive the NS limit, we require
several modifications.  We drop the part $\CF_{\CA,1/\sb}^{(\ell)}$
associated to the Borel singularities $\ell\sb\CA$ which will run away
to infinity.  On the other hand, we use a more generic boundary
condition in order to compare with \cite{Gu:2022fss},
\begin{equation}
  Z_{r,\CA} = \exp\left[\sum_{\ell\geq
      1}\tau_\ell\CF_{\CA,\sb}^{(\ell)}\right] =
 1+ \sum_{n> 0} C_n(\tau) \exp\left(-\frac{n\CA}{\sb g_s}\right),
\end{equation}
where $\tau_\ell$ are constants that parametrize the boundary
condition.  The generic non-perturbative component of the refined
partition function is
\begin{equation}
  Z_r =1 + \sum_{n > 0} C_n(\tau)\exp \Sigma_{n\sb^{-1}},
\end{equation}
from which the non-perturbative free energies are obtained
\begin{equation}
  F_r = \sum_{\ell\geq 1} F^{(\ell)} = \log Z_r = \log \left(1+
    \sum_{n > 0} C_n(\tau) \exp\Sigma_{n\sb^{-1}} \right).
\end{equation}
To find the NS limit of the non-perturbative free energies, we convert
$F_r$ to the $\alpha$-parametrisation, and then take the
$\alpha\rightarrow 0$ limit.
To further match with the convention in \cite{Gu:2022fss}, we also
need the rescaling
\begin{equation}
  \sD \rightarrow g_s \sD,\quad G(\sb) \rightarrow
  g_s^{-1}G(\alpha),
\end{equation}
where $G(\alpha=0)$ is the same $G$ in \cite{Gu:2022fss}.
We find that $F^{(\ell)}$ in the $\alpha\rightarrow 0$ limit indeed
has the correct asymptotics $\sim \CO(\alpha^{-1})$, and the leading
coefficients agree with NS non-perturbative series in
\cite{Gu:2022fss}.  For instance, we find\footnote{If we are to have
  numerical Stokes constants, the non-perturbative series in the
  $\CA$-frame in \cite{Gu:2022fss} should have an additional factor of
  $\hbar^2$, which implies that in the generic expressions of the
  non-perturbative series, the boundary parameters $\tau_\ell$ should
  also be scaled by a factor of $\hbar^2$.}
\begin{subequations}
  \label{eq:Fl-FC0NS}
  \begin{align}
    &F^{(1)}_{\text{NS}} =\hbar^2 \tau_1 \re^{-G/\hbar},\\
    &F^{(2)}_{\text{NS}} =\hbar^2\left(-\frac{\tau_2}{4} +\tau_1^2\frac{\sD
      G}{2}\right)\re^{-2G/\hbar},\\
    &F^{(3)}_{\text{NS}} = \hbar^2\left(\frac{\tau_3}{9} -
      \tau_1\tau_2\frac{\sD G}{2} + \tau_1^3
      \left(\frac{(\sD G)^2}{2} - \frac{\hbar \sD^2 G}{6}\right)\right)\re^{-3G/\hbar}.
  \end{align}
\end{subequations}

Notice also that according to \eqref{eq:S-Omega}, the Stokes constants
of the NS free energies are,
\begin{equation}
  \ms{S}_\gamma^{\text{NS}} =  \Omega(\gamma,-1)
\end{equation}
and they are the same as the Stokes constants of the conventional free
energy of unrefined topological string up to a sign \cite{Gu:2023wum},
if the spins of different BPS states of a fixed D-brane configuration
only differ by multiples of two.

We then consider the Wilson loops.  With the $\alpha$-parametrisation,
the generating series of the Wilson loop BPS sectors reads
\begin{align}
  F(\dt,\alpha;\hbar,M) =
  &\sum_{m\geq 0}\frac{M^m}{m!}
    F[m](\dt,\alpha;\hbar)\nn
    =
  &\sum_{m\geq 0} \frac{M^m}{m!} \sum_{g\geq 0}\hbar^{2g+2m-2}
    \sum_{n=0}^g(-1)^n\alpha^{g-n+m-1}(1-\alpha)^{2n}F^{(n,g-n)}[m].
\end{align}
In the NS limit with $\alpha\rightarrow 0$
\begin{equation}
  \label{eq:FNSlim}
  F(\dt,\alpha;\hbar,M) \sim \alpha^{-1}\hbar^{-2}
  \sum_{m\geq 0}
  \left(\frac{(\alpha \hbar^2 M)^{m}}{m!}F_{\text{NS}}[m] + \cO(\alpha^{m+1})\right),
\end{equation}
where
\begin{equation}
  F_{\text{NS}}[m] = \sum_{g\geq 0} \hbar^{2g}(-1)^g F^{(g,0)}[m].
\end{equation}
Therefore in the limit $\alpha\rightarrow 0$, each refined BPS sector
$F[m](\dt,\alpha;\hbar)$ has the asymptotics $\alpha^{m-1}$, and the
NS limit $F_{\text{NS}}[m]$ is the coefficient of the leading term.
It implies that non-perturbative corrections to the NS BPS sectors
$F_{\text{NS}}[m](\dt;\hbar)$ can be obtained by taking the
coefficient of the leading $\alpha^{m-1}$ term of the non-perturbative
corrections to the fully refined BPS sectors $F[m](\dt,\alpha;\hbar)$.
Alternatively, \eqref{eq:FNSlim} suggests that we define the
generating series of the NS BPS sectors
\begin{equation}
  \label{eq:WLNS-gen}
  F_{\text{NS}}(\dt;\hbar) \rightarrow  F_{\text{NS}}(\dt;\hbar,M) =
  \sum_{m\geq 0}\frac{M^m}{m!}  F_{\text{NS}}[m](\dt;\hbar).
\end{equation}
As the non-perturbative corrections to the NS free energy
$F_{\text{NS}}(\dt;\hbar)$ are functions of the free energy itself,
the non-perturbative corrections to the generating series
\eqref{eq:WLNS-gen} is obtained by the same expression but with the NS
free energy replaced by the NS generating series.  The
non-perturbative corrections to individual NS BPS sectors are
extracted as coefficients of $M^m/m!$.
For instance, from \eqref{eq:Fl-FC0NS}, we immediately get
\begin{subequations}
  \label{eq:Fl-FC1NS}
  \begin{align}
    &F^{(1)}_{\text{NS}}[1] = - \hbar\tau_1\sD F_{\text{NS}}[1] \re^{-G/\hbar},\\
    &F^{(2)}_{\text{NS}}[1] =\hbar\left(
      \frac{\tau_2}{2} \sD F_{\text{NS}}[1]  +\tau_1^2\Big(-\sD G
      \sD F_{\text{NS}}[1] +\frac{1}{2}\hbar \sD^2 F_{\text{NS}}[1]\Big)
      \right)\re^{-2G/\hbar},\\
    &F^{(3)}_{\text{NS}}[1] = \hbar\left(
      -\frac{\tau_3}{3}\sD F_{\text{NS}}[1] +
      \tau_1\tau_2\Big(\frac{3}{2}\sD G \sD F_{\text{NS}}[1]
      -\frac{1}{2}\hbar\sD^2 F_{\text{NS}}[1]\Big) \right.\nn
    &\left.\phantom{===}+ \tau_1^3
      \Big(-\frac{3}{2} (\sD G)^2\sD F_{\text{NS}}[1] + \frac{1}{2}\hbar
      \sD^2G\sD F_{\text{NS}}[1] + \hbar \sD G\sD^2 F_{\text{NS}}[1]
      -\frac{1}{6}\hbar^2\sD^3F_{\text{NS}}[1]
      \Big)
      \right)\re^{-3G/\hbar},
  \end{align}
\end{subequations}

In order to compare with \cite{Gu:2022fss}, we need to study the
Wilson loop vev $w$ in the NS limit, which is related to the first BPS
sector by
\begin{equation}
  \exp w = F[1]
\end{equation}
This relation is promoted to the trans-series and we have
\begin{equation}
  \exp\Big(w + \sum_{\ell\geq 1}\cC^\ell w^{(\ell)}\Big) = F_{\text{NS}}[1] +
  \sum_{\ell\geq 1} \cC^\ell F^{(\ell)}_{\text{NS}}[1].
\end{equation}
Using again the Fa\`{a} di Bruno's formula, we find
\begin{equation}
  w^{(\ell)} = \sum_{d(\dk) = \ell}(-1)^{|\dk|}(|\dk|-1)!\prod_{j\geq
    1} \frac{1}{k_j!}\Big(F^{(j)}_{\text{NS}}[1]/F_{\text{NS}}[1]\Big)^{k_j}.
\end{equation}
For instance,
\begin{subequations}
  \label{eq:wlNS}
  \begin{align}
    &w^{(1)} = - \hbar\tau_1\sD w \re^{-G/\hbar},\\
    &w^{(2)} = \hbar\left(
      \frac{\tau_2}{2} \sD w  +\tau_1^2\Big(- \sD G
      \sD w+\frac{1}{2}\hbar \sD^2 w\Big)
      \right)\re^{-2G/\hbar},\\
    &w^{(3)} = \hbar\left(
      -\frac{\tau_3}{3}\sD w +
      \tau_1\tau_2\Big(\frac{3}{2}\sD G \sD
      w-\frac{1}{2}\hbar\sD^2 w\Big) \right.\nn
    &\left.\phantom{===}+ \tau_1^3
      \Big(-\frac{3}{2} (\sD G)^2\sD w + \frac{1}{2}\hbar
      \sD^2G\sD w + \hbar \sD G\sD^2 w -\frac{1}{6}\hbar^2\sD^3w
      \Big)
      \right)\re^{-3G/\hbar},
  \end{align}
\end{subequations}
which agree with \cite{Gu:2022fss}.

Two important features of the resurgent structures of the Wilson loop
vevs were discovered in \cite{Gu:2022fss}: the Borel singularities
are integral periods which are not local flat coordinates, and that
the Stokes constants, if they are not vanishing, are the same as those
of the NS free energy.  And they were presented as independent results
from those of the NS free energies.  It is clear now that they are
directly related, as both the NS Wilson loop and the NS free energy
are parts of the same generating series, which enjoys the uniform
resurgent structure.

\section{Examples}
\label{sc:examples}

\subsection{Example: local $\IP^2$}
\label{sc:P2}

We consider the local $\mathbb{P}^2$, i.e.~the total space of the
canonical bundle of $\IP^2$, in this section. Local $\mathbb{P}^2$ is
a basic Calabi-Yau manifold, but with rich geometric structure, and it
has been discussed in great detail in the literature.  We follow the
convention in \cite{Haghighat:2008gw,Couso-Santamaria2015
} where its moduli space is parametrized by the global complex
coordinate $z$ such that the large radius singularity, the conifold
singularity, and the orbifold singularity are located respectively at
$z=0$, $z=-1/27$ and $z=\infty$.

The periods in local $\mathbb{P}^2$ are annihilated by the
Picard-Fuchs operator \cite{Chiang:1999tz}
\begin{equation}
\mathcal{L}=(1+60z)\partial_z+3z(1+36z)\partial_z^2+z^2(1+27z)\partial_z^3.
\end{equation}
Near the large radius point, the flat coordinate and its conjugate are
(see e.g.~\cite{Gu:2022sqc})
\begin{subequations}
\begin{align}
  t_{\mathrm{LR}}=
  &-\log(z)+6z{}_4F_3\left(1,1,\frac{4}{3},\frac{5}{3};2,2,2;-27z\right),\\
  \frac{\pd \CF^{(0,0)}_{\mathrm{LR}}}{\pd t_{\mathrm{LR}}}=
  &\frac{1}{3\sqrt{2}\pi}G^{3,2}_{3,3} \left(
  \begin{array}{c}
    \tfrac{2}{3},\tfrac{1}{3};1\\0,0,0
  \end{array}; 27z \right)
  -\frac{4\pi^2}{9},
\end{align}
\end{subequations}
while near the conifold point, the flat coordinate and its conjugate are
\begin{subequations}
  \begin{align}
    t_{\text{c}} =
    &-\frac{3}{2\pi \ri}\frac{\pd \CF^{(0,0)}_{\text{LR}}}{\pd t_{\text{LR}}},\\
      \frac{\pd \CF^{(0,0)}_{\text{c}}}{\pd t_{\text{c}}} =
    &-\frac{2\pi\ri}{3} t_{\text{LR}}.
  \end{align}
\end{subequations}

We will inspect the non-perturbative corrections for Wilson loop BPS
sectors. More specifically, we consider the loci in the moduli space
near the large radius point $z = 0$ and near the conifold point
$z=-1/27$.

\begin{figure}
  \centering
  \subfloat[$z=-10^{-6}$]{\includegraphics[height=5cm]{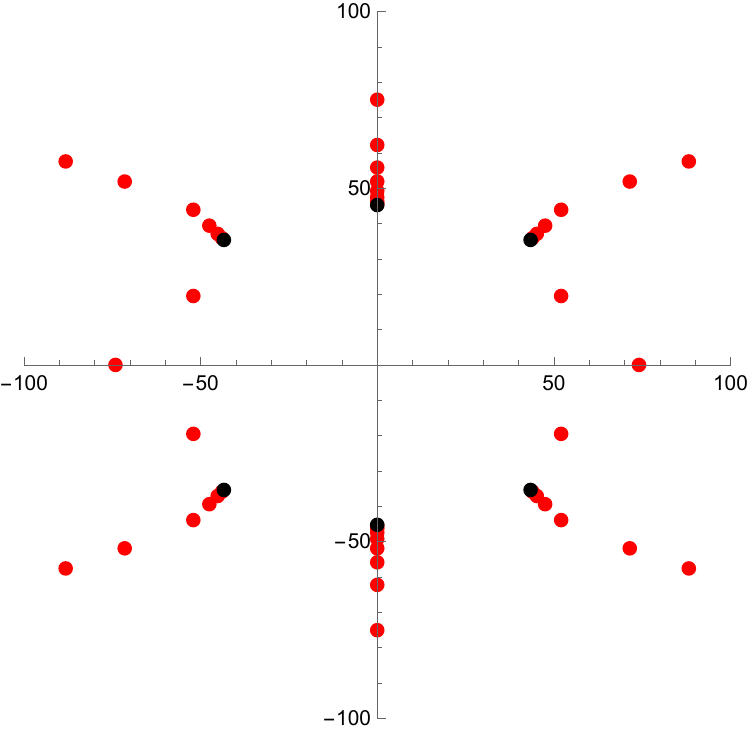}}\hspace{5ex}
  \subfloat[$z=-\frac{9}{270}$]{\includegraphics[height=5cm]{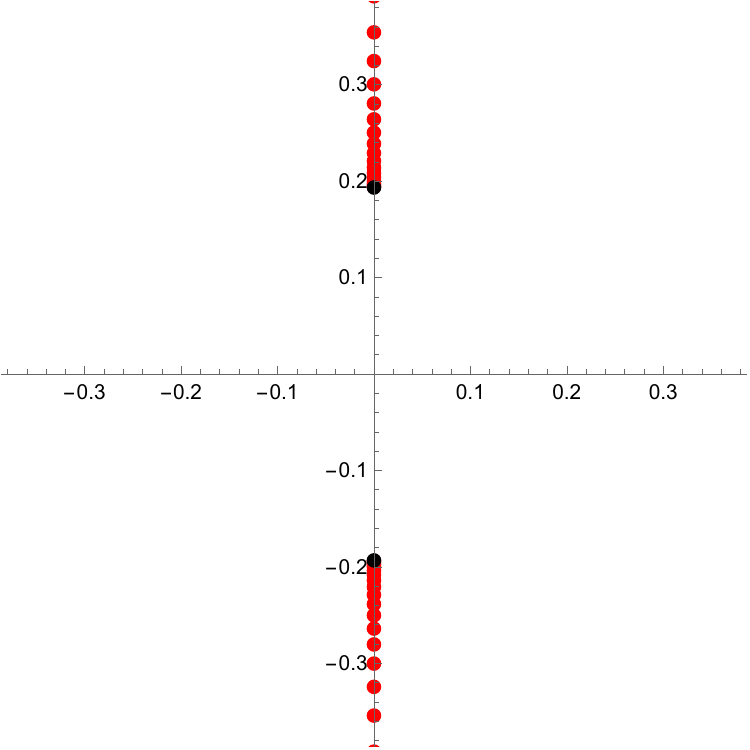}}
  \caption{Borel singularities of refined Wilson loop BPS sectors
    $\CF[1](\sb;g_s)$ for local $\IP^2$ with $\sb=2$ up to $g=50$ in
    the large radius frame, respectively (a) near the large radius
    point $z=0$ and (b) near the conifold point $z=-1/27$.  The red
    dots are approximate singularities from numerical calculations,
    which would accumulate to branch cuts.  The branch points (black
    dots) on the imaginary axis are
    $\sb^{-1}\CA_{\pm(-3,0,0)_\text{LR}}$, and those in the four
    quadrants are $\sb^{-1}\CA_{\pm(-3,1,0)_\text{LR}}$ and
    $\sb^{-1}\CA_{\pm(-3,-1,-1)_\text{LR}}$.}
  \label{fig:P2C1b2LR-brl}
\end{figure}

\begin{figure}
  \centering
  \subfloat[$z=-10^{-6}$]{\includegraphics[height=5cm]{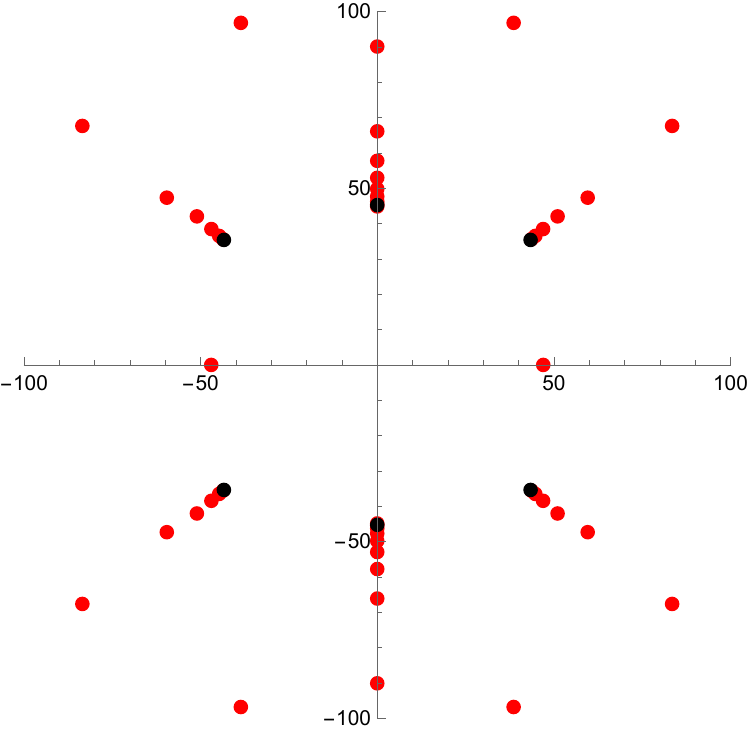}}\hspace{5ex}
  \subfloat[$z=-\frac{9}{270}$]{\includegraphics[height=5cm]{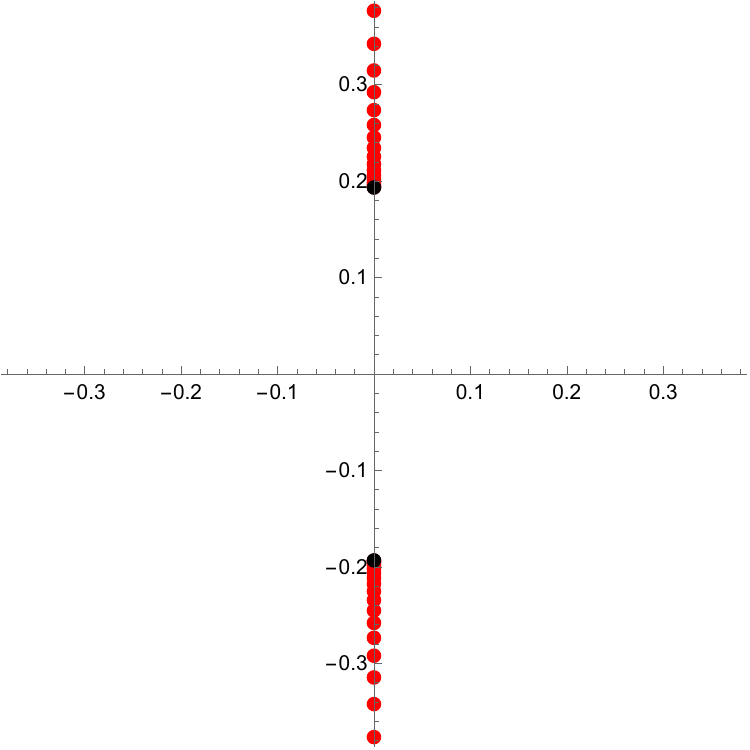}}
  \caption{Borel singularities of refined Wilson loop BPS sectors
    $\CF[2](\sb;g_s)$ for local $\IP^2$ with $\sb=2$ up to $g=50$ in
    the large radius frame, respectively (a) near the large radius
    point $z=0$ and (b) near the conifold point $z=-1/27$.  The red
    dots are approximate singularities from numerical calculations,
    which would accumulate to branch cuts.  The branch points (black
    dots) are the same as in Fig.~\ref{fig:P2C1b2LR-brl}.}
  \label{fig:P2C2b2LR-brl}
\end{figure}

\begin{figure}
  \centering
  \subfloat[$z=-10^{-6}$]{\includegraphics[height=5cm]{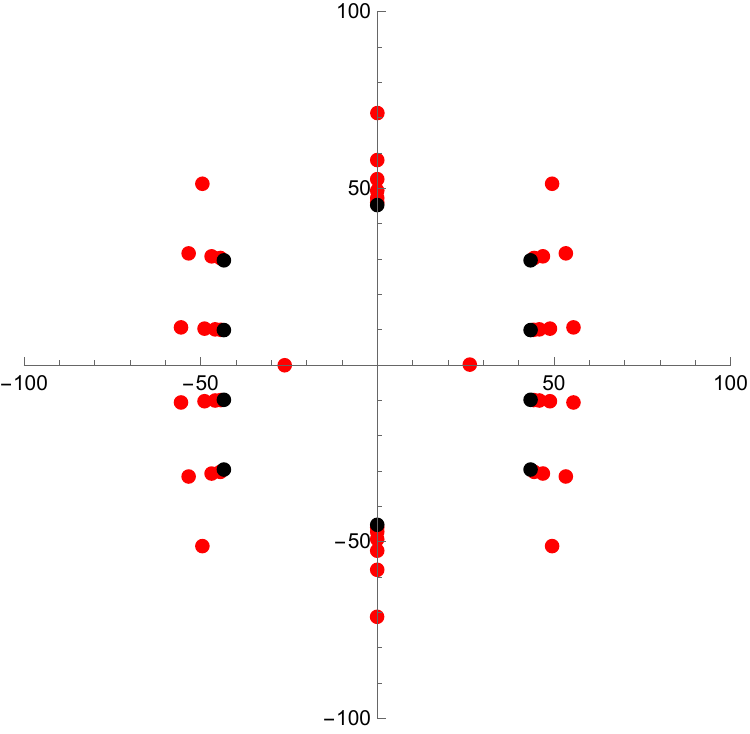}}\hspace{5ex}
  \subfloat[$z=-\frac{9}{270}$]{\includegraphics[height=5cm]{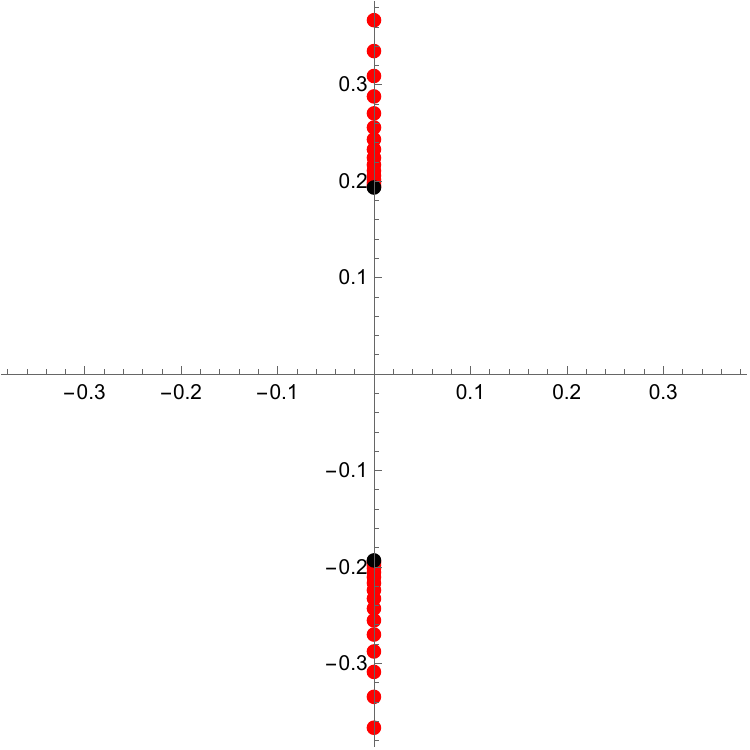}}
  \caption{Borel singularities of refined free energies
    $\CF[0](\sb;g_s)$ for local $\IP^2$ with $\sb=2$ up to $g=50$ in
    the large radius frame, respectively (a) near the large radius
    point $z=0$ and (b) near the conifold point $z=-1/27$.  The red
    dots are approximate singularities from numerical calculations.
    The branch points (black dots) on the imaginary axis are
    $\sb^{-1}\CA_{\pm(-3,0,0)_{\text{LR}}}$, and those away from the
    imaginary axis are
    $\sb^{-1}\CA_{\pm(0,1,n)_\text{LR}} (n=-1,0,1,2)$.}
  \label{fig:P2C0b2LR-brl}
\end{figure} 
  
We first study the location of Borel singularities, i.e. the singular
points of the Borel transform.  We evaluate the perturbative BPS
sectors $\CF[1]$ and $\CF[2]$ in the holomorphic limit of the large
radius frame, where $t_{\mathrm{LR}}$ is the flat coordinate, near
respectively the large radius point $z = 0$ and the conifold point
$z=-1/27$. The Borel singularities of $\CF[1]$ and $\CF[2]$ are
plotted respectively in Figs. \ref{fig:P2C1b2LR-brl} and
Figs. \ref{fig:P2C2b2LR-brl}.  The plots are similar for the two BPS
sectors.  Near the large radius point, the visible Borel singularities
are located at $\mathsf{b}^{-1}\mathcal{A}_{\gamma_\mathrm{LR}}$ (we
take $\mathsf{b} > 1$ so that $\mathsf{b}^{-1}\CA$ is smaller than
$\mathsf{b}\CA$) with the charge vectors
\begin{equation}
  \gamma_{\mathrm{LR}}=\pm(-3,0,0),\ \text{and},\ \pm(-3,1,0),\ \pm(-3,-1,-1),
\end{equation}
and we use the notation that 
\begin{equation}\label{eq:LRsigularity}
  \CA_{\gamma_{\text{LR}}} =
  -c\ri \pd_{t_{\text{LR}}} \CF^{(0,0)}_{\text{LR}} +2\pi d t_{\text{LR}} + 4\pi^2 \ri
  d_0,\quad
  \gamma_{\text{LR}} = (c,d,d_0).
\end{equation}
Near the conifold points, the visible Borel singularities are located
at $\mathsf{b}^{-1}\mathcal{A}_{\gamma_\mathrm{LR}}$ with
\begin{equation}
  \gamma_{\mathrm{LR}}=\pm (-3,0,0).
\end{equation} 
For comparison, we also give the location of Borel singularities for
the free energies\footnote{The constant map contributions to free
  energies are removed.} $\CF[0]$ in
Figs. \ref{fig:P2C0b2LR-brl}. Near the large radius point, the Borel
singularities are located at
$\mathsf{b}^{-1}\mathcal{A}_{\gamma_\mathrm{LR}}$ with
\begin{equation}
  \gamma_{\mathrm{LR}}=\pm(-3,0,0)_{\text{LR}},\;
  \pm (0,1,n),\,\,n=0,\pm 1,\pm 2,\ldots.
\end{equation}
Near the conifold point, the Borel singularities are located at
$\sb^{-1}\CA_{\pm(-3,0,0)_\text{LR}}$.  In contrast to the free energies,
the Borel singularities of Wilson loop BPS sectors never coincide with
the flat coordinate up to a constant, i.e.~the coefficient $c$ in the
charge vector $\gamma_{\mathrm{LR}}$ does not vanish.

\begin{figure}
  \centering
  \subfloat[$z=-10^{-6}$]{\includegraphics[width=0.3\linewidth]{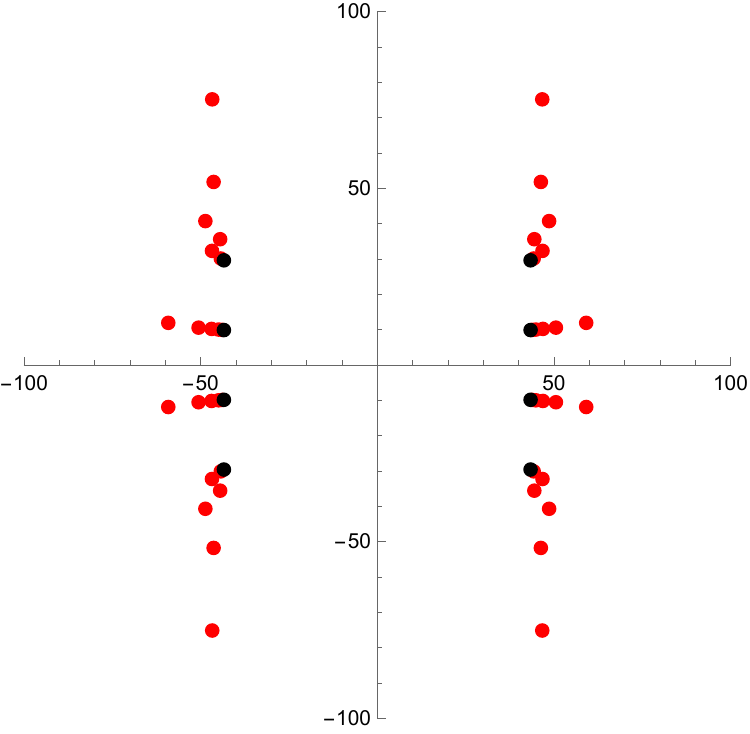}}\hspace{1ex}
  \subfloat[$z=-\frac{9}{270}$]{\includegraphics[width=0.3\linewidth]{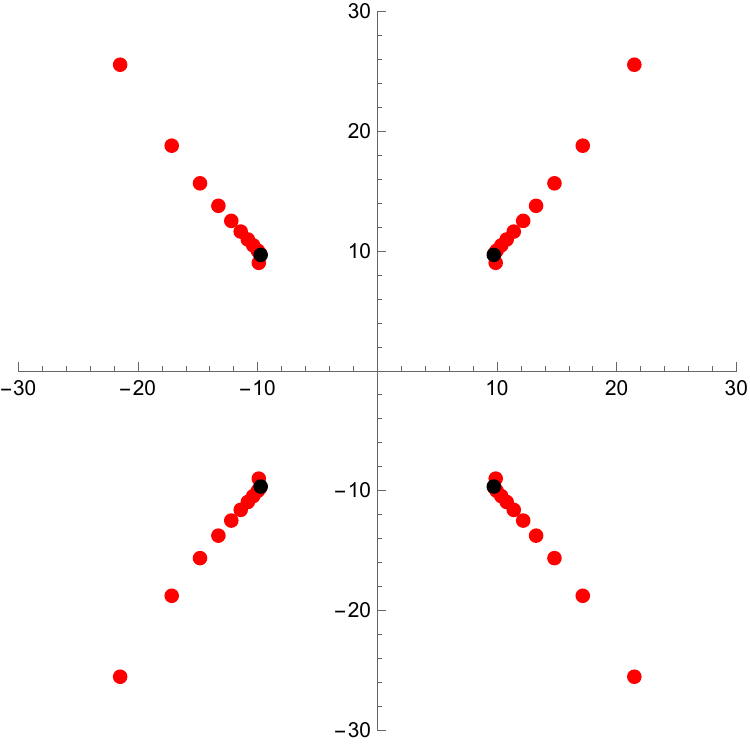}}
  \subfloat[$z=10^{-3}$]{\includegraphics[width=0.3\linewidth]{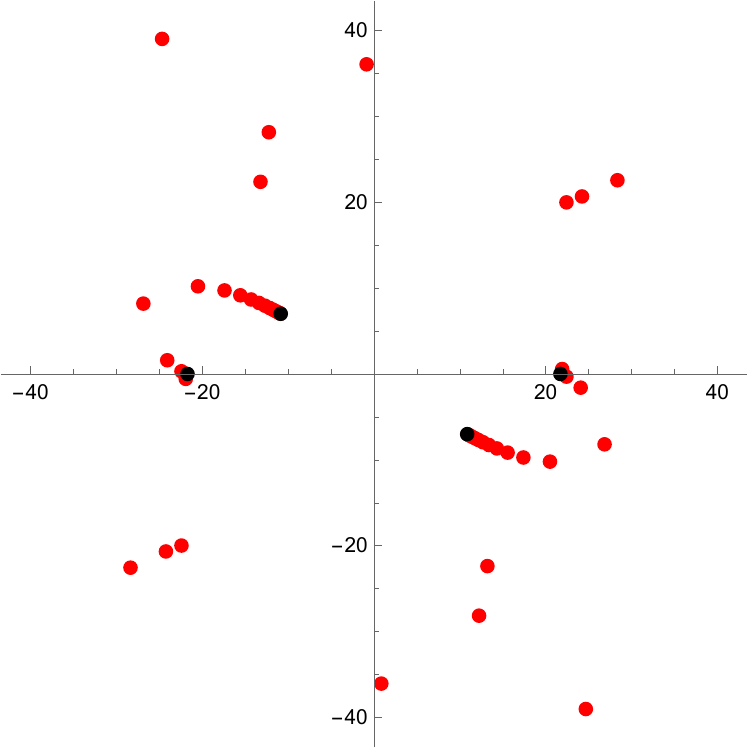}}\hspace{1ex}
  \caption{Borel singularities of refined Wilson loop BPS sectors
    $\CF[1](\sb;g_s)$ for local $\IP^2$ with $\sb=2$ up to $g=50$ in
    the conifold frame, respectively (a) near the large radius point
    with $z<0$, (c) near the conifold point, and (c) with $z>0$.  The
    red dots are approximate singularities from numerical
    calculations.  The branch points (black dots) are (a)
    $\sb^{-1}\CA_{\pm(-3,0,n)_\text{c}} (n=-1,0,1,2)$, (b)
    $\sb^{-1}\CA_{\pm(-3,1,0)_{\text{c}}}$,
    $\sb^{-1}\CA_{\pm(-3,-1,1)_{\text{c}}}$ and (c)
    $\sb^{-1}\CA_{\pm(-3,0,0)_{\text{c}}}$ (horizontal),
    $\sb^{-1}\CA_{\pm(-3,-1,0)_{\text{c}}}$ respectively.}
  \label{fig:P2C1b2Con-brl}
\end{figure}

\begin{figure}
  \centering
  \subfloat[$z=-10^{-6}$]{\includegraphics[width=0.3\linewidth]{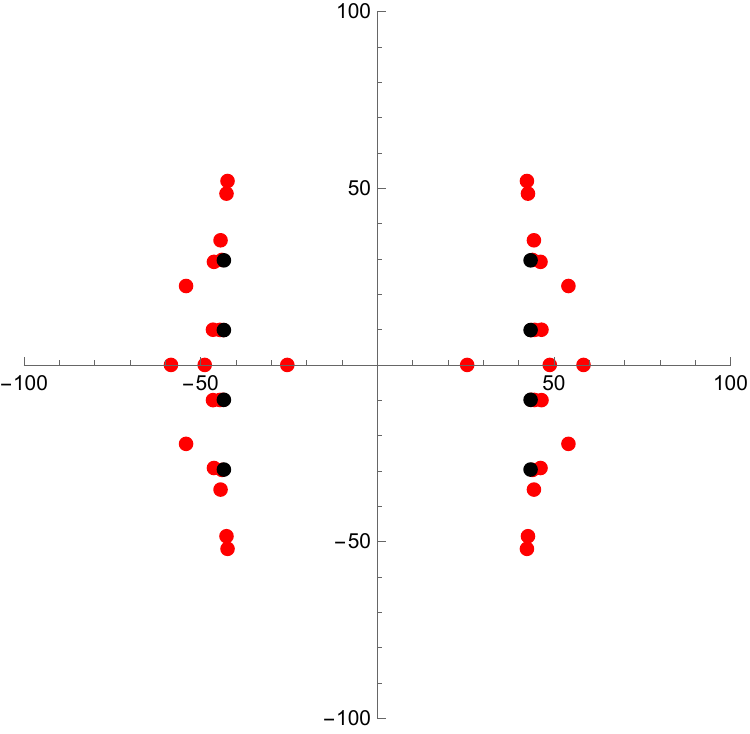}}\hspace{1ex}
  \subfloat[$z=-\frac{9}{270}$]{\includegraphics[width=0.3\linewidth]{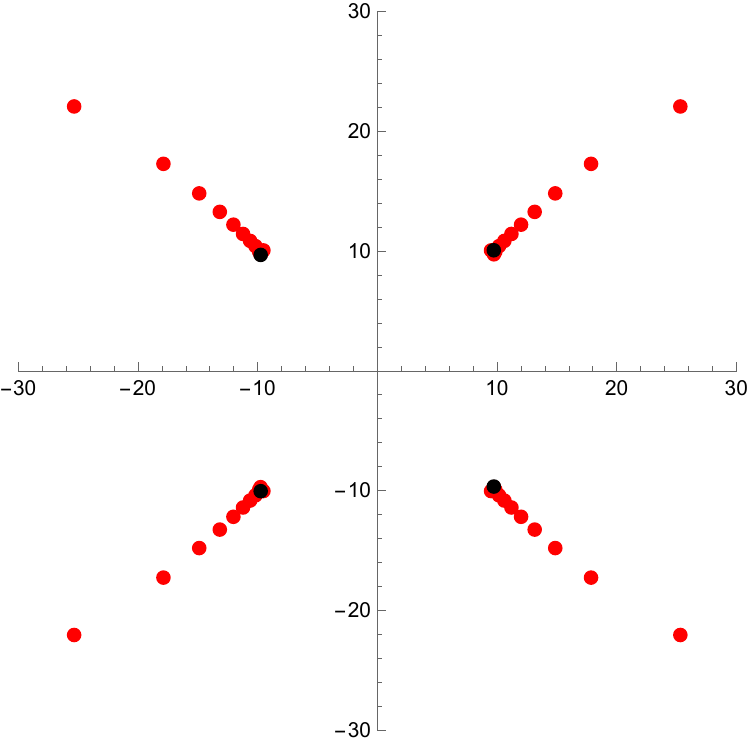}}
  \subfloat[$z=10^{-3}$]{\includegraphics[width=0.3\linewidth]{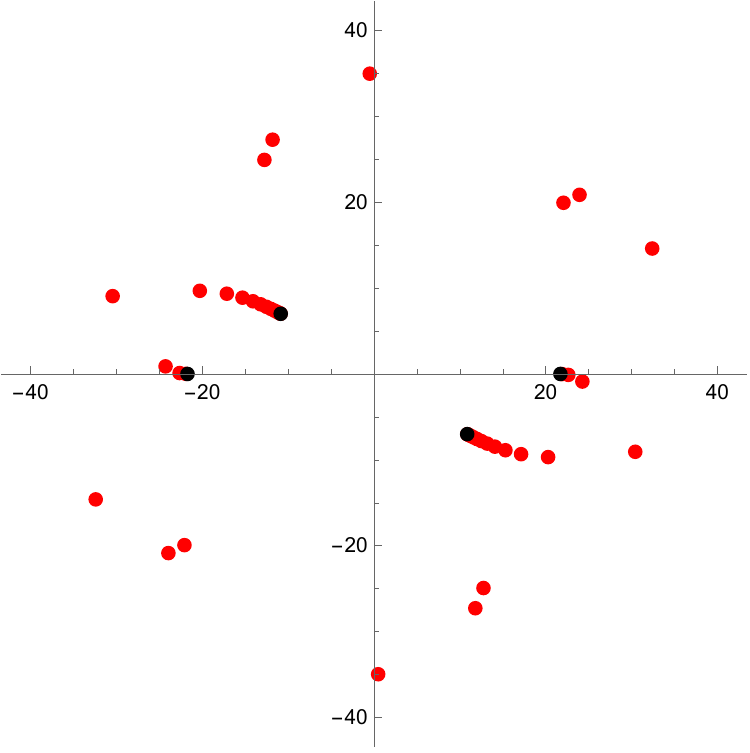}}\hspace{1ex}
  \caption{Borel singularities of refined Wilson loop BPS sectors
    $\CF[2](\sb;g_s)$ for local $\IP^2$ with $\sb=2$ up to $g=50$ in
    the conifold frame, respectively (a) near the large radius point
    with $z<0$, (b) near the conifold point $z=-1/27$, and (c) with
    $z>0$.  The red dots are approximate singularities from numerical
    calculations.  The branch points (black dots) are the same as in
    Fig.~\ref{fig:P2C1b2Con-brl}.}
  \label{fig:P2C2b2Con-brl}
\end{figure}

We also evaluate the perturbative BPS sectors $\CF[1]$ and $\CF[2]$ in
the holomorphic limit of the conifold frame, where $t_{\mathrm{c}}$ is
the flat coordinate.  Similarly, we focus on the loci near
respectively the large radius point $z = 0$ and the conifold point
$z=-1/27$.  The Borel singularities are shown in
Figs.~\ref{fig:P2C1b2Con-brl} and Figs.~\ref{fig:P2C2b2Con-brl}
respectively.
%
%
In both examples, we find that near the large radius point with $z<0$
(we take $\mathsf{b}>1$ so that $\mathsf{b}^{-1}\CA$ is smaller than
$\mathsf{b}\CA$), the visible Borel singularities are as usual located
at $\mathsf{b}^{-1}\CA_{\gamma_{\mathrm{c}}}$ with charge vectors
\begin{equation}
  \gamma_{\text{c}}=\pm(-3,0,\pm n),\quad n=1,2,3....
\end{equation}
Near the conifold point, the visible Borel singularities are located
at $\mathsf{b}^{-1}\CA_{\pm(-3,\pm1,0)}$.
Here we denote by $\CA_{\gamma_{\text{c}}}$
\begin{equation}\label{eq:Consingualrity}
  \CA_{\gamma_{\text{c}}} =
  -c'\ri\pd_{t_{\text{c}}}\CF^{(0,0)}_{\text{c}} + 2\pi d' t_{\text{c}}
  + 4\pi^2\ri d'_0,\quad
  \gamma_{\text{c}} = (c',d',d'_0),
\end{equation}
and the two charge conventions
$\gamma_{\text{LR}}$ and $\gamma_{\text{c}}$ are related to each other
via the relationship
\begin{equation}\label{eq:P2-relation}
  \gamma_{\text{c}} =
  \begin{pmatrix}
    0 & -3 & 0\\\frac{1}{3} & 0 & 0\\0&0&1
  \end{pmatrix}\gamma_{\text{LR}},\quad
  \gamma_{\text{LR}} =
  \begin{pmatrix}
    0 & 3 & 0\\-\frac{1}{3} & 0 & 0\\0&0&1
  \end{pmatrix}\gamma_{\text{c}}.
\end{equation}
In addition, we also consider the loci at $z>0$, and we find visible
Borel singularities
\begin{equation}
  \gamma_{\text{c}} = \pm (-3,1,0),\;\pm (-3,0,0),
\end{equation}
as shown in Figs.~\ref{fig:P2C1b2Con-brl} (c), \ref{fig:P2C2b2Con-brl} (c).
%
%
We find yet again that none of the Borel singularities coincide with
the flat coordinate up to a constant, i.e. the coefficient $c^\prime$
in $\gamma_{\text{c}}$ does not vanish.

Next, we study the non-perturbative series.  We focus on the
1-instanton sector, and check the coefficients of the non-perturbative
series \eqref{eq:Fm1inst-ref} in the generic refined case, and
\eqref{eq:Fm1inst-unref} in the unrefined limit with $\sb=1$.  In the
generic refined case, the 1-instanton non-perturbative series can be
written as
\begin{equation}
  \CF^{(1)}[m] = g_s^{m}\re^{-\tilde{\CA}/g_s}(\mu_0 + \mu_1 g_s + \mu_2 g_s^2 +
  \ldots)
\end{equation}
with $\tilde{A}$ being either $\sb^{-1}\CA$ or $\sb\CA$.  Compared
with the form of perturbative series \eqref{eq:Fm-pert}, standard 
resurgence analysis predicts the large order asymptotics of the
perturbative coefficients
\begin{equation}
  \label{eq:FgAsymp-ref}
  \CF_g[m] \sim \frac{\sS}{\pi\ri} \frac{\Gamma(2g+m-2)}{\tilde{\CA}^{2g+m-2}}\left(
    \mu_0 + \frac{\mu_1 \tilde{\CA}}{2g+m-3} +
    \frac{\mu_2\tilde{\CA}}{(2g+m-3)(2g+m-4)} + \ldots
  \right).
\end{equation}
where $\tilde{\CA}$ is the dominant Borel singularity, the closest to
the origin, and we have taken into account that both $\pm\tilde{\CA}$
sectors contribute equally to the asymptotic formula.  In the
unrefined limit with $\sb=1$, the 1-instanton non-perturbative series
can be written as
\begin{equation}
  \CF^{(1)}[m] = g_s^{m-1}\re^{-\tilde{\CA}/g_s}(\mu_0 + \mu_1 g_s + \mu_2 g_s^2 +
  \ldots)
\end{equation}
and the large order asymptotics should be modified to
\begin{equation}
  \label{eq:FgAsymp-unref}
  \CF_g[m] \sim \frac{\sS}{\pi\ri} \frac{\Gamma(2g+m-1)}{\tilde{\CA}^{2g+m-1}}\left(
    \mu_0 + \frac{\mu_1 \tilde{\CA}}{2g+m-2} +
    \frac{\mu_2\tilde{\CA}}{(2g+m-2)(2g+m-3)} + \ldots
  \right).
\end{equation}

\begin{figure}
  \centering
  \subfloat[$\mu_0$]{\includegraphics[width=0.3\linewidth]{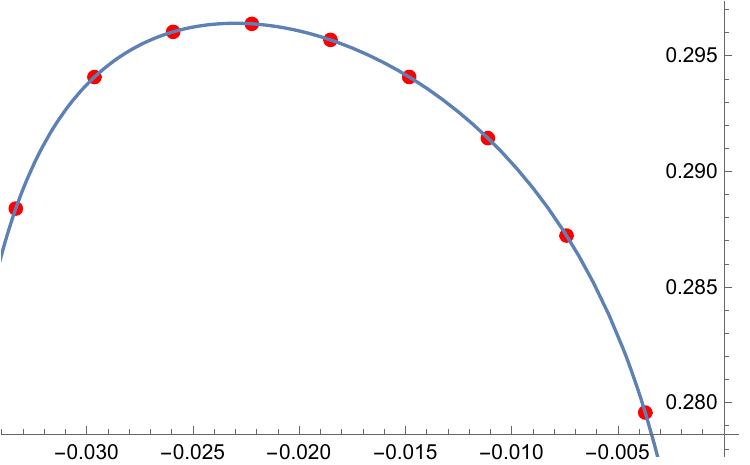}}\hspace{1ex}
  \subfloat[$\mu_1$]{\includegraphics[width=0.3\linewidth]{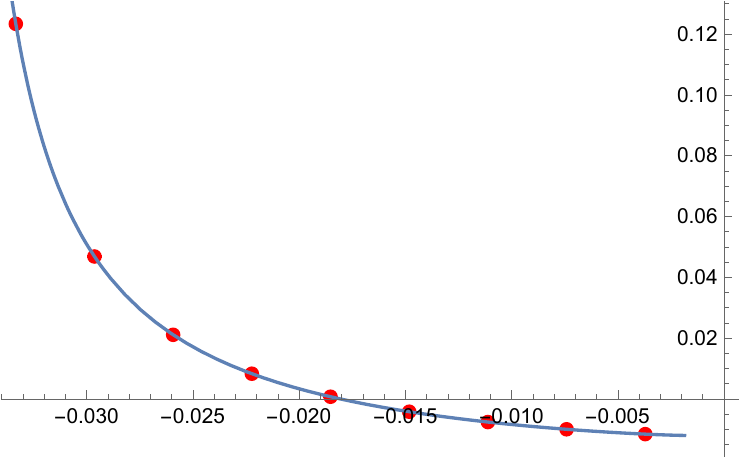}}\hspace{1ex}
  \subfloat[$\mu_2$]{\includegraphics[width=0.3\linewidth]{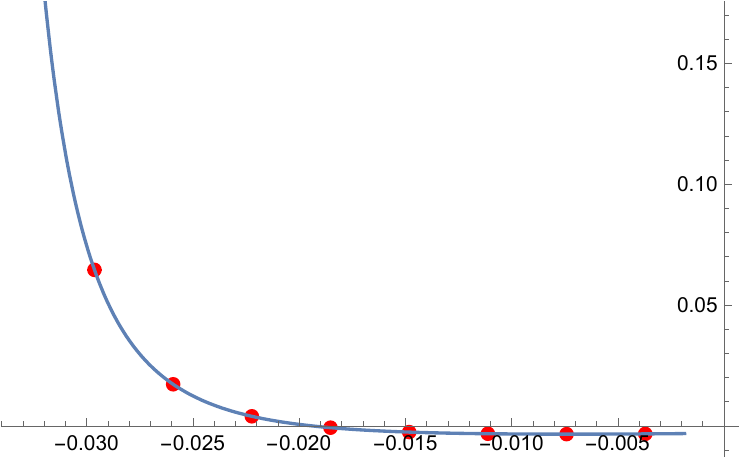}}
  \caption{Comparison for local $\IP^2$ between numerical results (red
    dots) of $\frac{1}{\pi\ri}\ms{S}\cdot\mu_{0,1,2}$ extracted from
    the large order asymptotics of $\CF_g[1]$ up to $g=50$ in the
    large radius frame at $\sb=2$ with error bars (vertical bars,
    virtually invisible) and trans-series solutions from HAE (solid
    line).  Richardson transformation of degree 10 is used to improve
    the numerics.  The horizontal axis is modulus $z$.}
  \label{fig:P2C1LRb2}
\end{figure}

\begin{figure}
  \centering
  \subfloat[$\mu_0$]{\includegraphics[width=0.3\linewidth]{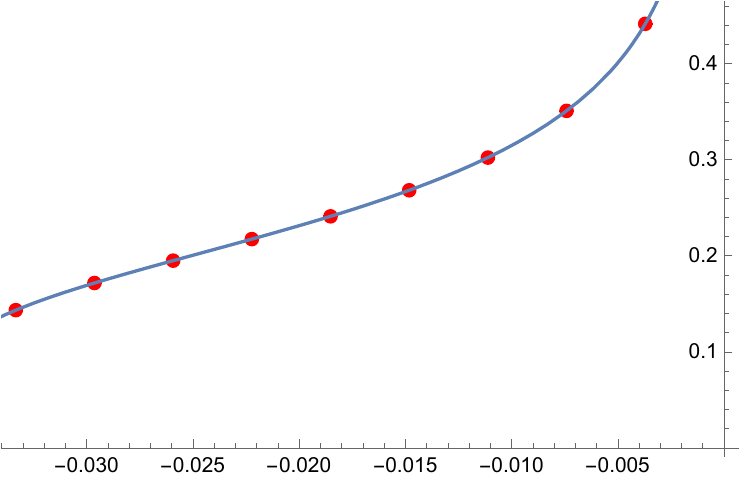}}\hspace{1ex}
  \subfloat[$\mu_1$]{\includegraphics[width=0.3\linewidth]{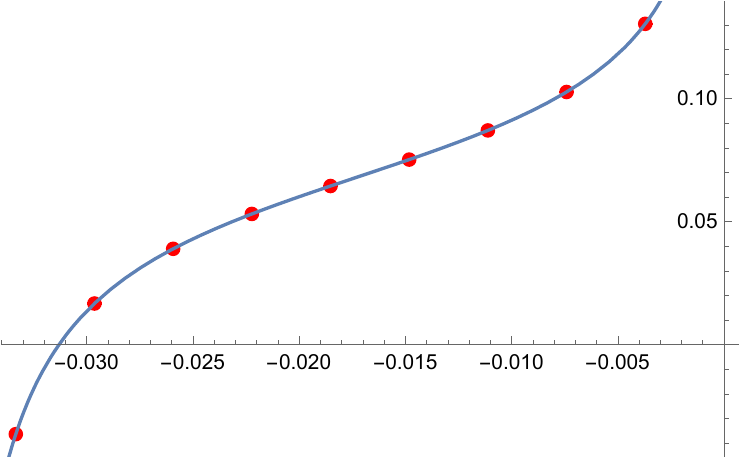}}\hspace{1ex}
  \subfloat[$\mu_2$]{\includegraphics[width=0.3\linewidth]{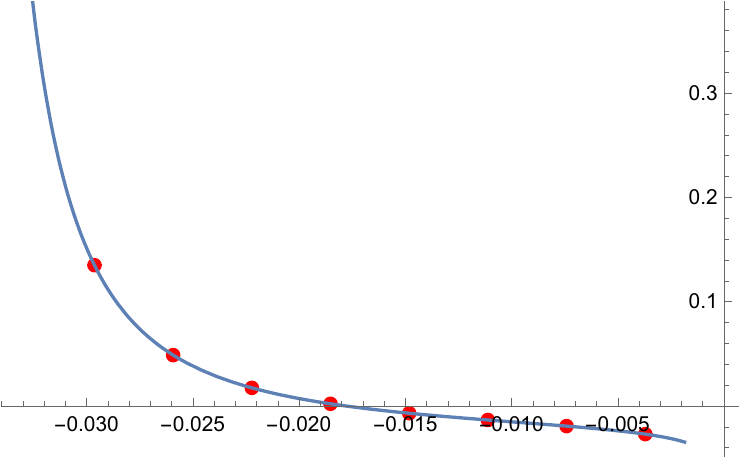}}
  \caption{Comparison for local $\IP^2$ between numerical results (red
    dots) of $\frac{1}{\pi\ri}\ms{S}\cdot\mu_{0,1,2}$ extracted from
    the large order asymptotics of $\CF_g[2]$ up to $g=50$ in the
    large radius frame at $\sb=2$ with error bars (vertical bars,
    virtually invisible) and trans-series solutions from HAE (solid
    line).  Richardson transformation of degree 10 is used to improve
    the numerics.  The horizontal axis is modulus $z$.}
  \label{fig:P2C2LRb2}
\end{figure}

\begin{figure}
  \centering
  \subfloat[$\mu_0$]{\includegraphics[width=0.3\linewidth]{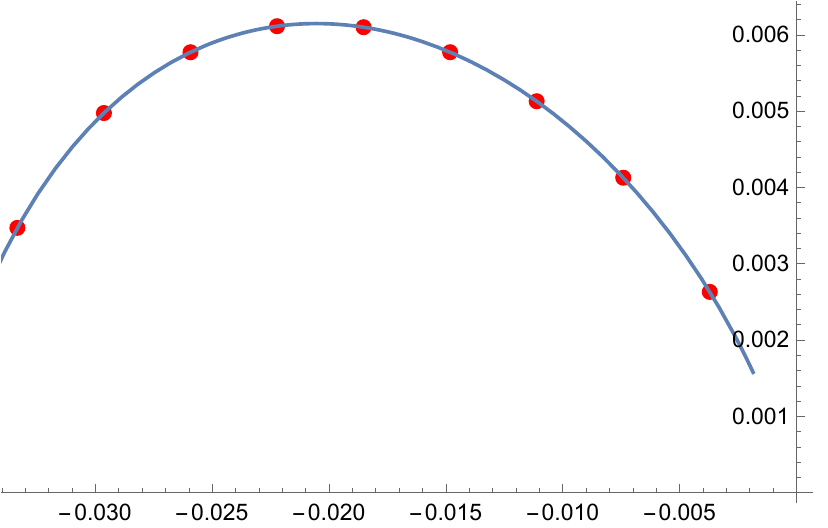}}\hspace{1ex}
  \subfloat[$\mu_1$]{\includegraphics[width=0.3\linewidth]{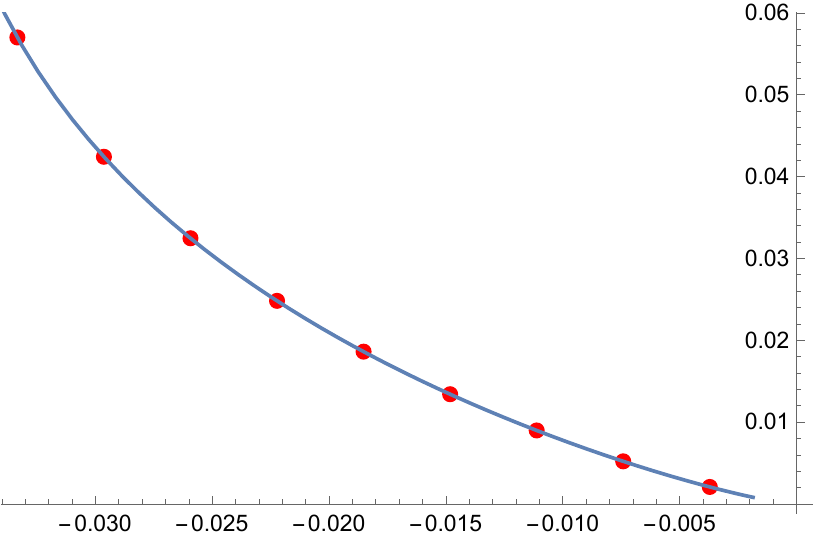}}\hspace{1ex}
  \subfloat[$\mu_2$]{\includegraphics[width=0.3\linewidth]{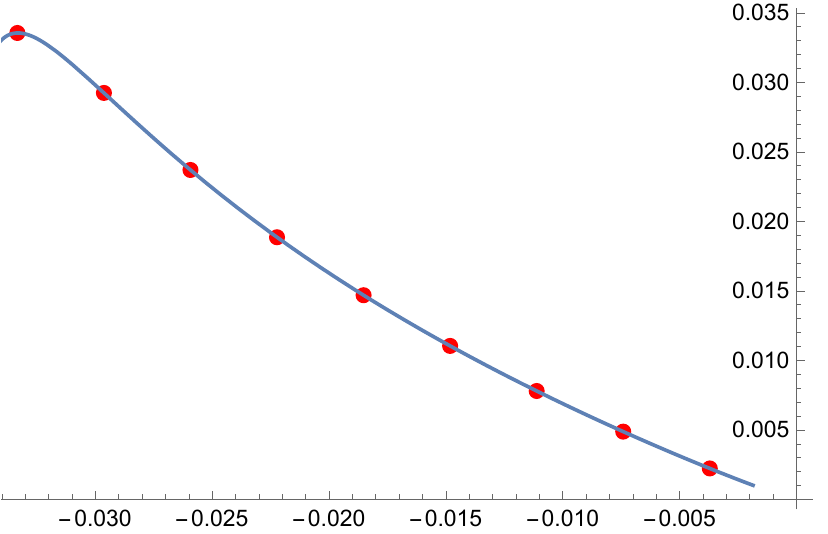}}
  \caption{Comparison for local $\IP^2$ between numerical results (red dots) of
    $\frac{1}{\pi\ri}\ms{S}\cdot\mu_{0,1,2}$ extracted from the large
    order asymptotics of $\CF_g[1]$ up to $g=100$ in the large radius frame at
    $\sb=1$ with error bars (vertical bars, virtually invisible) and
    trans-series solutions from HAE (solid line).  Richardson
    transformation of degree 15 is used to improve the numerics.  The
    horizontal axis is modulus $z$.}
  \label{fig:P2C1LRb1}
\end{figure}

\begin{figure}
  \centering
  \subfloat[$\mu_0$]{\includegraphics[width=0.3\linewidth]{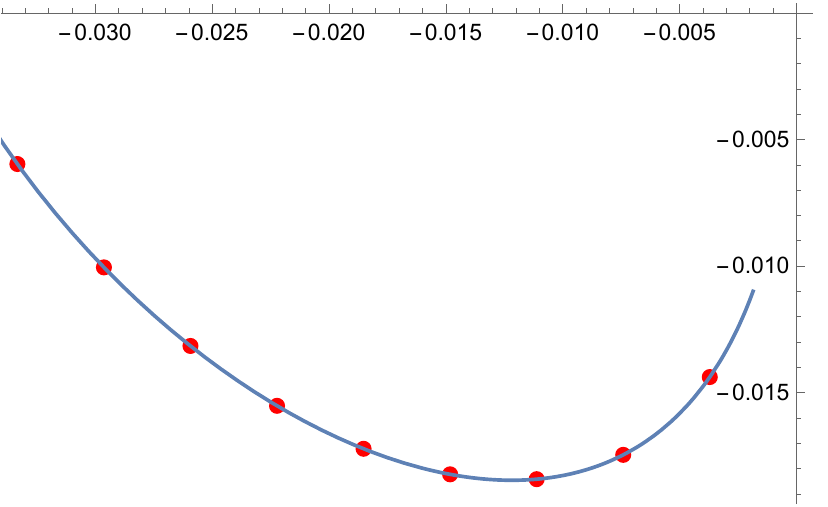}}\hspace{1ex}
  \subfloat[$\mu_1$]{\includegraphics[width=0.3\linewidth]{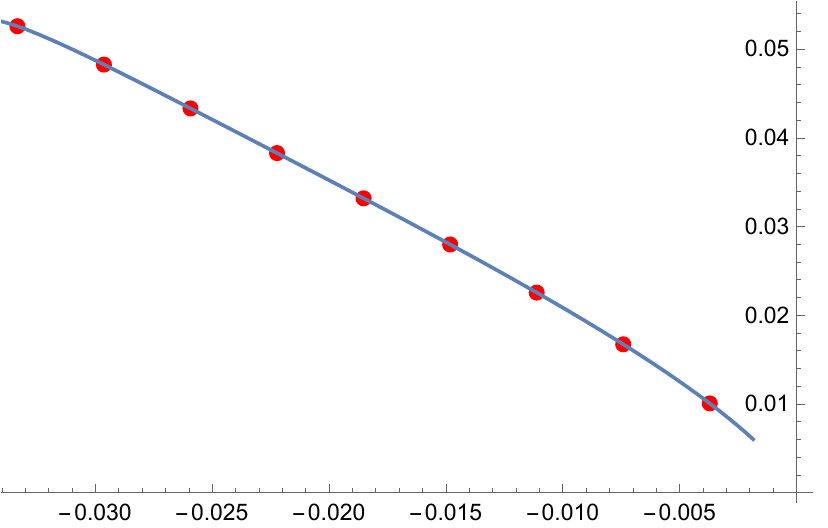}}\hspace{1ex}
  \subfloat[$\mu_2$]{\includegraphics[width=0.3\linewidth]{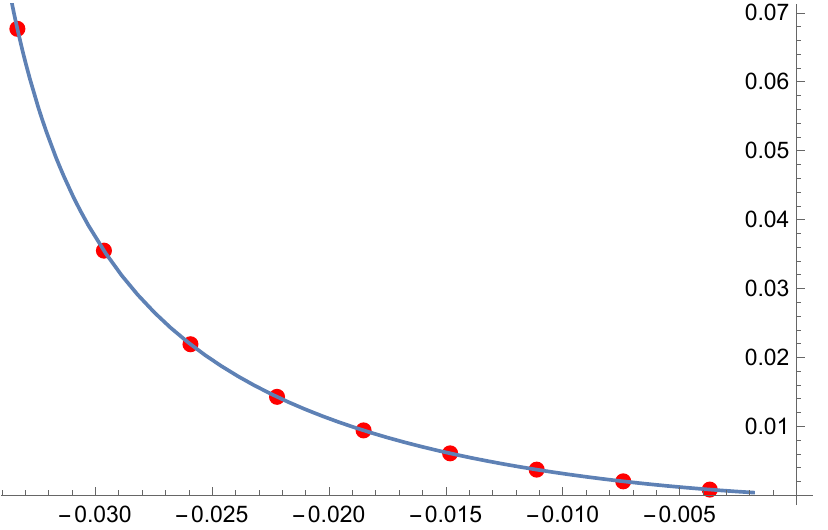}}
  \caption{Comparison for local $\IP^2$ between numerical results (red dots) of
    $\frac{1}{\pi\ri}\ms{S}\cdot\mu_{0,1,2}$ extracted from the large
    order asymptotics of $\CF_g[2]$ up to $g=100$ in the large radius frame at
    $\sb=1$ with error bars (vertical bars, virtually invisible) and
    trans-series solutions from HAE (solid line).  Richardson
    transformation of degree 15 is used to improve the numerics.  The
    horizontal axis is modulus $z$.}
  \label{fig:P2C2LRb1}
\end{figure}

We consider two different cases.  The first is the BPS sectors near
the conifold point in the large radius frame.  The dominant Borel
singularities are the pair of $\gamma_{\text{LR}} = \pm (-3,0,0)$ as
shown in Figs.~\ref{fig:P2C1b2LR-brl} (b), \ref{fig:P2C2b2LR-brl} (b).
The large order asymptotics formula \eqref{eq:FgAsymp-ref} (formula
\eqref{eq:FgAsymp-unref} in the unrefined limit) can be used to
extract the non-perturbative coefficients $\mu_0,\mu_1,\mu_2,\ldots$,
and we compare these numerical results with our prediction from
Sections~\ref{sc:Fm-nonp} and \ref{sc:Fm-nonp-unref} in
Figs.~\ref{fig:P2C1LRb2}, \ref{fig:P2C2LRb2} for generic $\sb$, and in
Figs.~\ref{fig:P2C1LRb1}, \ref{fig:P2C2LRb1} in the unrefined limit.
The numerical results and the theoretical predictions match perfectly,
as long as we choose the Stokes constants
\begin{equation}
  \sS_{\pm(-3,0,0)_{\text{LR}}}(\sb) = 1,
\end{equation}
corresponding to the spin 0 BPS state of D4 brane wrapping $\IP^2$ in
type IIA superstring.

\begin{figure}
  \centering
  \subfloat[$\mu_0$]{\includegraphics[width=0.3\linewidth]{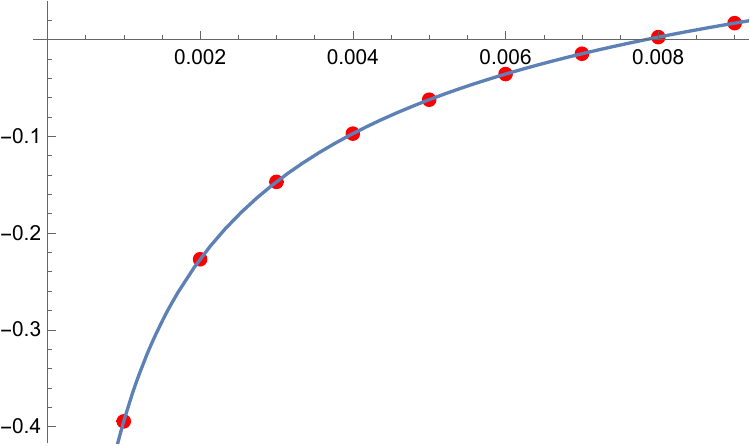}}\hspace{1ex}
  \subfloat[$\mu_1$]{\includegraphics[width=0.3\linewidth]{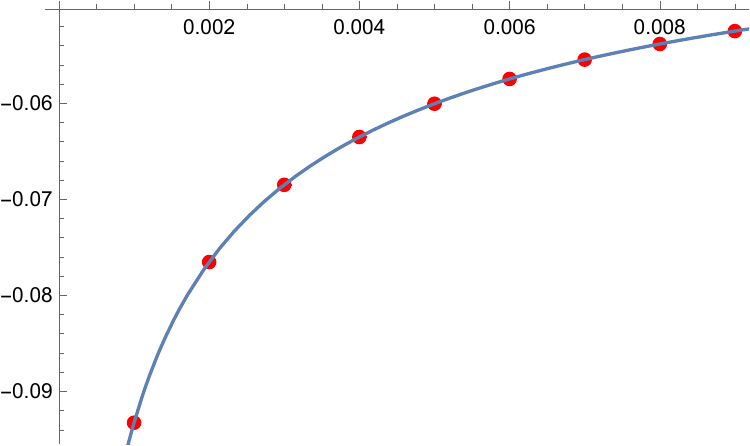}}\hspace{1ex}
  \subfloat[$\mu_2$]{\includegraphics[width=0.3\linewidth]{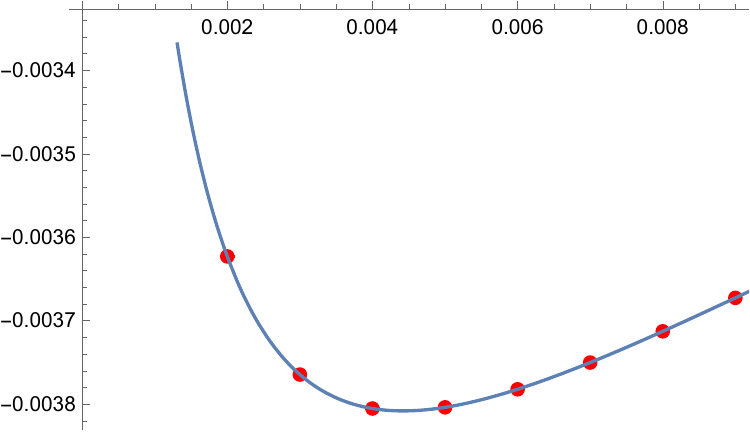}}
  \caption{Comparison for local $\IP^2$ in the conifold frame at
    $\sb=2$ with large positive $z$ between numerical results (red
    dots) of $\frac{1}{\pi\ri}\ms{S}\cdot\mu_{0,1,2}$ for
    $\CA_{\pm(-3,1,0)_{\text{c}}}$ extracted from the large order
    asymptotics of $\CF_g[1]$ up to $g=50$ with error bars (vertical
    bars, virtually invisible) and trans-series solutions from HAE
    (solid line).  Richardson transformation of degree 5 is used to
    improve the numerics.  The horizontal axis is modulus $z$.}
  \label{fig:P2C1Conb2}
\end{figure}

\begin{figure}
  \centering
  \subfloat[$\mu_0$]{\includegraphics[width=0.3\linewidth]{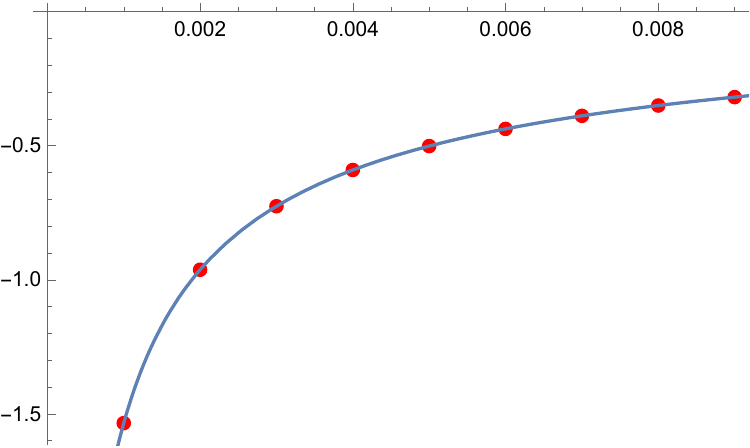}}\hspace{1ex}
  \subfloat[$\mu_1$]{\includegraphics[width=0.3\linewidth]{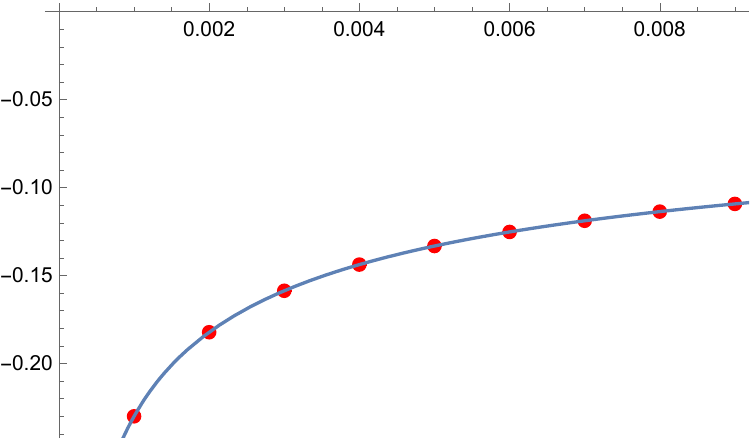}}\hspace{1ex}
  \subfloat[$\mu_2$]{\includegraphics[width=0.3\linewidth]{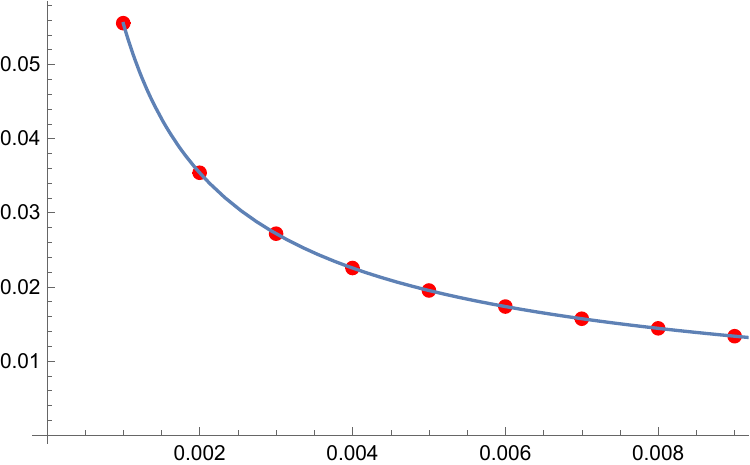}}
  \caption{Comparison for local $\IP^2$ in the conifold frame at
    $\sb=2$ with large positive $z$ between numerical results (red
    dots) of $\frac{1}{\pi\ri}\ms{S}\cdot\mu_{0,1,2}$ for
    $\CA_{\pm(-3,1,0)_{\text{c}}}$ extracted from the large order
    asymptotics of $\CF_g[2]$ up to $g=50$ with error bars (vertical
    bars, virtually invisible) and trans-series solutions from HAE
    (solid line).  Richardson transformation of degree 5 is used to
    improve the numerics.  The horizontal axis is modulus $z$.}
  \label{fig:P2C2Conb2}
\end{figure}

Similarly, we consider the BPS sectors at $z>0$ in the conifold frame.
Depending on the actual value of $z$, the two pairs of Borel
singularities $\gamma_{\text{c}} = \pm(-3,1,0)$ and
$\gamma_{\text{c}} = \pm(-3,0,0)$ compete in dominance, as shown in
Fig.~\ref{fig:P2C1b2Con-brl} (c), \ref{fig:P2C2b2Con-brl} (c).  If
$z \gtrsim 4\times 10^{-6}$, the pair of
$\gamma_{\text{c}}=\pm(-3,1,0)_{\text{c}}$ is dominant (closer to the
origin).  The comparison between the numerical results of
$\mu_0,\mu_1,\mu_2$ from large order asymptotics and theoretical
predictions are plotted in Figs.~\ref{fig:P2C1Conb2},
\ref{fig:P2C2Conb2}.  Here we have chosen the Stokes constants
\begin{equation}
  \sS_{\pm(-3,1,0)_{\text{c}}}(\sb) = \sS_{\pm (3,1,0)_{\text{LR}}}(\sb)  = 1,
\end{equation}
where we have used the charge vector relations \eqref{eq:P2-relation},
and they correspond to the spin 0 BPS states of D4 brane wrapping
$\IP^2$ together with a D2 brane wrapping $\IP^1\subset \IP^2$.

\begin{figure}
  \centering
  \subfloat[$\mu_0$]{\includegraphics[width=0.3\linewidth]{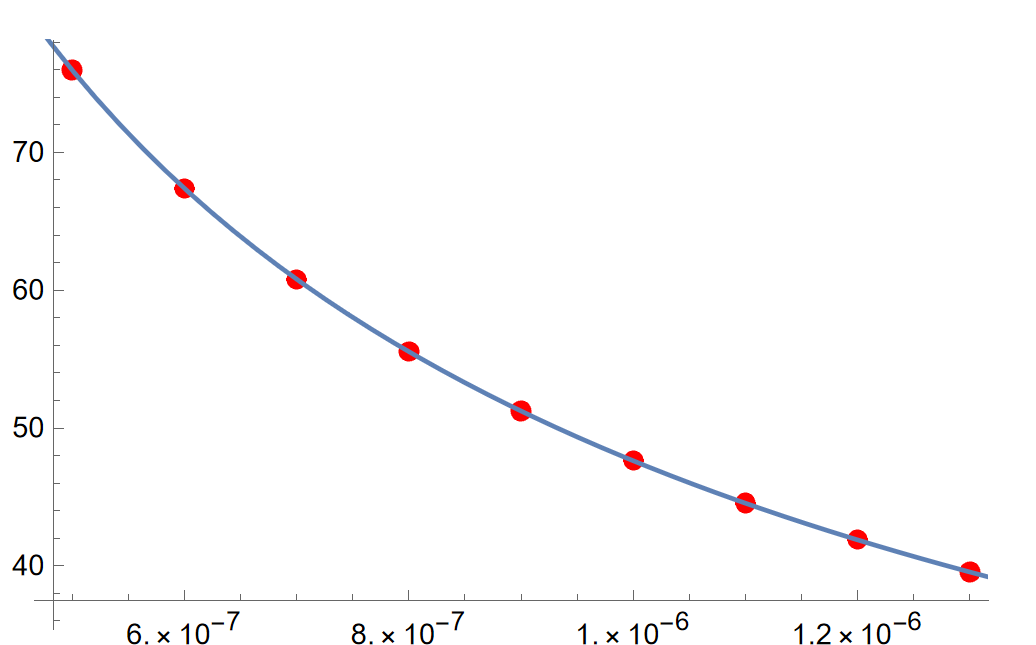}}\hspace{1ex}
  \subfloat[$\mu_1$]{\includegraphics[width=0.3\linewidth]{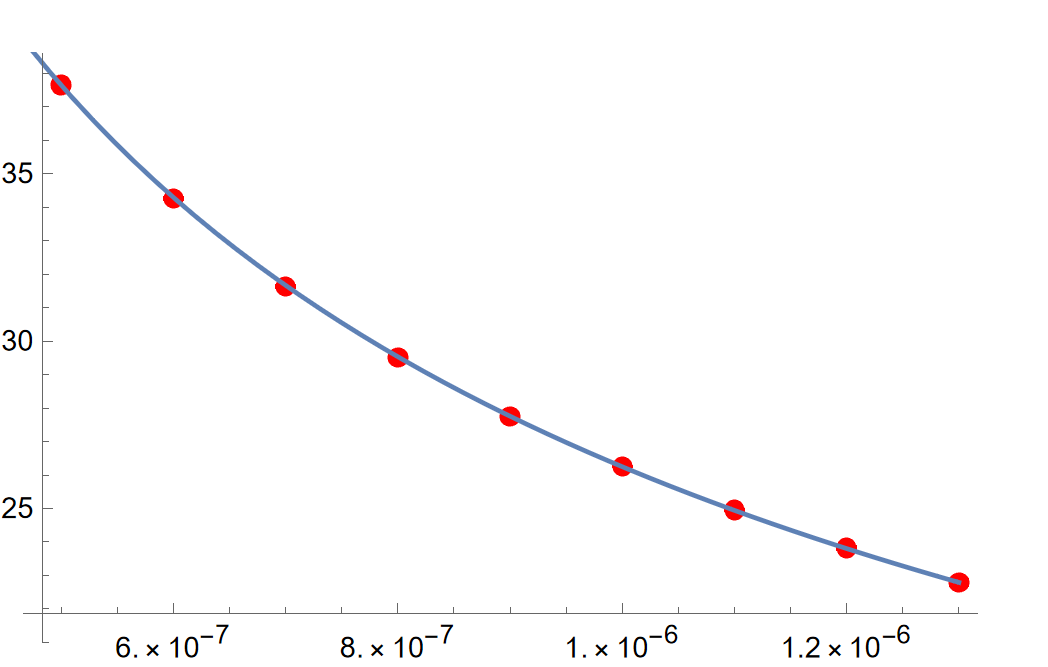}}\hspace{1ex}
  \subfloat[$\mu_2$]{\includegraphics[width=0.3\linewidth]{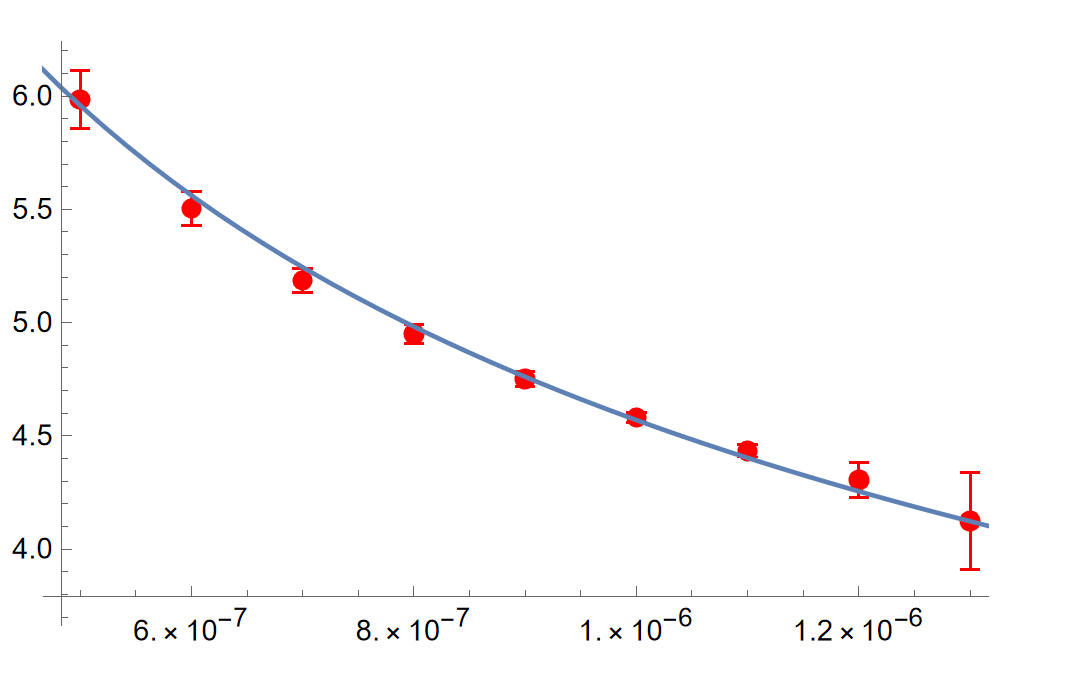}}
  \caption{Comparison for local $\IP^2$ in the conifold frame at
    $\sb=1$ with small and positive $z$ between numerical results (red
    dots) of $\frac{1}{\pi\ri}\ms{S}\cdot\mu_{0,1,2}$ for
    $\CA_{\pm(-3,0,0)_{\text{c}}}$ extracted from the large order
    asymptotics of $\CF_g[1]$ up to $g=100$ with error bars (vertical
    bars, virtually invisible) and trans-series solutions from HAE
    (solid line).  Richardson transformation of degree 5 is used to
    improve the numerics.  The horizontal axis is modulus $z$.}
  \label{fig:P2C1Conb1}
\end{figure}

\begin{figure}
  \centering
  \subfloat[$\mu_0$]{\includegraphics[width=0.3\linewidth]{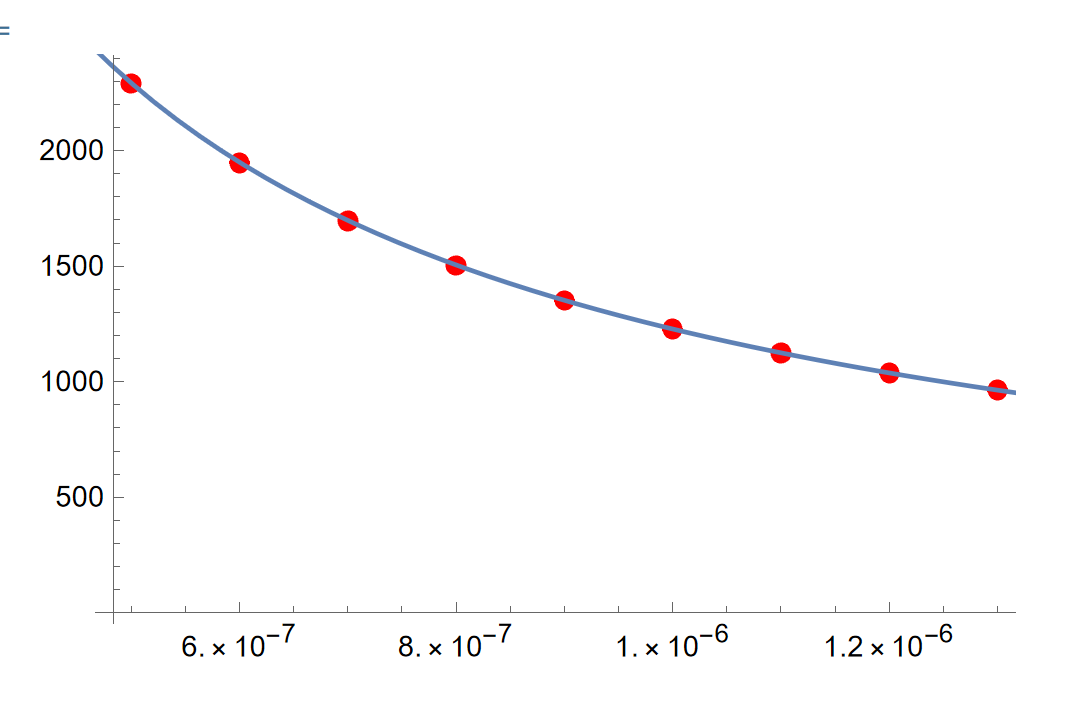}}\hspace{1ex}
  \subfloat[$\mu_1$]{\includegraphics[width=0.3\linewidth]{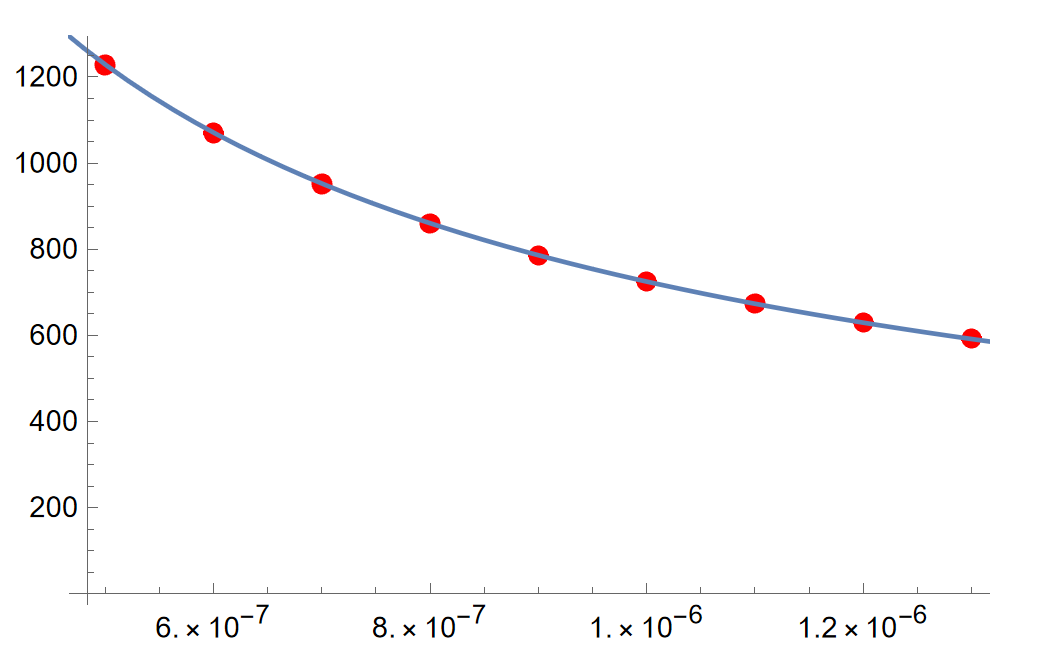}}\hspace{1ex}
  \subfloat[$\mu_2$]{\includegraphics[width=0.3\linewidth]{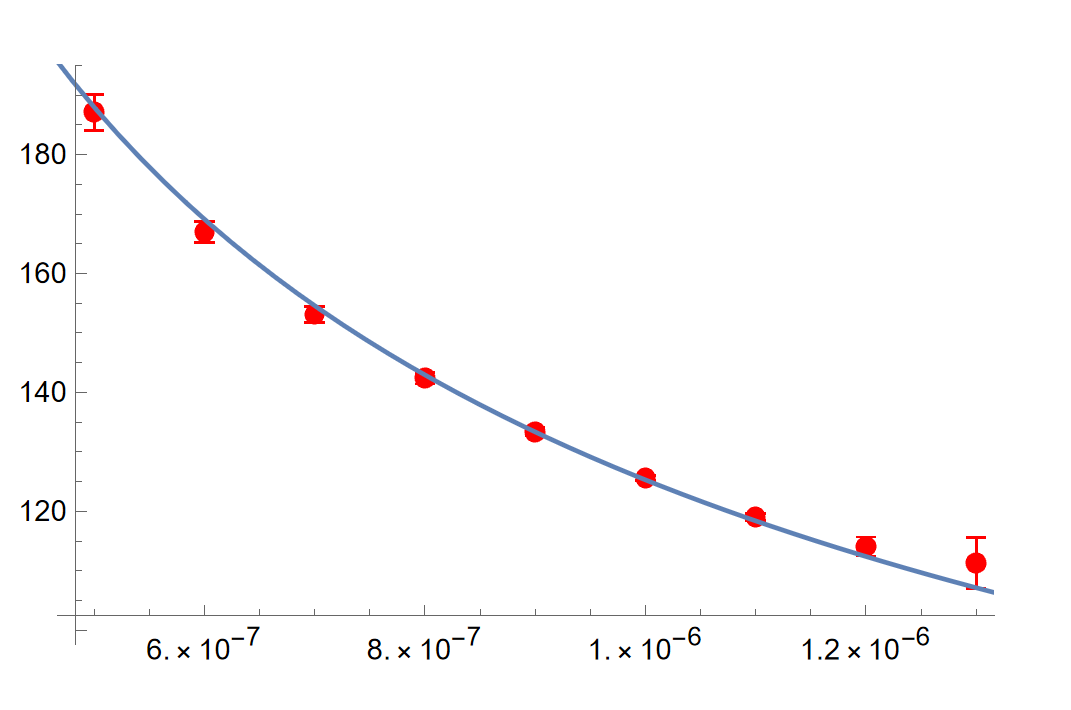}}
  \caption{Comparison for local $\IP^2$ in the conifold frame at
    $\sb=1$ with small and positive $z$ between numerical results (red
    dots) of $\frac{1}{\pi\ri}\ms{S}\cdot\mu_{0,1,2}$ for
    $\CA_{\pm(-3,0,0)_{\text{c}}}$ extracted from the large order
    asymptotics of $\CF_g[2]$ up to $g=100$ with error bars (vertical
    bars, virtually invisible) and trans-series solutions from HAE
    (solid line).  Richardson transformation of degree 5 is used to
    improve the numerics.  The horizontal axis is modulus $z$.}
  \label{fig:P2C2Conb1}
\end{figure}

Finally, if $0< z \lesssim 4\times 10^{-6}$, the pair of
$\gamma_{\text{c}} = \pm(-3,0,0)_{\text{c}}$ is dominant, and the
plots for $\mu_0,\mu_1,\mu_2$ are given in Figs.~\ref{fig:P2C1Conb1},
\ref{fig:P2C2Conb1} (in the unrefined limit).  Here we have chosen the Stokes constants
\begin{equation}
  \sS_{\pm (-3,0,0)_{\text{c}}}(\sb) =
  \sS_{\pm (0,1,0)_{\text{LR}}}(\sb)
  = 3\chi_{1/2}(-\re^{-\pi\ri/\sb^2}),
\end{equation}
corresponding to the well-known spin 1/2 BPS states of a single D2
brane wrapping $\IP^1\subset \IP^2$.

\subsection{Example: local $\IP^1\times \IP^1$}
\label{sc:F0}

Similar to Section \ref{sc:P2}, we consider the example of refined
topological string theory on local $\IP^1\times\IP^1$, i.e.~the total
space of the canonical bundle of $\IP^1\times \IP^1$, with the
constraint that the two $\IP^1$s have the same volume, also known as
the massless limit.  This theory has also been discussed in detail in
the literature.  It has a one dimensional moduli space with three
singular points of large radius, conifold, and orbifold types
\cite{Haghighat:2008gw}, and we take the convention that it is
parametrized by the global complex coordinate $z$, such that the three
singular points are located at $z=0$, $z=1/16$, and $z=\infty$
respectively\footnote{Note that in the case of massless local $\IF_0$,
  the orbifold point at $z=\infty$ has a conifold singularity
  superimposed upon it so that the free energies satisfy the gap
  conditions at both the conifold point and the orbifold point.}.

The periods of the theory are annihilated by the Picard-Fuchs operator
\cite{Chiang:1999tz}  
\begin{equation}
  \CL = (1-16z)\pd_z + z(3-64z)\pd_z^2 + z^2(1-16z)\pd_z^3.
\end{equation}
Near the large radius point, the flat coordinate and its conjugate
are (see e.g.~\cite{Marino:2015ixa})
\begin{subequations}
\begin{align}
  t_{\text{LR}}=
  &-\log(z) -
    4z{}_4F_3(1,1,\tfrac{3}{2},\tfrac{3}{2};2,2,2;16z),\nn
    \frac{\pd \CF^{(0,0)}_{\text{LR}}}{\pd t_{\text{LR}}} =
  &\frac{1}{\pi}G^{3,2}_{3,3}
    \left(
  \begin{array}{c}
    \tfrac{1}{2},\tfrac{1}{2};1\\0,0,0
  \end{array}; 16z \right)
  - \pi^2,
\end{align}
\end{subequations}
while near the conifold point, the flat coordinate and its conjugate
are
\begin{subequations}
  \begin{align}
    t_{\text{c}} =
    &\frac{1}{\pi \ri}\frac{\pd \CF^{(0,0)}_{\text{LR}}}{\pd t_{\text{LR}}},\nn
      \frac{\pd \CF^{(0,0)}_{\text{c}}}{\pd t_{\text{c}}} =
    &-\pi\ri t_{\text{LR}}.
  \end{align}
\end{subequations}


As in Section \ref{sc:P2}, we first inspect the non-perturbative
corrections for Wilson loop BPS sectors, which can be calculated
effectively using the algorithm in \cite{Wang:2023zcb}.  For
simplicity, we focus on the range between the large radius point
$z = 0$ and the conifold point $z = 1/16$. 

\begin{figure}
  \centering
  \subfloat[$z=10^{-6}$]{\includegraphics[height=5cm]{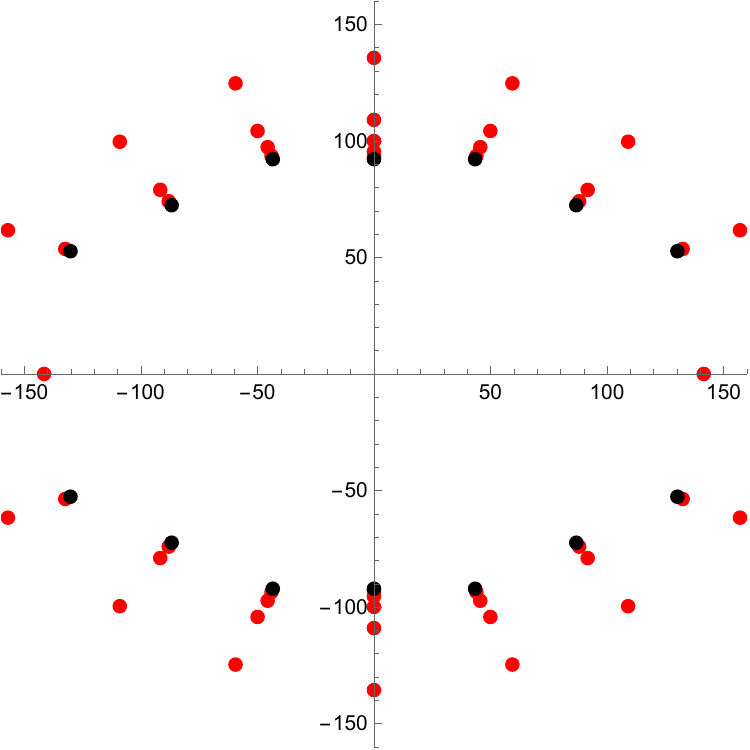}}\hspace{5ex}
  \subfloat[$z=\frac{9}{160}$]{\includegraphics[height=5cm]{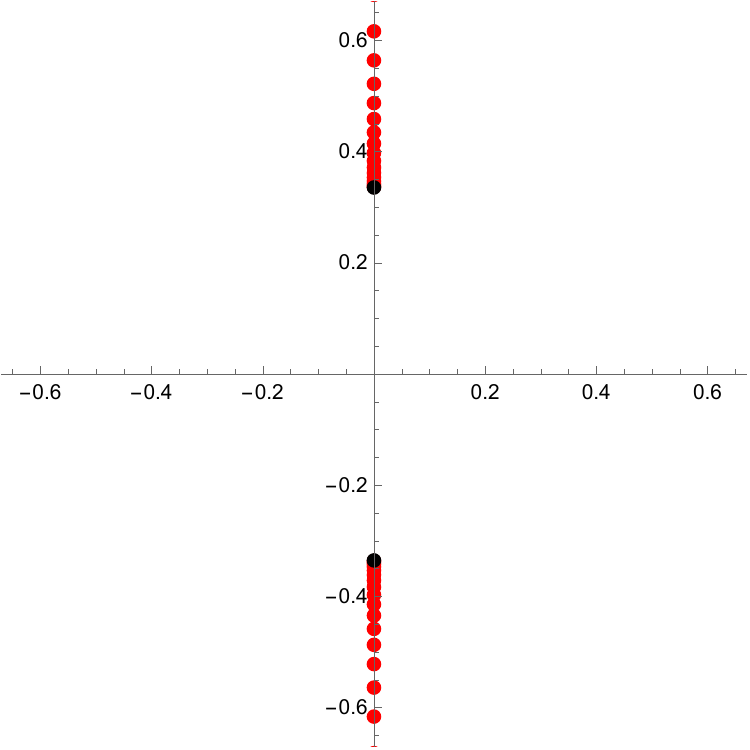}}
  \caption{Borel singularities of refined Wilson loop BPS sectors
    $\CF[1](\sb;g_s)$ for local $\IP^1\times\IP^1$ with $\sb=2$ up to
    $g=50$ in the large radius frame, respectively (a) near the large
    radius point $z=0$ and (b) near the conifold point $z=1/16$.  The
    red dots are approximate singularities from numerical
    calculations, which would accumulate to branch cuts.  The branch
    points (black dots) on the imaginary axis are
    $\sb^{-1}\CA_{\pm(-2,0,0)_\text{LR}}$, and those away from the
    imaginary axis are $\sb^{-1}\CA_{\pm(-2,1,0)_\text{LR}},%
    \sb^{-1}\CA_{\pm(-2,2,-1)_\text{LR}},%
    \sb^{-1}\CA_{\pm(-2,3,-2)_\text{LR}}$.}
  \label{fig:F0C1b2LR-brl}
\end{figure} 

\begin{figure}
  \centering
  \subfloat[$z=10^{-6}$]{\includegraphics[height=5cm]{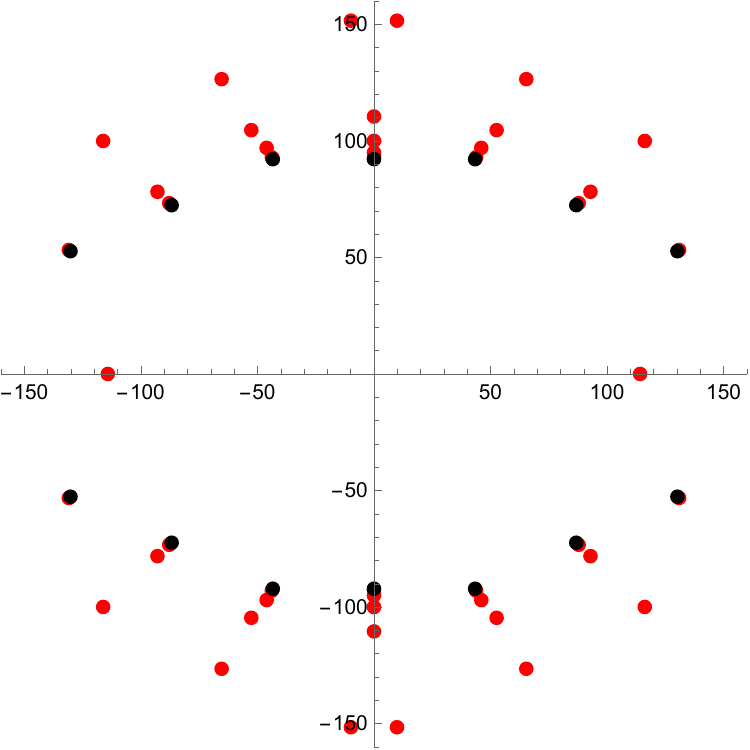}}\hspace{5ex}
  \subfloat[$z=\frac{9}{160}$]{\includegraphics[height=5cm]{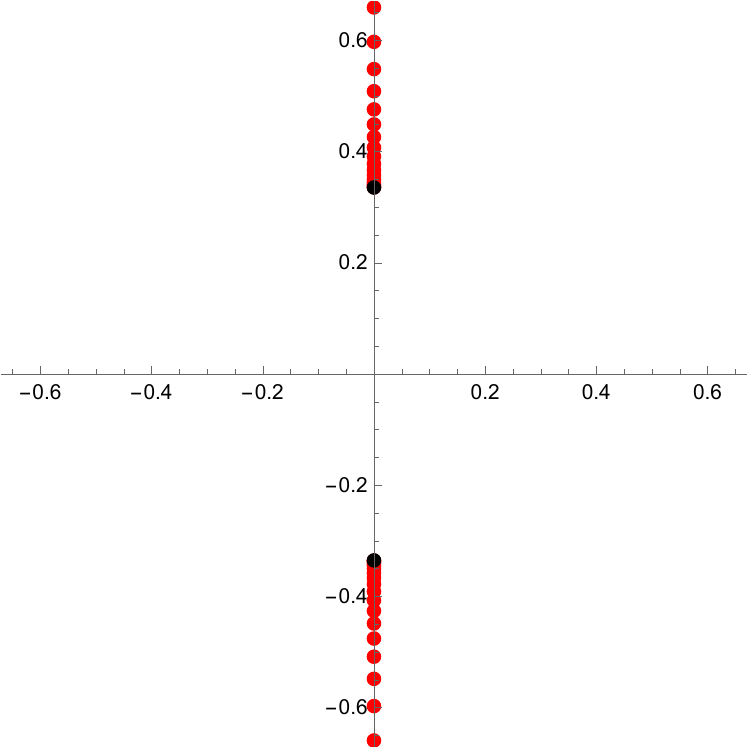}}
  \caption{Borel singularities of refined Wilson loop BPS sectors
    $\CF[2](\sb;g_s)$ for local $\IP^1\times\IP^1$ with $\sb=2$ up to
    $g=50$ in the large radius frame, respectively (a) near the large
    radius point $z=0$ and (b) near the conifold point $z=1/16$.  The
    red dots are approximate singularities from numerical
    calculations, which would accumulate to branch cuts.  The branch
    points (black dots) are the same as in
    Fig.~\ref{fig:F0C1b2LR-brl}.}
  \label{fig:F0C2b2LR-brl}
\end{figure}

\begin{figure}
  \centering
  \subfloat[$z=10^{-6}$]{\includegraphics[height=5cm]{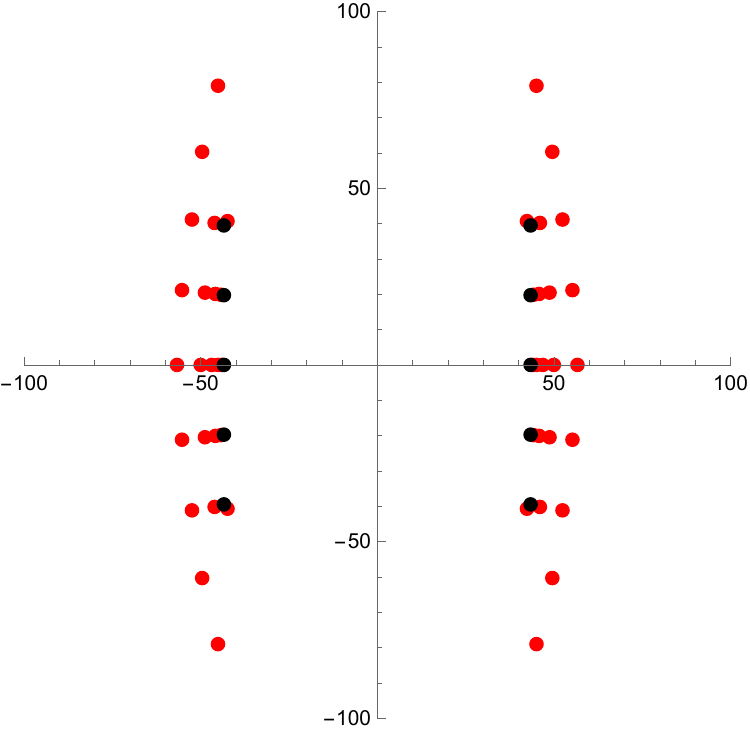}}\hspace{5ex}
  \subfloat[$z=\frac{9}{160}$]{\includegraphics[height=5cm]{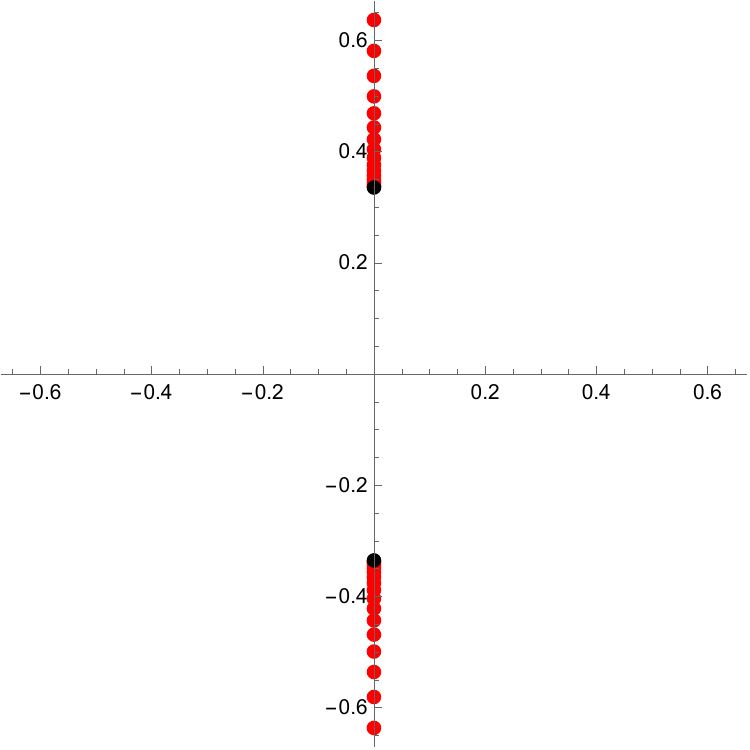}}
  \caption{Borel singularities of refined free energies
    $\CF[0](\sb;g_s)$ for local $\IP^1\times\IP^1$ with $\sb=2$ up to
    $g=50$ in the large radius frame, respectively (a) near the large
    radius point $z=0$ and (b) near the conifold point $z=1/16$.  The
    red dots are approximate singularities from numerical
    calculations.  The branch points (black dots) are (a)
    $\sb^{-1}\CA_{\pm(0,1,n)_{\text{LR}}}$ ($n=-1,-2,0,1,2$)
    and (b) $\sb^{-1}\CA_{\pm(-2,0,0)_\text{LR}}$ respectively.}
  \label{fig:F0C0b2LR-brl}
\end{figure}

Let us study the location of Borel singularities first.  We evaluate
the perturbative BPS sectors $\CF[1]$ and $\CF[2]$ in the holomorphic
limit of the large radius frame, where $t_{\text{LR}}$ is the flat
coordinate, near respectively the large radius point $z=0$ and the
conifold point $z= 1/16$.  The Borel singularities of $\CF[1]$ and
$\CF[2]$ are plotted in Fig.~\ref{fig:F0C1b2LR-brl} and
Fig.~\ref{fig:F0C2b2LR-brl}.  These two plots are similar.  Near the
large radius point, the visible Borel singularities are located at
$\sb^{-1}\CA_{\gamma_{\text{LR}}}$
(we take $\sb>1$ so that $\sb^{-1}\CA$ is smaller than $\sb\CA$) with
the charges 
\begin{equation}
  \label{eq:gammaF0}
  \gamma_{\text{LR}} = \pm (-2,0,0),\pm(-2,\pm(1+n),-n),\;\; n=0,1,2,\ldots.
\end{equation}
Near the conifold point, the visible Borel singularities are located
at $\sb^{-1}\CA_{\pm(-2,0,0)_{\text{LR}}}$.  For comparison, we also
give the same plots for the free energies\footnote{The constant map
  contributions to free energies are removed.} $\CF[0]$ in
Fig.~\ref{fig:F0C0b2LR-brl}.  The visible Borel singularities are
located at $\sb^{-1}\CA_{\gamma_{\text{LR}}}$ with
\begin{equation}
  \gamma_{\text{LR}} = \pm (0,1,n),\;\; n = 0,\pm 1,\pm 2,\ldots,
\end{equation}
near the large radius point and at
$\sb^{-1}\CA_{\pm(-2,0,0)_\text{LR}}$ near the conifold point.  The
same as local $\IP^2$, the Borel singularities of Wilson loop BPS
sectors never coincide with the flat coordinate up to a constant,
i.e.~the first coefficient in the charge vector $\gamma_{\text{LR}}$
does not vanish.

\begin{figure}
  \centering
  \subfloat[$z=10^{-6}$]{\includegraphics[height=5cm]{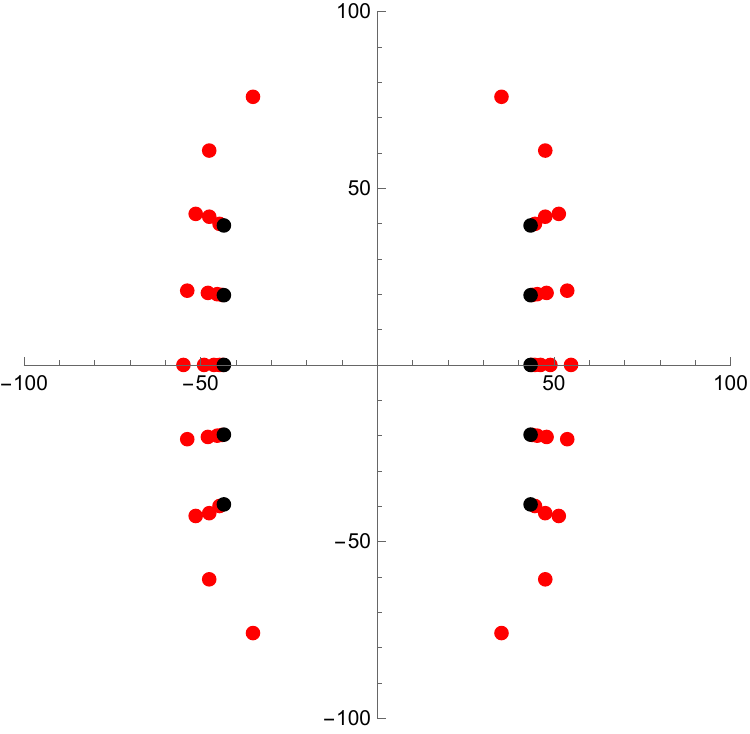}}\hspace{5ex}
  \subfloat[$z=10^{-2}$]{\includegraphics[height=5cm]{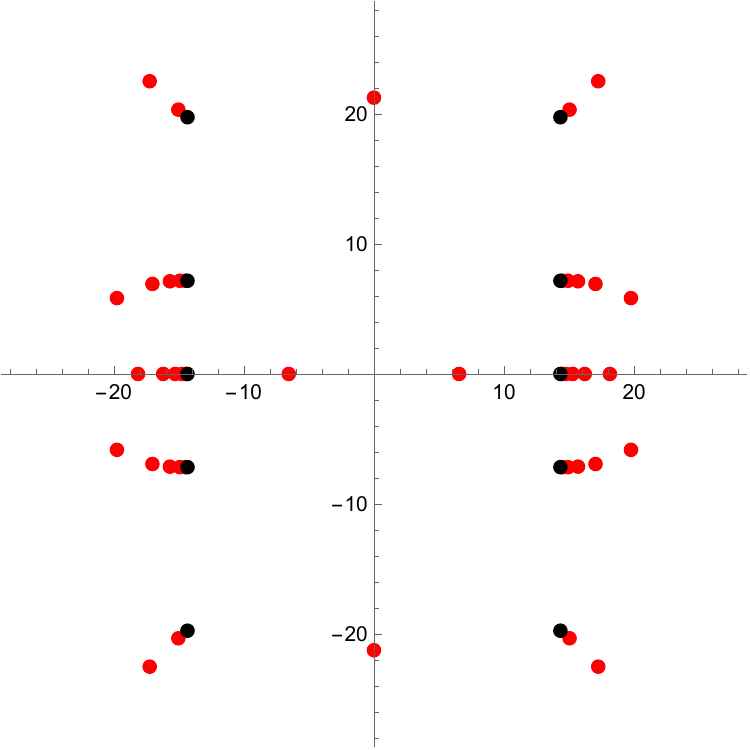}}
  \caption{Borel singularities of refined Wilson loop BPS sectors
    $\CF[1](\sb;g_s)$ for local $\IP^1\times\IP^1$ with $\sb=2$ up to
    $g=50$ in the conifold frame, respectively (a) very close to the
    large radius point and (b) away from it toward the conifold point.
    The red dots are approximate singularities from numerical
    calculations.  The branch points (black dots) are (a)
    $\sb^{-1}\CA_{\pm(0,1,n)_{\text{LR}}}$ ($n=-2,-1,0,1,2$) and (b)
    $\sb^{-1}\CA_{\pm (-2,0,0)_{\text{c}}}$ (horizontal),
    $\sb^{-1}\CA_{\pm(-2,\pm 1,0)_{\text{c}}}$ (slight away),
    $\sb^{-1}\CA_{\pm(-2,0,\pm 1)_{\text{c}}}$ (further away)
    respectively.}
  \label{fig:F0C1b2Con-brl}
\end{figure} 

\begin{figure}
  \centering
  \subfloat[$z=10^{-6}$]{\includegraphics[height=5cm]{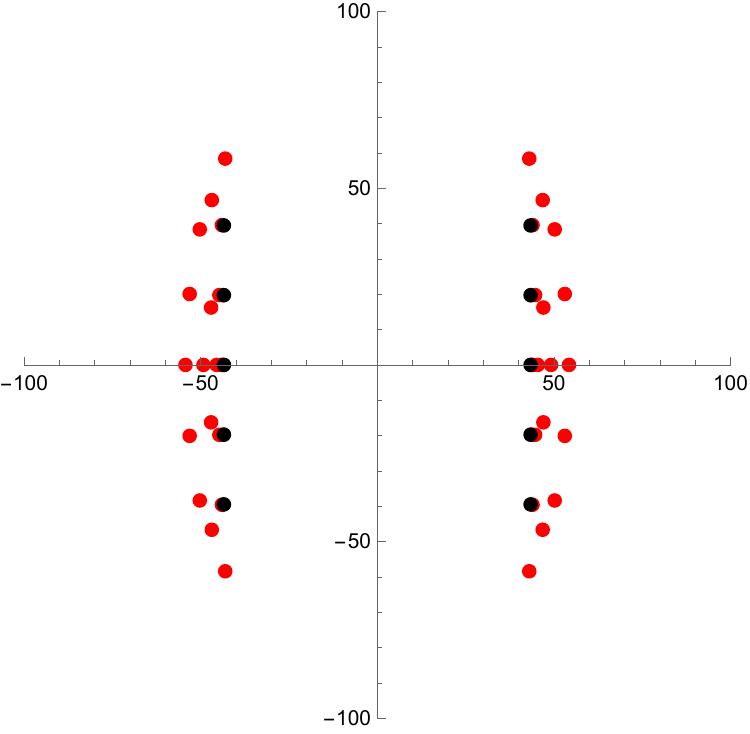}}\hspace{5ex}
  \subfloat[$z=10^{-2}$]{\includegraphics[height=5cm]{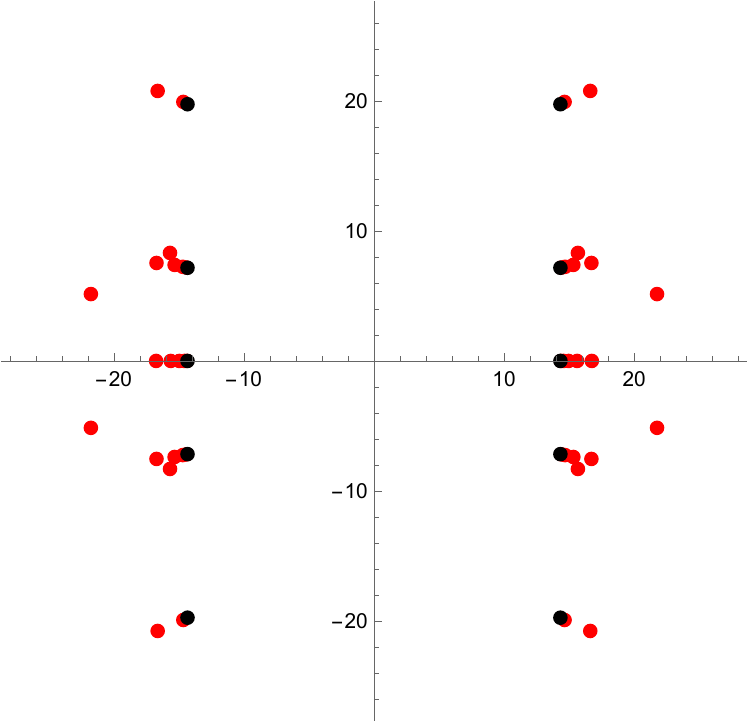}}
  \caption{Borel singularities of refined Wilson loop BPS sectors
    $\CF[2](\sb;g_s)$ for local $\IP^1\times\IP^1$ with $\sb=2$ up to
    $g=50$ in the conifold frame, respectively (a) very close to the
    large radius point and (b) away from it toward the conifold point.
    The red dots are approximate singularities from numerical
    calculations.  The branch points (black dots) are the same as in
    Fig.~\ref{fig:F0C1b2Con-brl}.}
  \label{fig:F0C2b2Con-brl}
\end{figure}

We also evaluate the perturbative BPS sectors $\CF[1]$ and $\CF[2]$ in
the holomorphic limit of the conifold frame, where $t_{\text{c}}$ is
the flat coordinate, respectively very close to the large radius
point, and away from it toward the conifold point.  The Borel
singularities are plotted respectively in
Figs.~\ref{fig:F0C1b2Con-brl} and Figs.~\ref{fig:F0C2b2Con-brl}.  In
both examples, the visible Borel singularities are located at
$\sb^{-1}\CA_{\gamma_{\text{c}}}$ with
\begin{equation}
  \gamma_{\text{c}} = \pm (-2,0,n),\;\; n = 0,\pm 1, \pm 2,\ldots,
\end{equation}
near the large radius point, and at $\sb^{-1}\CA_{\text{c}}$ with
\begin{equation}
  \gamma_{\text{c}} = \pm (-2,0,0),\;\pm (-2,\mp 1,0),\pm (-2,0,\pm 1),
\end{equation}
near the conifold point (we take $\sb>1$ so that $\sb^{-1}\CA$ is
smaller than $\sb\CA$).  Similarly, none of the Borel singularities
coincide with the flat coordinate up to a constant, i.e.~the first
coefficient in the charge vector $\gamma_{\text{c}}$ does not vanish.
Note that we have used two types of charge vectors
$\gamma_{\text{LR}}$ and $\gamma_{\text{c}}$ defined respectively in
\eqref{eq:LRsigularity} and \eqref{eq:Consingualrity} , which are
related to each other in the case of local $\IP^1\times\IP^1$ by
\begin{equation}
  \gamma_{\text{c}} =
  \begin{pmatrix}
    0 & -2 & 0\\\frac{1}{2} & 0 & 0\\0&0&1
  \end{pmatrix}\gamma_{\text{LR}},\quad
  \gamma_{\text{LR}} =
  \begin{pmatrix}
    0 & 2 & 0\\-\frac{1}{2} & 0 & 0\\0&0&1
  \end{pmatrix}\gamma_{\text{c}}.
\end{equation}

\begin{figure}
  \centering
  \subfloat[$\mu_0$]{\includegraphics[width=0.3\linewidth]{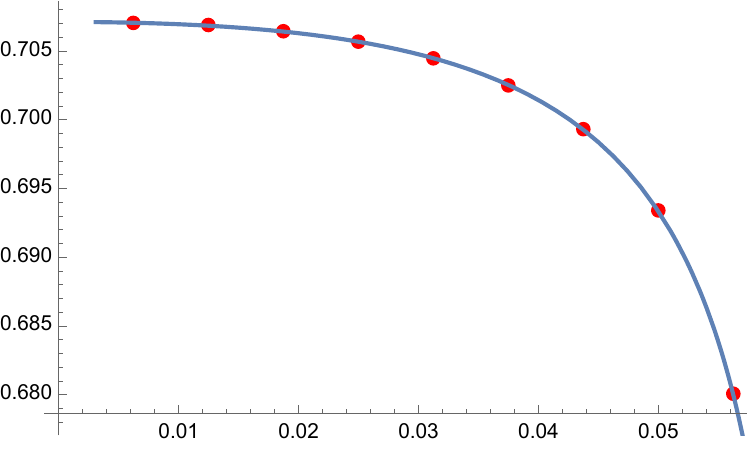}}\hspace{1ex}
  \subfloat[$\mu_1$]{\includegraphics[width=0.3\linewidth]{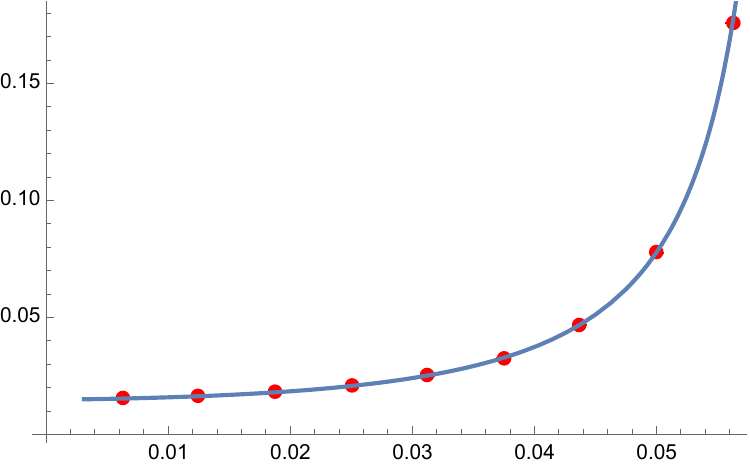}}\hspace{1ex}
  \subfloat[$\mu_2$]{\includegraphics[width=0.3\linewidth]{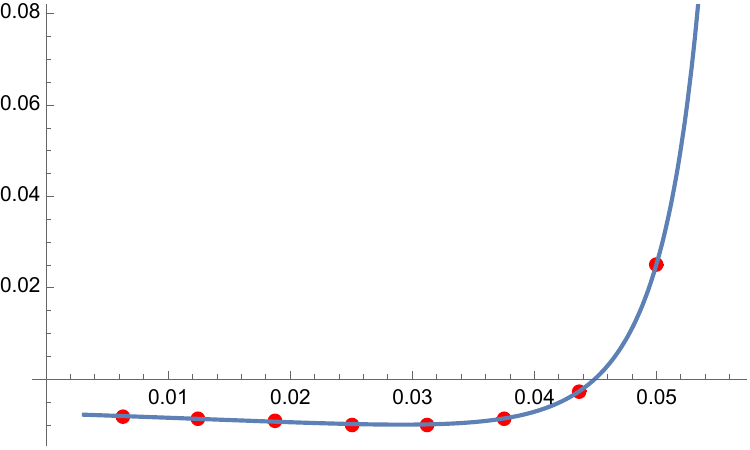}}
  \caption{Comparison for local $\IP^1\times\IP^1$ between numerical
    results (red dots) of $\frac{1}{\pi\ri}\ms{S}\cdot\mu_{0,1,2}$
    extracted from the large order asymptotics of $\CF_g[1]$ up to
    $g=50$ in the large radius frame at $\sb=2$ with error bars
    (vertical bars, virtually invisible) and trans-series solutions
    from HAE (solid line).  Richardson transformation of degree 10 is
    used to improve the numerics.  The horizontal axis is modulus
    $z$.}
  \label{fig:F0C1LRb2}
\end{figure}

\begin{figure}
  \centering
  \subfloat[$\mu_0$]{\includegraphics[width=0.3\linewidth]{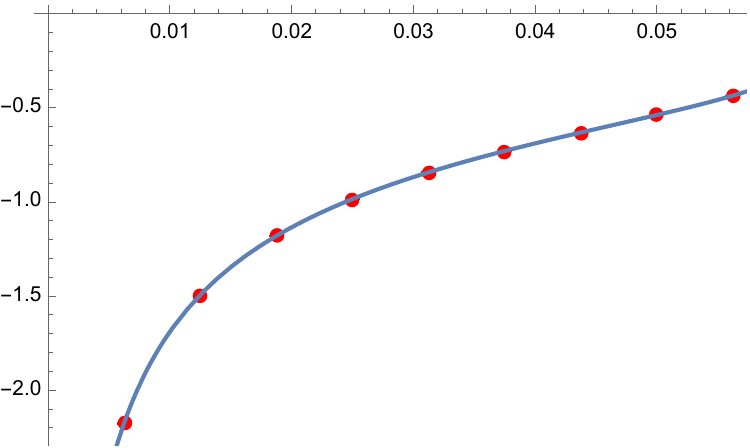}}\hspace{1ex}
  \subfloat[$\mu_1$]{\includegraphics[width=0.3\linewidth]{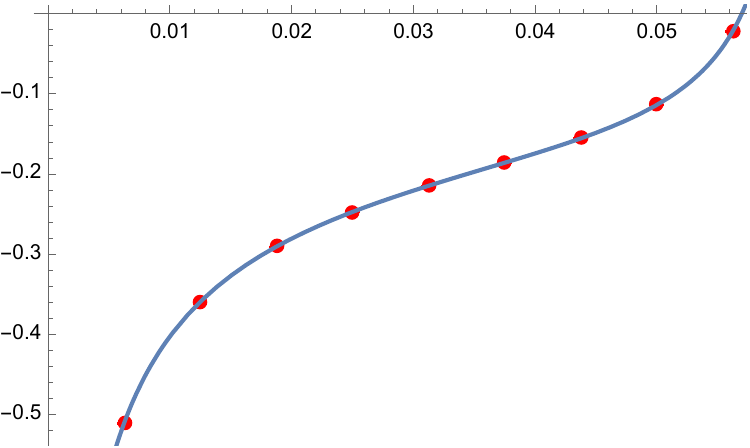}}\hspace{1ex}
  \subfloat[$\mu_2$]{\includegraphics[width=0.3\linewidth]{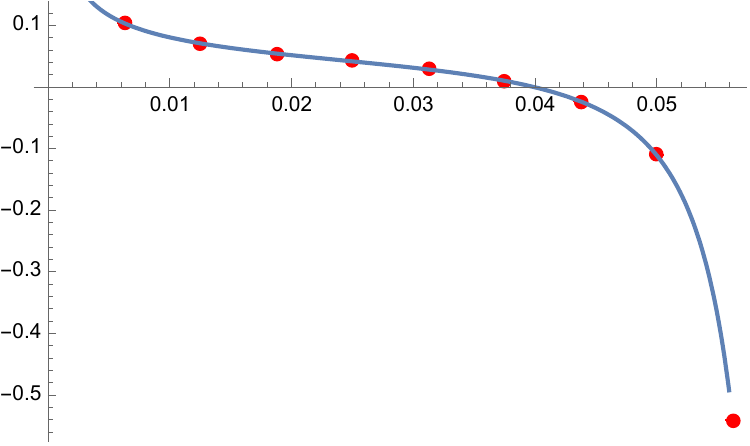}}
  \caption{Comparison for local $\IP^1\times\IP^1$ between numerical
    results (red dots) of $\frac{1}{\pi\ri}\ms{S}\cdot\mu_{0,1,2}$
    extracted from the large order asymptotics of $\CF_g[2]$ up to
    $g=50$ in the large radius frame at $\sb=2$ with error bars
    (vertical bars, virtually invisible) and trans-series solutions
    from HAE (solid line).  Richardson transformation of degree 10 is
    used to improve the numerics.  The horizontal axis is modulus
    $z$.}
  \label{fig:F0C2LRb2}
\end{figure}

\begin{figure}
  \centering
  \subfloat[$\mu_0$]{\includegraphics[width=0.3\linewidth]{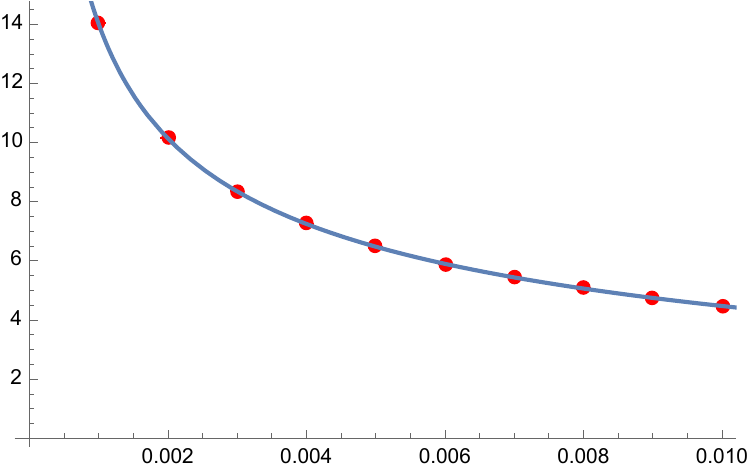}}\hspace{1ex}
  \subfloat[$\mu_1$]{\includegraphics[width=0.3\linewidth]{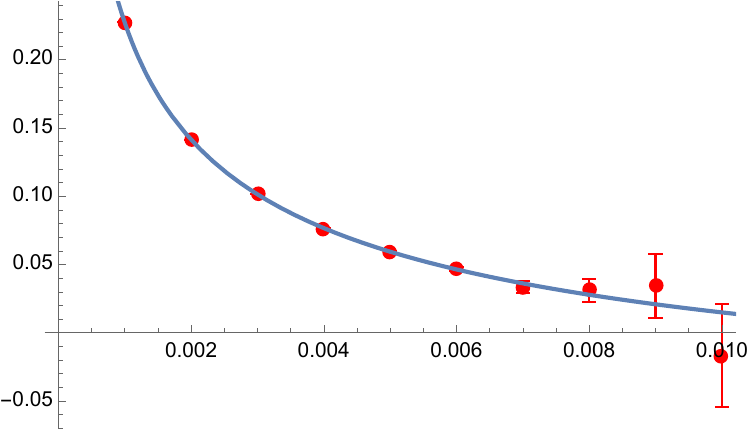}}\hspace{1ex}
  \subfloat[$\mu_2$]{\includegraphics[width=0.3\linewidth]{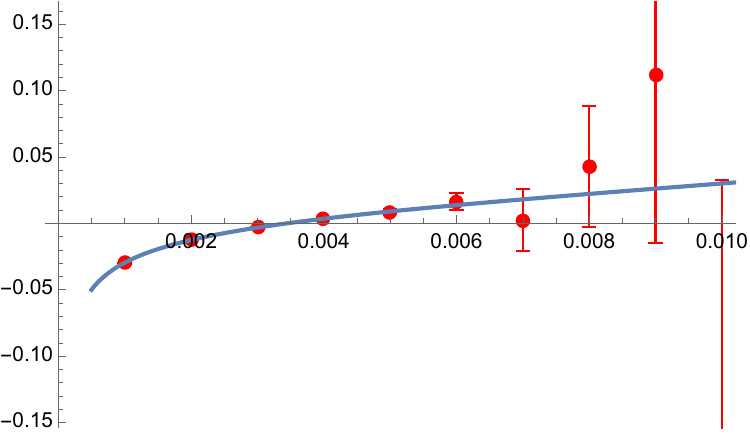}}
  \caption{Comparison for local $\IP^1\times\IP^1$ between numerical
    results (red dots) of $\frac{1}{\pi\ri}\ms{S}\cdot\mu_{0,1,2}$
    extracted from the large order asymptotics of $\CF_g[1]$ up to
    $g=50$ in conifold frame at $\sb=2$ with error bars (vertical
    bars) and trans-series solutions from HAE (solid line).
    Richardson transformation of degree 10 is used to improve the
    numerics.  The horizontal axis is modulus $z$.}
  \label{fig:F0C1Conb2}
\end{figure}

\begin{figure}
  \centering
  \subfloat[$\mu_0$]{\includegraphics[width=0.3\linewidth]{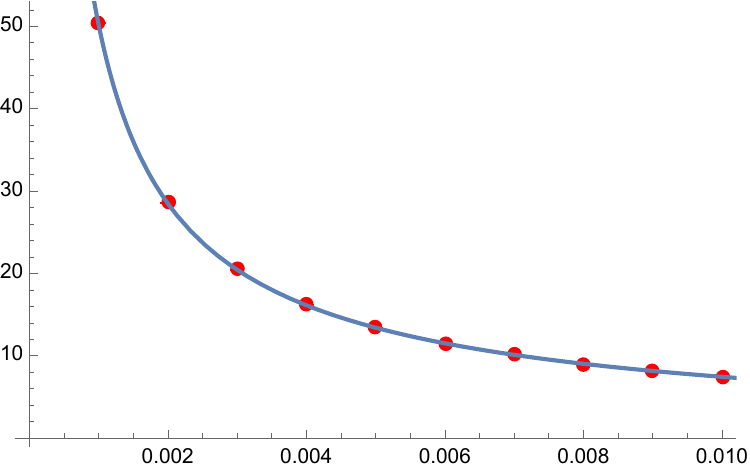}}\hspace{1ex}
  \subfloat[$\mu_1$]{\includegraphics[width=0.3\linewidth]{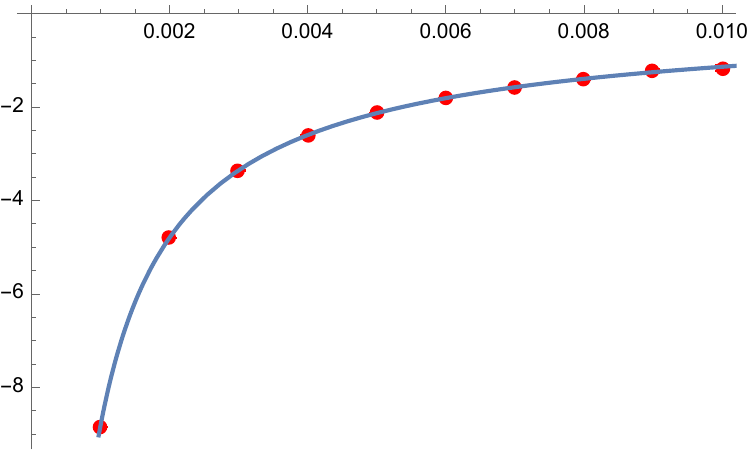}}\hspace{1ex}
  \subfloat[$\mu_2$]{\includegraphics[width=0.3\linewidth]{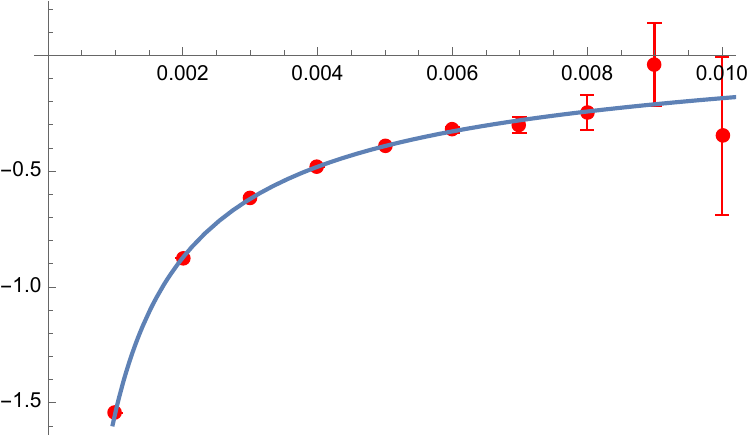}}
  \caption{Comparison for local $\IP^1\times\IP^1$ between numerical
    results (red dots) of $\frac{1}{\pi\ri}\ms{S}\cdot\mu_{0,1,2}$
    extracted from the large order asymptotics of $\CF_g[2]$ up to
    $g=50$ in conifold frame at $\sb=2$ with error bars (vertical
    bars) and trans-series solutions from HAE (solid line).
    Richardson transformation of degree 10 is used to improve the
    numerics.  The horizontal axis is modulus $z$.}
  \label{fig:F0C2Conb2}
\end{figure}

\begin{figure}
  \centering
  \subfloat[$\mu_0$]{\includegraphics[width=0.3\linewidth]{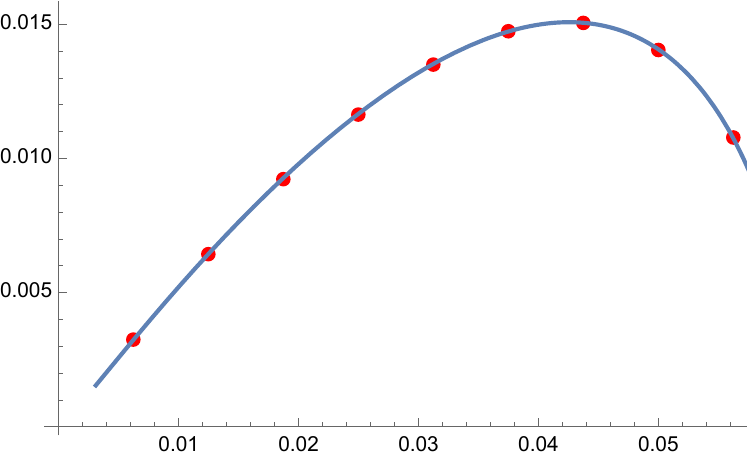}}\hspace{1ex}
  \subfloat[$\mu_1$]{\includegraphics[width=0.3\linewidth]{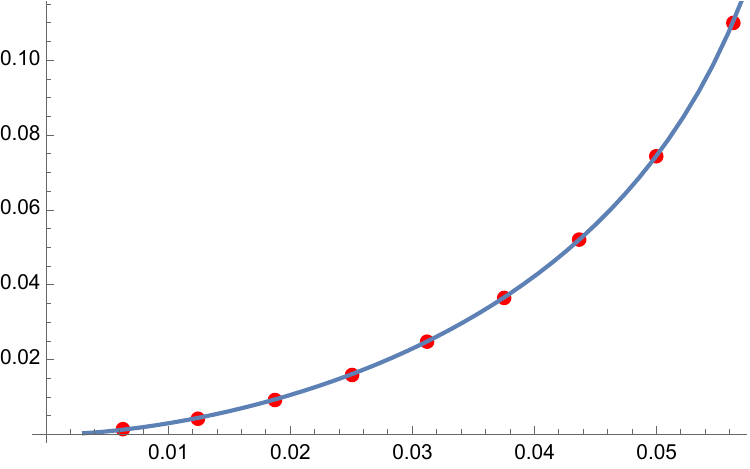}}\hspace{1ex}
  \subfloat[$\mu_2$]{\includegraphics[width=0.3\linewidth]{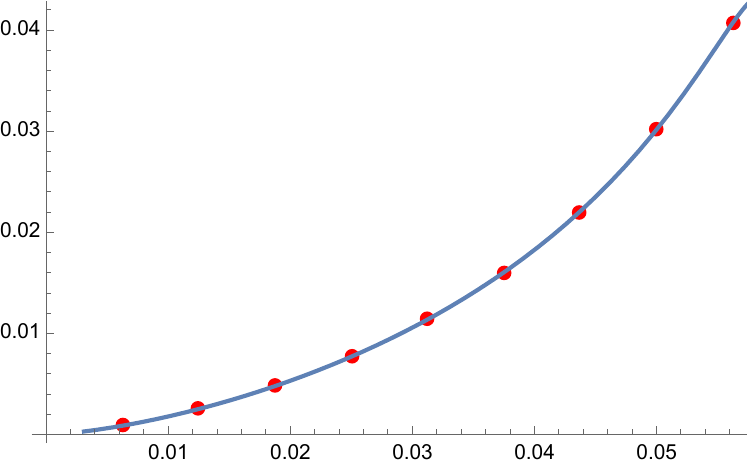}}
  \caption{Comparison for local $\IP^1\times\IP^1$ between numerical
    results (red dots) of $\frac{1}{\pi\ri}\ms{S}\cdot\mu_{0,1,2}$
    extracted from the large order asymptotics of $\CF_g[1]$ up to
    $g=100$ in the large radius frame at $\sb=1$ with error bars
    (vertical bars, virtually invisible) and trans-series solutions
    from HAE (solid line).  Richardson transformation of degree 10 is
    used to improve the numerics.  The horizontal axis is modulus
    $z$.}
  \label{fig:F0C1LRb1}
\end{figure}

\begin{figure}
  \centering
  \subfloat[$\mu_0$]{\includegraphics[width=0.3\linewidth]{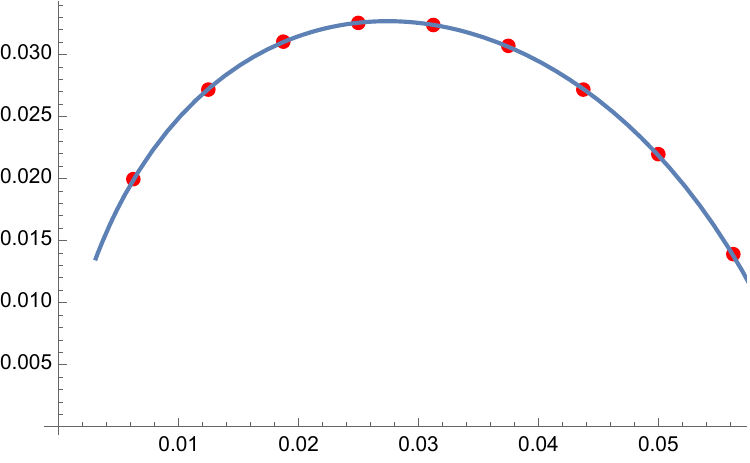}}\hspace{1ex}
  \subfloat[$\mu_1$]{\includegraphics[width=0.3\linewidth]{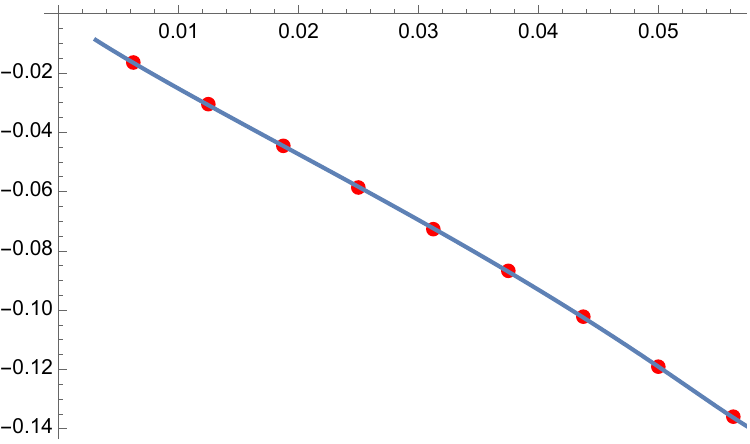}}\hspace{1ex}
  \subfloat[$\mu_2$]{\includegraphics[width=0.3\linewidth]{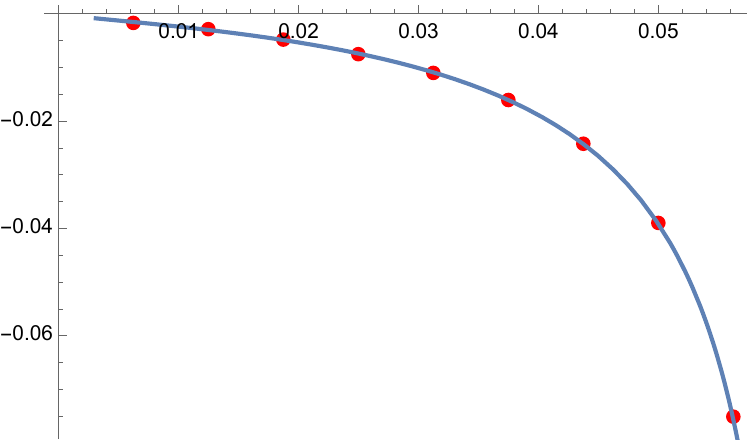}}
  \caption{Comparison for local $\IP^1\times\IP^1$ between numerical
    results (red dots) of $\frac{1}{\pi\ri}\ms{S}\cdot\mu_{0,1,2}$
    extracted from the large order asymptotics of $\CF_g[2]$ up to
    $g=100$ in the large radius frame at $\sb=1$ with error bars
    (vertical bars, virtually invisible) and trans-series solutions
    from HAE (solid line).  Richardson transformation of degree 10 is
    used to improve the numerics.  The horizontal axis is modulus
    $z$.}
  \label{fig:F0C2LRb1}
\end{figure}

\begin{figure}
  \centering
  \subfloat[$\mu_0$]{\includegraphics[width=0.3\linewidth]{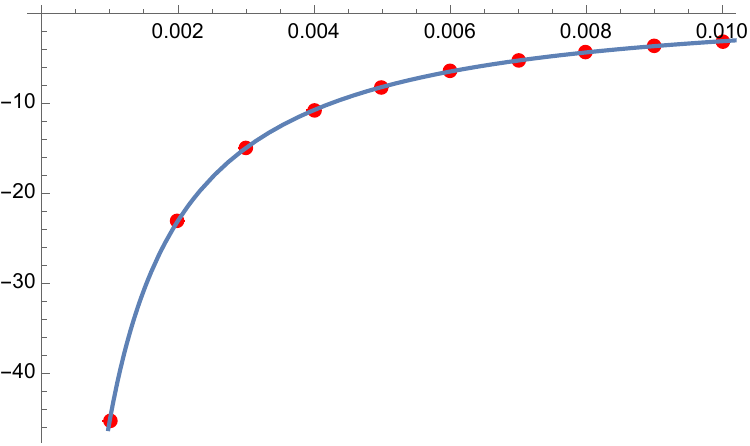}}\hspace{1ex}
  \subfloat[$\mu_1$]{\includegraphics[width=0.3\linewidth]{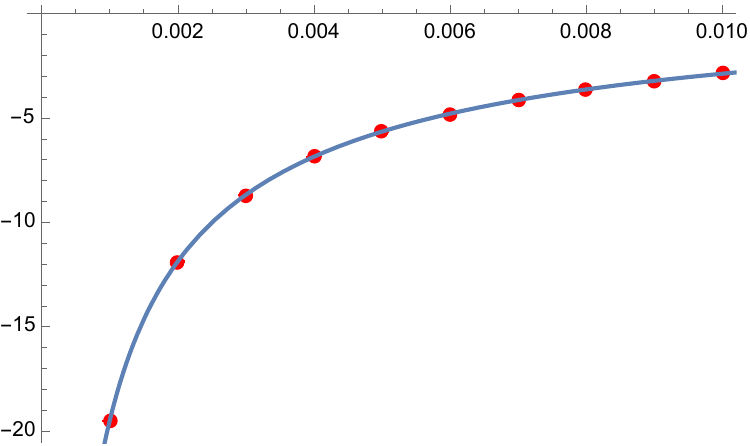}}\hspace{1ex}
  \subfloat[$\mu_2$]{\includegraphics[width=0.3\linewidth]{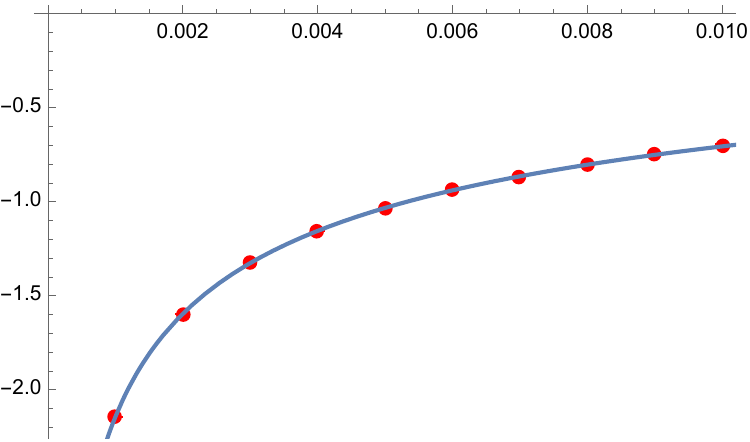}}
  \caption{Comparison for local $\IP^1\times\IP^1$ between numerical
    results (red dots) of $\frac{1}{\pi\ri}\ms{S}\cdot\mu_{0,1,2}$
    extracted from the large order asymptotics of $\CF_g[1]$ up to
    $g=100$ in the conifold frame at $\sb=1$ with error bars (vertical
    bars, virtually invisible) and trans-series solutions from HAE
    (solid line).  Richardson transformation of degree 10 is used to
    improve the numerics.  The horizontal axis is modulus $z$.}
  \label{fig:F0C1Conb1}
\end{figure}

\begin{figure}
  \centering
  \subfloat[$\mu_0$]{\includegraphics[width=0.3\linewidth]{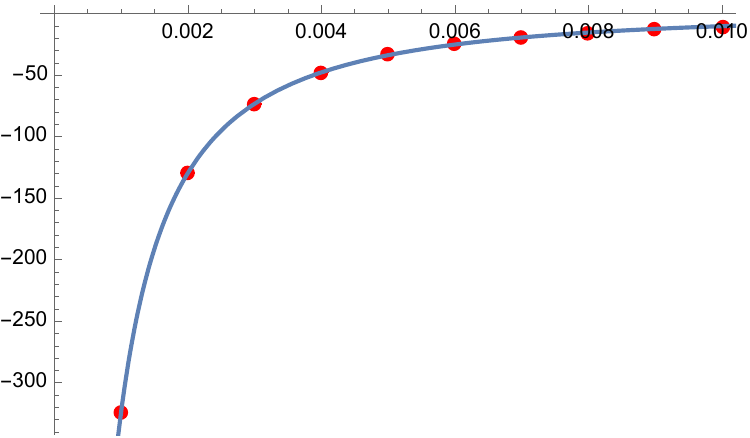}}\hspace{1ex}
  \subfloat[$\mu_1$]{\includegraphics[width=0.3\linewidth]{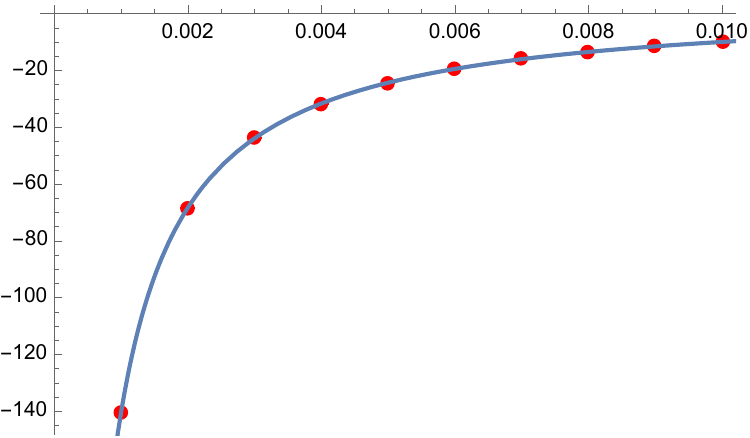}}\hspace{1ex}
  \subfloat[$\mu_2$]{\includegraphics[width=0.3\linewidth]{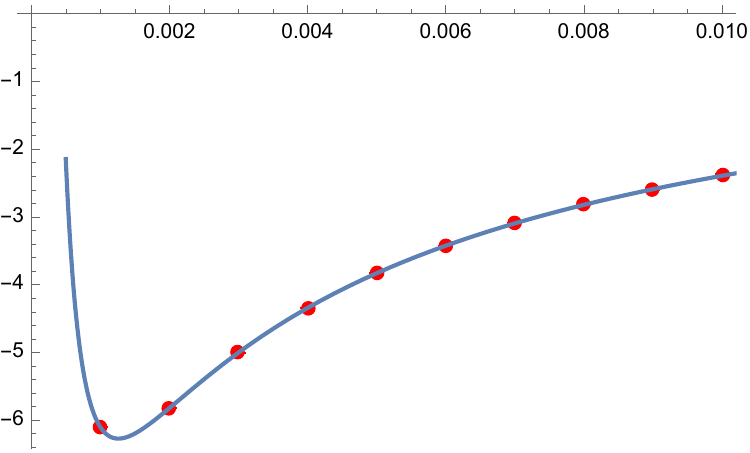}}
  \caption{Comparison for local $\IP^1\times\IP^1$ between numerical
    results (red dots) of $\frac{1}{\pi\ri}\ms{S}\cdot\mu_{0,1,2}$
    extracted from the large order asymptotics of $\CF_g[2]$ up to
    $g=100$ in the conifold frame at $\sb=1$ with error bars (vertical
    bars, virtually invisible) and trans-series solutions from HAE
    (solid line).  Richardson transformation of degree 10 is used to
    improve the numerics.  The horizontal axis is modulus $z$.}
  \label{fig:F0C2Conb1}
\end{figure}

Next, we study the non-perturbative series. We focus on the
1-instanton sector as Section in \ref{sc:F0}. The large order
asymptotics of the perturbative coefficients are the same as
(\ref{eq:FgAsymp-ref}) and (\ref{eq:FgAsymp-unref}).
We consider two cases, the BPS sectors near the conifold point in the
large radius frame, as well as in the conifold frame.  The dominant
Borel singularities are respectively $\gamma_{\text{LR}} =\pm(-2,0,0)$
and $\gamma_{\text{c}} =\pm(-2,0,0)$, as shown in the plots of
Figs.~\ref{fig:F0C1b2LR-brl} (b), \ref{fig:F0C2b2LR-brl} (b) and
Figs.~\ref{fig:F0C1b2Con-brl} (b), \ref{fig:F0C2b2Con-brl} (b).
we compare these numerical results of $\mu_0,\mu_1,\mu_2,\ldots$
extracted from perturbative data using the large order formulas with
the theoretical prediction from Sections~\ref{sc:Fm-nonp} and
\ref{sc:Fm-nonp-unref} in Figs.~\ref{fig:F0C1LRb2},
\ref{fig:F0C2LRb2}, \ref{fig:F0C1Conb2}, \ref{fig:F0C2Conb2} for
generic $\sb$, and in Figs.~\ref{fig:F0C1LRb1}, \ref{fig:F0C2LRb1},
\ref{fig:F0C1Conb1}, \ref{fig:F0C2Conb1} for the unrefined limit
$\sb=1$.

Finally, we note that the numerical results and the theoertical
prediction can match very well, as shown in the above Figures, only if
we have taken the Stokes constant associated to $\CA_{\pm(-2,0,0)_{\text{LR}}}$ to be
\begin{equation}
  \sS_{\pm(-2,0,0)_{\text{LR}}}(\sb) = 1,
\end{equation}
and the Stokes constant associated to
$\CA_{\pm(-2,0,0)_{\text{c}}}$ to be
\begin{equation}
  \sS_{\pm(-2,0,0)_{\text{c}}}(\sb) = 2\chi_{1/2}(-\re^{-\pi\ri/\sb^2}).
\end{equation}
The Borel singularities $\CA_{\pm\gamma_1}$ with
$\gamma_1 = (2,0,0)_{\text{LR}}$ are associated to the spin 0 BPS
state of D4 brane wrapping $\IP^1\times\IP^1$ in type II superstring,
and the refined DT-invariant is
\begin{equation}
  \Omega(\gamma_1,y) = 1.
\end{equation}
The Borel singularities $\CA_{\pm\gamma_2}$ with
$\gamma_2 = (-2,0,0)_{\text{c}} = (0,1,0)_{\text{LR}}$ are associated
to the spin $1/2$ BPS state of D2 brane wrapping either $\IP^1$ in
type IIA superstring, and the refined DT-invarint is
\begin{equation}
  \Omega(\gamma_2,y) = 2\chi_{1/2}(y).
\end{equation}
Therefore the Stokes constants agree with the refined DT-invariants,
in accord with our prediction from Section~\ref{sc:Fm-nonp}.

\section{Conclusion}


In this paper, we study the resurgent structures of refined Wilson
loops in topological string theory on a local Calabi-Yau threefold.
The refined Wilson loops are treated as asymptotic series in $g_s$
with deformation parameter $\sb$, using the parametrisation
\eqref{eq:ep12gs}.  We find that they are very similar to those of
refined free energies.  The non-perturbative actions are integral
periods, but they cannot be local flat coordinates or equivalent
A-periods in the B-model.  The non-perturbative trans-series can be
solved in closed form from the holomorphic anomaly equations for
Wilson loops, and finally, the Stokes constants are identified with
refined DT invariants.

There are many interesting open problems related to this work.  First
of all, Wilson loop is a concept borrowed from 5d $\cN=1$ gauge
theory, related to topological string via geometric engineering
\cite{Katz:1996fh}.  Here we consider Wilson loops in 5d gauge
theories, which are codimension four defects.  Defects of other
codimensions and of other natures exist.  One other important type of
defects in 5d gauge theories is codimension two defects, and their
partition functions play the role of wave-functions in quantum mirror
curve.  It is argued in \cite{Grassi:2022zuk} with the simple example
of topological string on $\IC^3$ or the resolved conifold that the
Borel singularities of these wave-functions should correspond to BPS
states of 3d/5d coupled systems.  Similarly it was found that the
Borel singularities of wave-functions of quantum Seiberg-Witten curves
of 4d $\cN=2$ gauge theories correspond to BPS states of 2d/4d coupled
systems.  It would be interesting to generalise these results to the
generic setup in topological string, and to also find out the
non-perturbative series associated to these BPS states of the coupled
systems.

Second, there has been now convincing evidence that the Stokes
constants of both the refined free energies and the refined Wilson
loops are the refined DT invariants.  It would be certainly nice to
work out a rigorous proof.  Another interesting direction is to use
Stokes constants to help with the calculation of DT invariants, or to
study the stability walls.  In this regard, Wilson loops can sometimes
yield more information than free energies, as shown by the Borel
singularities of charge vectors \eqref{eq:gammaF0}, which should
correspond to non-trivial BPS states\footnote{One should be able to
  compare with the BPS spectrum produced in \cite{Longhi:2021qvz},
  albeit in different stability chambers.} for local
$\IP^1\times \IP^1$.  The explicit calculation of Stokes constants
associated to these singularities, and beyond, is numerically 
challenging, but the results in \cite{Gu:2021ize} in the special
conifold limit could be a promising start.

Third, the evaluation of either the refined free energy or the refined
Wilson loops depends on a choice of frame.  It has been observed in
\cite{Gu:2022sqc,Gu:2023mgf} and also in this paper that the
calculation of Stokes constants is independent of this choice.  This
is natural as the DT invariants, which are conjectured to identify
with the Stokes constants, know nothing of the frame.  It would
nevertheless be reassuring if one can find a proof of this
observation.

Finally, the DT invariants which are conjectured to coincide with
Stokes constants are counting of stable bound states of D-branes,
either D6-D4-D2-D0 branes in type IIA superstring, or D5-D3-D1-D(-1)
banes in type IIB superstring\footnote{In the case of local CY3, D6
  and D5 are respectively missing in type IIA and type IIB.}.  It was
suggested in \cite{Couso-Santamaria2017a} that NS5 brane effects may
be found after the resummation of D-brane effects.  It might be
interesting to verify this idea, given that we now have a good
understanding of the non-perturbative series for the
D-branes. 

\printindex

\bibliographystyle{amsmod}
\bibliography{wilson-res} 

\ifx\undefined\bysame
\newcommand{\bysame}{\leavevmode\hbox to3em{\hrulefill}\,}
\fi
\begin{thebibliography}{10}

\bibitem{Ecalle}
J.~{\'E}calle, {\em {Les fonctions r{\'e}surgentes. Vols. I-III}},
  Universit{\'e} de Paris-Sud, D{\'e}partement de Math{\'e}matiques, Bât. 425,
  1981.

\bibitem{Marino:2006hs}
M.~Marino, {\em {Open string amplitudes and large order behavior in topological
  string theory}}, JHEP {\bf 03} (2008) 060, {\tt arXiv:hep-th/0612127} {\tt
  [hep-th]}.

\bibitem{Marino:2007te}
M.~Marino, R.~Schiappa, and M.~Weiss, {\em {Nonperturbative effects and the
  large-order behavior of matrix models and topological strings}}, Commun. Num.
  Theor. Phys. {\bf 2} (2008) 349--419, {\tt arXiv:0711.1954} {\tt [hep-th]}.

\bibitem{Marino:2008ya}
M.~Marino, {\em {Nonperturbative effects and nonperturbative definitions in
  matrix models and topological strings}}, JHEP {\bf 12} (2008) 114, {\tt
  arXiv:0805.3033} {\tt [hep-th]}.

\bibitem{Marino2009}
M.~Marino, R.~Schiappa, and M.~Weiss, {\em {Multi-instantons and multi-cuts}},
  J. Math. Phys. {\bf 50} (2009) 052301, {\tt arXiv:0809.2619} {\tt [hep-th]}.

\bibitem{Pasquetti2010}
S.~Pasquetti and R.~Schiappa, {\em {Borel and Stokes nonperturbative phenomena
  in topological string theory and c=1 matrix models}}, Annales Henri Poincare
  {\bf 11} (2010) 351--431, {\tt arXiv:0907.4082} {\tt [hep-th]}.

\bibitem{Drukker:2011zy}
N.~Drukker, M.~Marino, and P.~Putrov, {\em {Nonperturbative aspects of ABJM
  theory}}, JHEP {\bf 11} (2011) 141, {\tt arXiv:1103.4844} {\tt [hep-th]}.

\bibitem{Aniceto2012}
I.~Aniceto, R.~Schiappa, and M.~Vonk, {\em {The resurgence of instantons in
  string theory}}, Commun. Num. Theor. Phys. {\bf 6} (2012) 339--496, {\tt
  arXiv:1106.5922} {\tt [hep-th]}.

\bibitem{Couso-Santamaria2016}
R.~Couso-Santamar\'{i}a, J.~D. Edelstein, R.~Schiappa, and M.~Vonk, {\em
  {Resurgent transseries and the holomorphic anomaly}}, Annales Henri Poincare
  {\bf 17} (2016) 331--399, {\tt arXiv:1308.1695} {\tt [hep-th]}.

\bibitem{Couso-Santamaria2015}
R.~Couso-Santamar\'{i}a, J.~D. Edelstein, R.~Schiappa, and M.~Vonk, {\em
  {Resurgent transseries and the holomorphic anomaly: Nonperturbative closed
  strings in local ${\mathbb{C}\mathbb{P}^2}$}}, Commun. Math. Phys. {\bf 338}
  (2015) 285--346, {\tt arXiv:1407.4821} {\tt [hep-th]}.

\bibitem{Gu:2023mgf}
J.~Gu, A.-K. Kashani-Poor, A.~Klemm, and M.~Marino, {\em {Non-perturbative
  topological string theory on compact Calabi-Yau 3-folds}}, {\tt
  arXiv:2305.19916} {\tt [hep-th]}.

\bibitem{Kazakov:2004du}
V.~A. Kazakov and I.~K. Kostov, {\em {Instantons in noncritical strings from
  the two matrix model}}, {From Fields to Strings: Circumnavigating Theoretical
  Physics: A Conference in Tribute to Ian Kogan}, 3 2004, pp.~1864--1894, {\tt
  arXiv:hep-th/0403152}.

\bibitem{Couso-Santamaria2017a}
R.~Couso-Santamar\'{i}a, {\em {Universality of the topological string at large
  radius and NS-brane resurgence}}, Lett. Math. Phys. {\bf 107} (2017)
  343--366, {\tt arXiv:1507.04013} {\tt [hep-th]}.

\bibitem{Couso-Santamaria:2016vwq}
R.~Couso-Santamar\'\i{}a, M.~Marino, and R.~Schiappa, {\em {Resurgence Matches
  Quantization}}, J. Phys. A {\bf 50} (2017) 145402, {\tt arXiv:1610.06782}
  {\tt [hep-th]}.

\bibitem{Grassi:2014zfa}
A.~Grassi, Y.~Hatsuda, and M.~Marino, {\em {Topological strings from quantum
  mechanics}}, {\tt arXiv:1410.3382} {\tt [hep-th]}.

\bibitem{Bershadsky:1993ta}
M.~Bershadsky, S.~Cecotti, H.~Ooguri, and C.~Vafa, {\em {Holomorphic anomalies
  in topological field theories}}, Nucl. Phys. {\bf B405} (1993) 279--304, {\tt
  arXiv:hep-th/9302103} {\tt [hep-th]}.

\bibitem{Bershadsky:1993cx}
M.~Bershadsky, S.~Cecotti, H.~Ooguri, and C.~Vafa, {\em {Kodaira-Spencer theory
  of gravity and exact results for quantum string amplitudes}}, Commun. Math.
  Phys. {\bf 165} (1994) 311--428, {\tt arXiv:hep-th/9309140} {\tt [hep-th]}.

\bibitem{Gu:2022sqc}
J.~Gu and M.~Marino, {\em {Exact multi-instantons in topological string
  theory}}, {\tt arXiv:2211.01403} {\tt [hep-th]}.

\bibitem{Gu:2021ize}
J.~Gu and M.~Marino, {\em {Peacock patterns and new integer invariants in
  topological string theory}}, SciPost Phys. {\bf 12} (2022) 058, {\tt
  arXiv:2104.07437} {\tt [hep-th]}.

\bibitem{Bridgeland:2016nqw}
T.~Bridgeland, {\em {Riemann-Hilbert problems from Donaldson-Thomas theory}},
  Invent. Math. {\bf 216} (2019) 69--124, {\tt arXiv:1611.03697} {\tt
  [math.AG]}.

\bibitem{Bridgeland:2017vbr}
T.~Bridgeland, {\em {Riemann-Hilbert problems for the resolved conifold}}, {\tt
  arXiv:1703.02776} {\tt [math.AG]}.

\bibitem{Alim:2021mhp}
M.~Alim, A.~Saha, J.~Teschner, and I.~Tulli, {\em {Mathematical structures of
  non-perturbative topological string theory: from GW to DT invariants}}, {\tt
  arXiv:2109.06878} {\tt [hep-th]}.

\bibitem{Gu:2023wum}
J.~Gu, {\em {Relations between Stokes constants of unrefined and
  Nekrasov-Shatashvili topological strings}}, {\tt arXiv:2307.02079} {\tt
  [hep-th]}.

\bibitem{Iwaki:2023rst}
K.~Iwaki and M.~Marino, {\em {Resurgent Structure of the Topological String and
  the First Painlev\'e Equation}}, {\tt arXiv:2307.02080} {\tt [hep-th]}.

\bibitem{Alexandrov:2023wdj}
S.~Alexandrov, M.~Marino, and B.~Pioline, {\em {Resurgence of refined
  topological strings and dual partition functions}}, {\tt arXiv:2311.17638}
  {\tt [hep-th]}.

\bibitem{Alim:2022oll}
M.~Alim, L.~Hollands, and I.~Tulli, {\em {Quantum Curves, Resurgence and Exact
  WKB}}, SIGMA {\bf 19} (2023) 009, {\tt arXiv:2203.08249} {\tt [hep-th]}.

\bibitem{Grassi:2022zuk}
A.~Grassi, Q.~Hao, and A.~Neitzke, {\em {Exponential Networks, WKB and the
  Topological String}}, {\tt arXiv:2201.11594} {\tt [hep-th]}.

\bibitem{Codesido:2017jwp}
S.~Codesido, M.~Marino, and R.~Schiappa, {\em {Non-Perturbative Quantum
  Mechanics from Non-Perturbative Strings}}, Annales Henri Poincare {\bf 20}
  (2019) 543--603, {\tt arXiv:1712.02603} {\tt [hep-th]}.

\bibitem{CodesidoSanchez:2018vor}
S.~Codesido~Sanchez, {\em {A geometric approach to non-perturbative quantum
  mechanics}}, Ph.D. thesis, Geneva U., 2018.

\bibitem{Gu:2022fss}
J.~Gu and M.~Marino, {\em {On the resurgent structure of quantum periods}},
  {\tt arXiv:2211.03871} {\tt [hep-th]}.

\bibitem{Huang:2010kf}
M.-x. Huang and A.~Klemm, {\em {Direct integration for general $\Omega$
  backgrounds}}, Adv. Theor. Math. Phys. {\bf 16} (2012) 805--849, {\tt
  arXiv:1009.1126} {\tt [hep-th]}.

\bibitem{Aganagic:2011mi}
M.~Aganagic, M.~C.~N. Cheng, R.~Dijkgraaf, D.~Krefl, and C.~Vafa, {\em {Quantum
  geometry of refined topological strings}}, JHEP {\bf 11} (2012) 019, {\tt
  arXiv:1105.0630} {\tt [hep-th]}.

\bibitem{Nekrasov:2015wsu}
N.~Nekrasov, {\em {BPS/CFT correspondence: non-perturbative Dyson-Schwinger
  equations and qq-characters}}, JHEP {\bf 03} (2016) 181, {\tt
  arXiv:1512.05388} {\tt [hep-th]}.

\bibitem{Kim:2021gyj}
H.-C. Kim, M.~Kim, and S.-S. Kim, {\em {5d/6d Wilson loops from blowups}}, {\tt
  arXiv:2106.04731} {\tt [hep-th]}.

\bibitem{Huang:2022hdo}
M.-x. Huang, K.~Lee, and X.~Wang, {\em {Topological strings and Wilson loops}},
  {\tt arXiv:2205.02366} {\tt [hep-th]}.

\bibitem{Wang:2023zcb}
X.~Wang, {\em {Wilson loops, holomorphic anomaly equations and blowup
  equations}}, {\tt arXiv:2305.09171} {\tt [hep-th]}.

\bibitem{Guo:2024wlg}
S.~Guo, X.~Wang, and L.~Wu, {\em In preparation}.

\bibitem{Nekrasov:2002qd}
N.~A. Nekrasov, {\em {Seiberg-Witten prepotential from instanton counting}},
  Adv. Theor. Math. Phys. {\bf 7} (2003) 831--864, {\tt arXiv:hep-th/0206161}
  {\tt [hep-th]}.

\bibitem{Iqbal:2007ii}
A.~Iqbal, C.~Koz\c{c}az, and C.~Vafa, {\em {The refined topological vertex}},
  JHEP {\bf 10} (2009) 069, {\tt arXiv:hep-th/0701156} {\tt [hep-th]}.

\bibitem{Huang:2017mis}
M.-x. Huang, K.~Sun, and X.~Wang, {\em {Blowup equations for refined
  topological strings}}, {\tt arXiv:1711.09884} {\tt [hep-th]}.

\bibitem{Grassi:2016nnt}
A.~Grassi and J.~Gu, {\em {BPS relations from spectral problems and blowup
  equations}}, Lett. Math. Phys. {\bf 109} (2019) 1271--1302, {\tt
  arXiv:1609.05914} {\tt [hep-th]}.

\bibitem{Gu:2017ccq}
J.~Gu, M.-x. Huang, A.-K. Kashani-Poor, and A.~Klemm, {\em {Refined BPS
  invariants of 6d SCFTs from anomalies and modularity}}, JHEP {\bf 05} (2017)
  130, {\tt arXiv:1701.00764} {\tt [hep-th]}.

\bibitem{klemm2018b}
A.~Klemm, {\em The b-model approach to topological string theory on calabi-yau
  n-folds}, B-Model Gromov-Witten Theory, Springer, 2018, pp.~79--397.

\bibitem{Krefl:2010fm}
D.~Krefl and J.~Walcher, {\em {Extended holomorphic anomaly in gauge theory}},
  Lett. Math. Phys. {\bf 95} (2011) 67--88, {\tt arXiv:1007.0263} {\tt
  [hep-th]}.

\bibitem{Hollowood:2003cv}
T.~J. Hollowood, A.~Iqbal, and C.~Vafa, {\em {Matrix models, geometric
  engineering and elliptic genera}}, JHEP {\bf 03} (2008) 069, {\tt
  arXiv:hep-th/0310272} {\tt [hep-th]}.

\bibitem{Gopakumar:1998jq}
R.~Gopakumar and C.~Vafa, {\em {M theory and topological strings. 2.}}, {\tt
  arXiv:hep-th/9812127} {\tt [hep-th]}.

\bibitem{Ghoshal:1995wm}
D.~Ghoshal and C.~Vafa, {\em {C = 1 string as the topological theory of the
  conifold}}, Nucl. Phys. B {\bf 453} (1995) 121--128, {\tt
  arXiv:hep-th/9506122}.

\bibitem{Katz:1996fh}
S.~H. Katz, A.~Klemm, and C.~Vafa, {\em {Geometric engineering of quantum field
  theories}}, Nucl. Phys. {\bf B497} (1997) 173--195, {\tt
  arXiv:hep-th/9609239} {\tt [hep-th]}.

\bibitem{Katz:1997eq}
S.~Katz, P.~Mayr, and C.~Vafa, {\em {Mirror symmetry and exact solution of 4-D
  N=2 gauge theories: 1.}}, Adv. Theor. Math. Phys. {\bf 1} (1998) 53--114,
  {\tt arXiv:hep-th/9706110} {\tt [hep-th]}.

\bibitem{Young:2011aa}
D.~Young, {\em {Wilson Loops in Five-Dimensional Super-Yang-Mills}}, JHEP {\bf
  02} (2012) 052, {\tt arXiv:1112.3309} {\tt [hep-th]}.

\bibitem{Assel:2012nf}
B.~Assel, J.~Estes, and M.~Yamazaki, {\em {Wilson Loops in 5d N=1 SCFTs and
  AdS/CFT}}, Annales Henri Poincare {\bf 15} (2014) 589--632, {\tt
  arXiv:1212.1202} {\tt [hep-th]}.

\bibitem{Haghighat:2008gw}
B.~Haghighat, A.~Klemm, and M.~Rauch, {\em {Integrability of the holomorphic
  anomaly equations}}, JHEP {\bf 0810} (2008) 097, {\tt arXiv:0809.1674} {\tt
  [hep-th]}.

\bibitem{Chiang:1999tz}
T.~M. Chiang, A.~Klemm, S.-T. Yau, and E.~Zaslow, {\em {Local mirror symmetry:
  Calculations and interpretations}}, Adv. Theor. Math. Phys. {\bf 3} (1999)
  495--565, {\tt arXiv:hep-th/9903053} {\tt [hep-th]}.

\bibitem{Marino:2015ixa}
M.~Marino and S.~Zakany, {\em {Matrix models from operators and topological
  strings}}, Annales Henri Poincare {\bf 17} (2016) 1075--1108, {\tt
  arXiv:1502.02958} {\tt [hep-th]}.

\bibitem{Longhi:2021qvz}
P.~Longhi, {\em {On the BPS spectrum of 5d SU(2) super-Yang-Mills}}, {\tt
  arXiv:2101.01681} {\tt [hep-th]}.

\end{thebibliography}

\end{document}